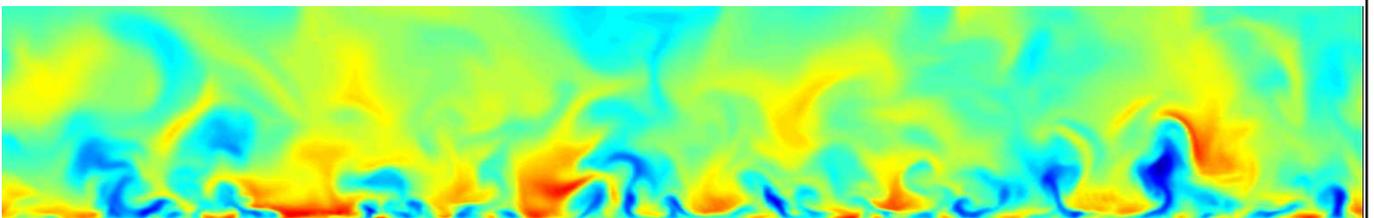

# Direct Numerical Simulation of Turbulent Open Channel Flow

Master's Thesis of

## Christian Bauer

at the Department of Civil Engineering
Institute for Hydromechanics (IfH)

Reviewer:            Prof. Dr. Markus Uhlmann
Second reviewer:  Prof. Dr.-Ing. Bettina Frohnapfel
Advisor:             M.Sc. Yoshiyuki Sakai

October 8, 2015

I herewith declare that I completed this document independently and did not use any other than the cited references and aids. I cited exact quotations and paraphrases as such, and I followed the rules of good scientific practice in the general regulations of the Karlsruhe Institute of Technology (KIT) (Regeln zur Sicherung guter wissenschaftlicher Praxis im Karlsruher Institut für Technologie (KIT)) in its most current edition. I did not submit this document to any other institution for the purpose of being awarded any other academic degree. **Karlsruhe, 12.10.2015**

. . . . . . . . . . . . . . . . . . . . . . . . . . . . . . . . . . . .
(Christian Bauer)

# Abstract

Direct numerical simulations of turbulent open channel flow with friction Reynolds numbers of $Re_\tau = 200, 400, 600$ are performed. Their results are compared with closed channel data in order to investigate the influence of the free surface on turbulent channel flows. The free surface effects fully developed turbulence statistics in so far that velocities and vorticities become highly anisotropic as they approach it. While the vorticity anisotropy layer scales in wall units, the velocity one is found to scale with outer flow units and exhibits much larger extension. Furthermore the influence of the free surface on coherent very large-scale motions (VLSM) is discussed through spectral statistics. It is found that the spanwise extension increases by a factor of two compared to closed channel flows independent of the Reynolds number. Finally instantaneous realizations of the flow fields are investigated in order to elucidate the turbulent mechanisms, which are responsible for the behaviour of a free surface in channel flows.

# Zusammenfassung

Eine offene Kanalströmung wird mit Hilfe der direkten numerischen Simulation für die auf der Schubspannungsgeschwindigkeit basierenden Reynoldszahlen $Re_\tau = 200, 400, 600$ berechnet. Die erzielten Ergebnisse werden anschließend mit Literaturergebnissen für geschlossene Kanalströmungen verglichen. Die Einführung der freien Oberfläche beeinflusst die voll ausgebildete turbulente Statistik insofern, dass Geschwindigkeits- und Wirbelstärkefluktuationen unterhalb der Oberfläche stark anisotrop werden. Während die Anisotropiegrenzschicht der Wirbelstärke in wandnahen Einheiten skaliert, wird für die Grenzschicht der Geschwindigkeiten eine Skalierung mit globalen Einheiten festgestellt. Desweiteren wird der Einfluss der freien Oberfläche auf sehr großskalige Bewegungen (engl. very large-scale motions, VLSM) mit Hilfe von spektralen Statistiken untersucht. Eine Verdopplung der Größe der VLSM quer zur Strömungsrichtung im Verhätnis zu Strukturen in geschlossenen Kanalströmungen kann unabhänigig von der Reynoldszahl in der offenen Kanalströmung nachgewiesen werden. Abschließend werden Momentaufnahmen von dreidimensionalen Strömungsfeldern untersucht, um die turbulenten Mechanismen unterhalb einer freien Kanaloberfläche zu verdeutlichen.

# Acknowledgments

Firstly, i wish to express my sincere thanks to Prof. Dr. Markus Uhlmann for providing me with all the necessary facilities to perform this research work. I am also grateful for the continuous support of my supervisor M.Sc. Yoshiyuki Sakai, who never became tired of answering my more or less stupid questions. Furthermore, i would like to thank all institute members in general and Tiago Pestana in specific for aiding me to not get lost on this turbulent journey through spectral space and vortex dynamics. The computational facilities provided by the Steinbuch Centre for Computing at the Karlsruhe Institut of Technology shall also be highly acknowledged. Finally, i express big gratitude to my family for their support throughout my entire study time.

# Contents

# Nomenclature

$\alpha$      lowest wave number w.r.t the streamwise direction

$\alpha_k$      Runge-Kutta coefficients

$\bar{\tau}$      integrated time scale

$\beta$      lowest wave number w.r.t the spanwise direction

$\boldsymbol{\kappa}$      wave number vector

$\boldsymbol{\omega}$      vorticity vector

$\Delta T$      total physical simulation time

$\Delta t$      time step

$\Delta t_{exec}$      execution time per time step

$\Delta x^+$      streamwise grid spacing in wall units

$\Delta y_c^+$      wall-normal grid spacing at the centerline in wall units

$\Delta y_{max}^+$      maximum wall-normal grid spacing in wall units

$\Delta z^+$      spanwise grid spacing in wall units

$\delta$      boundary layer thickness

$\delta_\nu$      viscous length scale

$\delta_\omega$      vorticity anisotropy layer

$\delta_{ij}$      Kronecker delta

$\delta_v$      velocity anisotropy layer

$\eta$      invariant of the Reynolds stress anisotropy tensor

$\kappa$      Von Kármán constant

$\kappa_i$      wave number w.r.t. the i-direction

$\lambda_i$      wavelength w.r.t. the i-direction

$\lambda_m$      mean free path

$\mathcal{L}$      characteristic length scale

$\mathcal{P}$      characteristic pressure scale

$\mathcal{P}$      production term in turbulent kinetic energy equation

$\mathcal{T}$      characteristic time scale

$\mathcal{U}$      characteristic velocity scale

$\nabla$      Nabla operator

$\nu$      kinematic viscosity

$\omega_i$      i-component of the vorticity

$\Omega_{ij}$      vorticity tensor

$\Phi_{ij}(\boldsymbol{\kappa}, t)$      velocity spectrum tensor

$\rho$      density

$\rho(s)$      auto-correlation coefficient

$\rho_{12}$      correlation coefficient between $u_1$ and $u_2$

$\tau_w$      wall shear stress

$\tau_{ij}$      viscous stress tensor

$\tau_{tot}$      total shear stress

$\mathbf{f}$      outer body forces

$\mathbf{R}$      Reynolds stress tensor

$\mathbf{r}$      separation vector

$\mathbf{u}$      velocity vector

$\varepsilon$      dissipation term in turbulent kinetic energy equation

$\xi$      invariant of the Reynolds stress anisotropy tensor

$B$      log-law constant

$b_{ij}$      Reynolds stress anisotropy tensor

$Fr$      Froude number

$h$      channel height

$H_i$      convective term of the i-component

$I$          imaginary number

$k$          turbulent kinetic energy

$Kn$        Knudsen number

$L_x$        streamwise extension of the numerical box

$L_y$        vertical extension of the numerical box

$L_z$        spanwise extension of the numerical box

$N_x$        number of modes in streamwise direction

$N_y$        number of modes in vertical direction

$N_z$        number of modes in spanwise direction

$n_{proc}$    number of processors

$n_{tot}$     number of time steps

$p$          pressure

$Q$          second invariant of the velocity gradient tensor

$R$          pipe radius

$R(s)$       auto-covariance

$R_{ij}(\mathbf{r}, \mathbf{x}, t)$  two-point velocity correlation

$Re$         Reynolds number

$Re_\tau$     friction Reynolds number

$S'_{ij}$     fluctuating rate-of-strain tensor

$S_{ij}$      rate-of-strain tensor

$t$          time

$T_q(y)$     Chebyshev polynomial

$t_{exec}$    total execution time (wall time)

$u$          streamwise velocity component

$u'_i$        i-component of fluctuating velocity

$u^+$        mean velocity in wall units

$u_b$        bulk velocity

$u_i$     i-component of the velocity

$u_\tau$     friction velocity

$u_{i,rms}$  i-component of the root mean square velocity fluctuation

$V$     sample space variable corresponding to $u$

$v$     vertical velocity component

$w$     spanwise velocity component

$We$     Weber number

$x_i$     cartesian coordinate of the i-direction

$y^+$     wall distance in wall units

CFL    Courant-Friedrich-Lewy number

# List of Figures

# List of Tables

# 1. Introduction

This thesis exhibits a direct numerical simulation of a turbulent open channel flow. For the deeper understanding we will introduce turbulence as a phenomenon first, followed by the description of direct numerical simulation (DNS) as a numerical tool for the solution of turbulent problems. Hereafter the flow characteristics in open channels and their differences with respect to closed channel flows shall be described. Finally this chapter will be concluded by the outline and goals of this work.

## 1.1. Turbulence

Turbulent flows can be observed in every day life in many different shapes. Whether it is smoke from a chimney, water current in a river flowing around cliffs, waterfalls or large scale weather phenomena such as typhoons.

Turbulence becomes especially important to engineering applications like pipe or channel flows or flows around vehicles such as air planes or auto mobiles. Also mixing, e.g. in chemical engineering processes exhibits turbulent flows.

The first to study turbulence extensively in pipe flows was Lord Reynolds in 1883, cf. Reynolds (1883). He injected dye on the centerline of a pipe water flow and later on discovered the dependence of the flow behaviour on a single non-dimensional parameter, which further should be denoted as the Reynolds number (Reynolds, 1894):

$$Re = \frac{\mathcal{U}\mathcal{L}}{\nu} \tag{1.1}$$

where $\mathcal{U}$ and $\mathcal{L}$ are characteristic velocity and length scale of the flow and $\nu$ is the kinematic viscosity. The Reynolds number can be interpreted as the ratio of inertial to viscous forces in a flow. By increasing the Reynolds number in the pipe flow experiment, the dye leaves its original shape of a straight line in the pipe centerline (laminar state) through a transition state and finally becomes completely mixed (turbulent state $Re \gtrsim 2000$ for pipe flows).

Since there is no clear consensus of a scientific definition of turbulence, it shall be characterized by exhibiting the following features, as done by Mathieu and Scott (2000):

- **Turbulence is a random process**
  Turbulent flow is space and time dependent with many degrees of freedom and in so far unpredictable in detail. Nevertheless statistics appear to be reproducible.

- **Turbulence contains a wide range of scales**
  The largest scales of turbulent motion are determined by the flow geometry, whereas the smallest scales, which will be denoted as Kolmogorov scales, after Kolmogorov (1941), depend on the viscosity. The range between largest and smallest scales rises when raising the Reynolds number.

- **Turbulence has small-scale random vorticity**
  High Reynolds number turbulent flows exhibit high intensive complex vortical structures varying both in space an time.

- **Turbulence arises at high Reynolds numbers**
  Turbulence occurs due to instability of a laminar flow. By increasing the Reynolds number the non-linear term of the governing equation becomes more important compared to the instability damping viscous term and the flow tends to become turbulent.

- **Turbulence dissipates energy**
  Due to locally large velocity gradients in turbulent flows as a consequence of small-scale motions, the rate of dissipation of kinetic energy into heat, which is proportional to the velocity gradients, is large in turbulent flows in comparison with laminar ones. Statistical stationary turbulence requires continuous supply of energy to large-scale motion, which is then successively transferred to the smallest scales where it dissipates. The latter is known as the turbulent energy cascade introduced by Richardson (2007).

- **Turbulence is a continuum phenomenon**
  Smallest scales of turbulent motion, which are dictated by viscosity, shall be some order of magnitude larger than the molecular mean free path, which is true for flows considered in this thesis, cf. 2.2.1.

## 1.2. Direct Numerical Simulation

With the upcoming chaos theory in the 1970s a new field of turbulence research emerged. Navier-Stokes equations, which are the governing equations for motions of fluids (cf. section 2.2), are amongst others example of non-linear equations exhibiting strong sensitivity with respect to small variations in the initial conditions. Infinitesimal change in the initial conditions can lead to a striking different solution, which is denoted as chaotic behaviour. A good description of the emergence of chaos in non-linear dynamic systems can be found in Jiménez (2004).

Since the full Navier-Stokes equations without radical simplification can not be solved analytically, numerical techniques have to be required. In case of fluid mechanics they are subsumed under the field of computational fluid dynamics (CFD). For turbulent flows the most common approaches are the direct numerical simulation (DNS), large eddy simulation (LES) and Reynolds averaged Navier Stokes (RANS). While the latter two solve either only a part of the turbulent scales (LES) or the time averaged equations introducing new unknown variables in form of correlation terms (RANS), turbulence modelling is required. DNS, on the contrary, solve the Navier-Stokes equations resolving the whole range of spatial and temporal scales.

Due to the large scale separation in turbulent flows the latter technique, which shall be used in this work, is extremely costly and its computational requirements increase with the Reynolds number, as Pope (2000) describes. Since the simulation data will be treated statistically afterwards, the equations have to be integrated over a substantial time interval in order to receive converged statistics. The numerical methodology will be described in chapter 3 in detail.

## 1.3. Open Channel Flow

Turbulent plane channel flows with periodic boundaries in stream- and spanwise direction and impermeable no-slip walls at the bottom and top, which will henceforth be denoted as closed channel flows, have been extensively studied over the last decades both in experiments and numerical simulations. Those studies were particularly aiming at the better understanding of wall-bounded turbulent flows.

First experiments with fully developed turbulent channel flows have been carried out by Nikuradse (1929). While Nikuradse's work was limited to the mean flow, Reichardt (1938) was the first to report turbulent velocity fluctuations. Besides numerous experimental studies involving turbulence statistics, such as Laufer (1948), Comte-Bellot (1963), Clark (1968) or Hussain and Reynolds (1975), a large number of numerical investigations has been performed over the last decades.

Moin and Kim (1982) performed a LES involving an eddy viscosity model for the smallest turbulent scales and reproduced turbulent wall-layer structures as they have been reported by experimental investigations. With rising computational power it became possible to resolve all spatial and temporal scales in fully developed turbulent channel flows at least in the marginal Reynolds number regime and DNS, such as Kim, Moin, et al. (1987), have been performed. The numerical technique of the current work is based on the latter study. A more recent study by Moser et al. (1999), based on the same technique, shall be the reference case for this work. Numerous studies have been reported by the group of Professor Jimenez from UPM, Madrid, such as Jiménez and Moin (1991), Jiménez and Pinelli (1999), Hoyas and Jiménez (2006) or Jiménez (2012), aiming at the better understanding of turbulent near-wall mechanisms, as the regeneration cycle of near-wall velocity streaks, and involving Reynolds numbers, based on the friction velocity and channel half-height, up to $Re_\tau$ = 2000 (cf. equation 2.15). Since near-wall dynamics are greatly understood by now, the interest of the community shifts more towards logarithmic layer dynamics, involving large-scale coherent structure investigation, such as Jiménez (1998), Del Álamo and Jiménez (2003) or Del Álamo, Jiménez, et al. (2004).

Less attention than on the above mentioned closed channel flows has been paid on open channel flows, where one of the no-slip walls is replaced by a free-slip boundary condition. Nevertheless is the slightly different configuration of practical relevance to a number of civil engineering applications, such as river engineering. An experimental study of fully developed turbulent open channel flows involving Reynolds numbers in the range of $439 \leq Re_\tau \leq 6139$ has been carried out by Rodi and Asce (1986). The few direct numerical simulations performed on open channel flow exhibit low Reynolds numbers ($Re_\tau < 200$) and marginal spatial resolutions (Swean et al. (1991), Lam and Banerjee (1992), Handler et al. (1993), Komori et al. (1993), Pan and Banerjee (1995) or Nagaosa (1999)). Due to the lack of a high-resolution DNS database for open channel flows, closed channel data is often taken as a numerical reference instead, which is feasible in the vicinity of the wall. By approaching the free surface, in contrary, turbulent open channel flows differ significantly from closed channel ones, since Reynolds stresses become highly anisotropic. Furthermore, open channel flows are believed to exhibit different behaviour in terms of large-scale coherent motion, where numerical data in the regime of higher than marginal Reynolds numbers is still to be required.

## 1.4. Goals and Outline

The goal of this study is to investigate the turbulent coherent structures near the free surface of an open channel flow by means of a direct numerical simulation. Therefore different Reynolds number cases with $Re_\tau = [200, 400, 600]$ shall be set up, corresponding to Kim, Moin, et al. (1987) and Moser et al. (1999) for closed channel flow, in computational domains which are sufficiently large (cf. table 4.2). The obtained turbulent statistics will be compared with the latter mentioned closed channel reference data. Furthermore shall the structure of coherent large scale motions of wall-bounded turbulent flows and their Reynolds number dependence be discussed and compared with existing closed and open channel data, having a special emphasis on the influence of the free-slip surface. Besides statistics, turbulent mechanisms appearing in the vicinity of the free surface shall be elucidated by instantaneous flow visualizations.

Chapter 2 of this work introduces the flow geometry first as well as deriving the governing equations for turbulent open channel flows and introducing basic turbulent statistics. We then move on to the numerical methodology of this work in chapter 3 before the results are going to be discussed in chapter 4. Outcome and further researches are presented in chapter 5.

# 2. Fundamentals

In this chapter the governing equations necessary to describe the flow will be derived. Furthermore the statistical and spectral mathematics needed for the computation and analysis of a turbulence flow will be introduced.

## 2.1. Flow Geometry

The flow geometry is described by an open channel with periodic boundaries in stream- and spanwise directions, since we assume the flow to be homogeneous in these directions. There is a smooth no-slip wall at the bottom and a free-slip boundary condition applied at the top of the channel. The latter is an approximation of a natural liquid-gas interface which resists surface deformation due to the free shear condition. In case of negligible surface tension the assumption holds if the Froude number is sufficiently small, such that gravity effectively suppresses any significant deformation of the surface. Froude and Weber number are non-dimensional flow parameters, which describe the ratio between inertial and gravitational forces or inertial forces and surface tension respectively. The open channel simulation of Komori et al. (1993) allowed small surface deformations yielding the same results as similar marginal Reynolds number open channel simulations with flat surfaces such as Lam and Banerjee (1992) or Handler et al. (1993).

The channel height is characterized by the parameter $h$, its extension in streamwise direction by $L_x$ and in spanwise direction by $L_z$ accordingly. Figure 2.1 shows both open and closed channel geometries. The results of the current work will be compared with closed channel simulations performed by Moser et al. (1999).

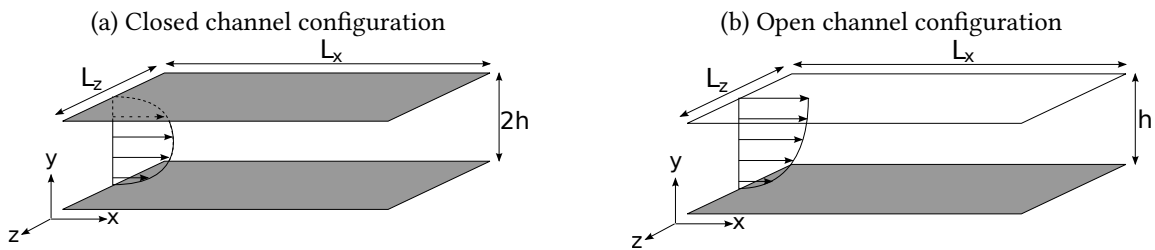

Figure 2.1.: Closed and open channel geometries.

In case of a gas-liquid interface such as air-water at the free surface, the reciprocal of the Weber number and the Froude number computed with the bulk velocity (cf. equation

3.17) yield:

$$\frac{1}{We} = \frac{\sigma}{\rho u_b^2 h} \approx 4.4 \cdot 10^{-5} \ll 1 \tag{2.1}$$

$$Fr = \frac{u_b^2}{gh} \approx 0.04 \ll 1 \tag{2.2}$$

As the values indicate, the aforementioned conditions of negligible surface tension ($1/We \ll 1$) and a sufficient low Froude number are fulfilled so that the flat free surface condition is applicable.

## 2.2. Equations of Fluid Motion

### 2.2.1. Assumptions

First, we will have a look at the assumptions that can be made for the present flow case.

1. Having the Knudsen number $Kn = \frac{\lambda_m}{\mathcal{L}} \ll 1$, with the mean free path $\lambda_m$ as a microscopic and the characteristic length $\mathcal{L}$ as a macroscopic scale, the flow of interest can be treated as a continuum.

2. The flow of interest in this work will also be incompressible, which means that density stays constant along the path of a fluid particle.

3. The fluid will be considered as a Newtonian fluid having the feature that its strain rate will be linear proportional to viscous stress, which is true for most of the fluids in technical applications. The Newtonian ansatz is

$$\tau_{ij} = \rho \nu \left( \frac{\partial u_i}{\partial x_j} + \frac{\partial u_j}{\partial x_i} \right) \tag{2.3}$$

having $\tau_{ij}$, the viscous stress tensor and $\nu$, the kinematic viscosity.

### 2.2.2. Governing Equations

Considering the conservation of mass and momentum we receive the governing equations as they are derived e.g. in Pope (2000).

$$\frac{\partial \rho}{\partial t} + \nabla \cdot (\rho \mathbf{u}) = 0 \tag{2.4a}$$

$$\frac{\partial \rho \mathbf{u}}{\partial t} + \nabla (\rho (\mathbf{u} \otimes \mathbf{u})) = -\nabla p + \nabla \tau + \rho \mathbf{f} \tag{2.4b}$$

with the velocity $\mathbf{u} = (u, v, w)$, density $\rho$, pressure $p$ and outer body forces $\mathbf{f}$ (e.g. gravitation). Applying the aforementioned assumptions on the equations, we receive the transport equations for mass and momentum for an incompressible, continuous flow of a Newtonian fluid.

$$\nabla \cdot \mathbf{u} = 0 \tag{2.5a}$$

$$\frac{\partial \mathbf{u}}{\partial t} + (\mathbf{u} \cdot \nabla)\mathbf{u} = -\frac{1}{\rho}\nabla p + \nu\nabla^2\mathbf{u} + \mathbf{f} \tag{2.5b}$$

For a constant density (i.e. no thermal effects) it is possible to absorb the gravity into a modified pressure variable, see Kundu and Cohen (2008).

$$\nabla \cdot \mathbf{u} = 0 \tag{2.6a}$$

$$\frac{\partial \mathbf{u}}{\partial t} + (\mathbf{u} \cdot \nabla)\mathbf{u} = -\frac{1}{\rho}\nabla p + \nu\nabla^2\mathbf{u} \tag{2.6b}$$

The equations can as well be expressed in indicial notation using the Einstein summation convention.

$$\frac{\partial u_i}{\partial x_i} = 0 \tag{2.7a}$$

$$\frac{\partial u_i}{\partial t} + u_j\frac{\partial u_i}{\partial x_j} = -\frac{1}{\rho}\frac{\partial p}{\partial x_i} + \nu\frac{\partial^2 u_i}{\partial x_j^2}, \ i = 1,2,3 \tag{2.7b}$$

Furthermore, the equations will be non-dimensionalized. In order to do that a characteristic length scale $\mathcal{L}$ and a characteristic velocity scale $\mathcal{U}$ is to be chosen. The reference values for time and pressure will then be $\mathcal{T} = \mathcal{L}/\mathcal{U}$ and $\mathcal{P} = \rho\mathcal{U}^2$. By multiplying equation 2.7a with $\mathcal{L}/\mathcal{U}$ and equation 2.7b with $\mathcal{L}/\mathcal{U}^2$ we receive the non-dimensional equations

$$\frac{\partial \tilde{u}_i}{\partial \tilde{x}_i} = 0 \tag{2.8a}$$

$$\frac{\partial \tilde{u}_i}{\partial \tilde{t}} + \tilde{u}_j\frac{\partial \tilde{u}_i}{\partial \tilde{x}_j} = -\frac{\partial \tilde{p}}{\partial \tilde{x}_i} + \frac{1}{Re}\frac{\partial^2 \tilde{u}_i}{\partial \tilde{x}_j^2}, \ i = 1,2,3 \tag{2.8b}$$

with the Reynolds number $Re = \mathcal{L}\mathcal{U}/\nu$ as a dimensionless parameter and the non-dimensional variables

$$\tilde{\mathbf{u}} = \frac{\mathbf{u}}{\mathcal{U}}, \ \tilde{p} = \frac{p}{\mathcal{P}}, \ \tilde{\mathbf{x}} = \frac{\mathbf{x}}{\mathcal{L}}, \ \tilde{t} = \frac{t}{\mathcal{T}}.$$

Henceforth, we will drop the tilde for dimensionless parameters and impose transport equations to be dimensionless. Now, vorticity shall be introduced as $\boldsymbol{\omega} = \nabla \times \mathbf{u} = (\omega_x, \omega_y, \omega_z)$. Following the approach of Kim, Moin, et al. (1987) equations 2.8a and 2.8b can be reduced to a fourth-order equation for the vertical velocity component $v$ and a second-order equation for the vertical component of vorticity $\omega_y$ respectively. In this formulation the pressure term has been eliminated and only two transport equations have to be solved instead of four, which reduces storage requirements compared to the primitive-variables formulation (Bhaganagar et al., 2002).

$$\frac{\partial \omega_y}{\partial t} = h_\omega + \frac{1}{Re}\nabla^2 \omega_y \tag{2.9a}$$

$$\frac{\partial \nabla^2 v}{\partial t} = h_v + \frac{1}{Re}\nabla^4 v \tag{2.9b}$$

$$q + \frac{\partial v}{\partial y} = 0 \tag{2.9c}$$

where

$$q = \frac{\partial u}{\partial x} + \frac{\partial w}{\partial z}, \ \omega_y = \frac{\partial u}{\partial z} - \frac{\partial w}{\partial x},$$

$$h_\omega = \left(\frac{\partial H_1}{\partial z} - \frac{\partial H_3}{\partial x}\right),$$

$$h_v = -\frac{\partial}{\partial y}\left(\frac{\partial H_1}{\partial x} + \frac{\partial H_3}{\partial z}\right) + \left(\frac{\partial^2}{\partial x^2} + \frac{\partial^2}{\partial z^2}\right)H_2,$$

$$H_i = -u_j \frac{\partial u_i}{\partial u_j}.$$

The following boundary conditions need to be provided: A no-slip condition at the bottom wall

$$u(x, y = 0, z) = 0, \ w(x, y = 0, z) = 0, \tag{2.10}$$

a free-slip condition at the free surface

$$\left(\frac{\partial u}{\partial y}\right)_{x, y=h, z} = 0, \ \left(\frac{\partial w}{\partial y}\right)_{x, y=h, z} = 0, \tag{2.11}$$

and impermeability conditions both at the wall and the free surface

$$v(x, y = 0, z) = v(x, y = h, z) = 0. \tag{2.12}$$

## 2.3. Viscous Scales

In this section some characteristic scales of the near-wall region shall be defined according to Schlichting and Gersten (2000). The characteristic velocity scale of the near-wall region, also known as friction velocity, reads as follows

$$u_\tau = \sqrt{\frac{\tau_w}{\rho}}, \tag{2.13}$$

where $\tau_w = \rho \nu \left|\frac{\partial \langle u \rangle}{\partial y}\right|_{y=0}$ is the wall shear-stress. Furthermore, the viscous length scale is defined as

$$\delta_\nu = \frac{\nu}{u_\tau}. \tag{2.14}$$

The Reynolds number based on $u_\tau$ and $h$ shall be denoted as friction Reynolds number and expresses the ratio between the outer flow length scale $h$ and the viscous length scale $\delta_v$:

$$Re_\tau = \frac{u_\tau h}{\nu} = \frac{h}{\delta_v}. \tag{2.15}$$

Quantities normalized with viscous scales, hereinafter also referred to as "in wall units", will be denoted with the superscript "+". Accordingly the wall distance and the mean velocity in wall units are defined as

$$y^+ = \frac{y}{\delta_v}, \tag{2.16}$$

$$u^+ = \frac{\langle u \rangle}{u_\tau}. \tag{2.17}$$

## 2.4. Statistics

Since turbulent flows exhibit a high sensitivity to small variations of the initial conditions, as described in chapter 1, turbulent quantities have to be treated statistically. Some basic statistical tools shall be described in this section. Further reading on the statistical description of turbulent flows can be found in Pope (2000). First, a random variable $u$, as a representative of a turbulent quantity whose value is unpredictable, shall be introduced. Note that the fact that turbulence quantities are treated as random variables does not mean turbulence is a random phenomenon itself. It is rather related to the unpredictability of turbulent quantities, which is caused by the unavoidable appearance of perturbations in initial- or boundary conditions or material properties respectively together with the sensitivity of non-linear equations to these perturbations. The latter makes statistical treatment necessary, even though the Navier-Stokes equations are deterministic themselves. However, in addition to the random variable the sample space variable $V$, characterizing the range of values possibly taken by $u$, shall be introduced. Considering an event $B$, that the random variable $u$ has a smaller value than one specific value $V_b$, viz. $B = \{u < B_b\}$ leads to an expression for the probability that the event $B$ is true:

$$p = P(B) = P\{u < V_b\}. \tag{2.18}$$

with $0 \leq p \leq 1$. The probability of any event within the sample space is determined by the cumulative distribution function (CDF) defined as

$$F(V) = P\{u < V\} \tag{2.19}$$

whose derivative is denoted as the probability density function (PDF), viz.

$$f(V) = \frac{dF(V)}{dV}. \tag{2.20}$$

The mean of the random variable shall be defined as

$$\langle u \rangle = \int_{-\infty}^{\infty} V f(V) dV \tag{2.21}$$

or more generally the mean of a function $Q(u)$

$$\langle Q(u) \rangle = \int_{-\infty}^{\infty} Q(V) f(V) dV. \tag{2.22}$$

Furthermore, we define the fluctuation of $u$ with respect to the mean as

$$u' = u - \langle u \rangle \tag{2.23}$$

and the variance to be the mean square fluctuation

$$var(u) = \langle (u')^2 \rangle = \int_{-\infty}^{\infty} (V - \langle u \rangle)^2 f(V) dV \tag{2.24}$$

whose square root is the standard deviation $\sigma_u$, which is also referred to as the root mean square (r.m.s.) of $u$:

$$u_{rms} = \sigma_u = \sqrt{var(u)} = \sqrt{\langle u' u' \rangle}. \tag{2.25}$$

For problems involving more random variables, which are not necessarily stochastically independent, joint probability functions (JPDF) need to be defined, e.g. for two random variables $u_1, u_2$ as follows

$$f_{12}(V_1, V_2) = \frac{\partial^2}{\partial V_1 \partial V_2} F_{12}(V_1, V_2) \tag{2.26}$$

where $F_{12}(V_1, V_2) = P\{u_1 < V_1, u_2 < V_2\}$ is the CDF of the two random variables. The mean of any function $Q$ of the random variable reads as

$$\langle Q(u_1, u_2) \rangle = \int_{-\infty}^{\infty} \int_{-\infty}^{\infty} Q(V_1, V_2)_{12} f(V_1, V_2) dV_1 dV_2 \tag{2.27}$$

and means or variances can be determined from this equation. One important quantity is the covariance of $u_1$ and $u_2$

$$cov(u_1, u_2) = \langle u_1' u_2' \rangle = \int_{-\infty}^{\infty} \int_{-\infty}^{\infty} (V_1 - \langle u_1 \rangle)(V_2 - \langle u_2 \rangle) f(V_1, V_2) dV_1 dV_2 \tag{2.28}$$

or the correlation coefficient respectively

$$\rho_{12} = \frac{\langle u_1' u_2' \rangle}{\sqrt{\langle u_1' u_1' \rangle \langle u_2' u_2' \rangle}}, \tag{2.29}$$

as a measure for the statistical dependency between two random variables. When considering time-dependent random variables, so-called random processes the CDF and PDF get extended by the dimension of time. Information about temporal correlation of the random variables can then be extracted from $N$-time joint CDF $F_N$ or PDF $f_N$ respectively, including information at $N$ different time points. However, in case of statistically stationary random processes $F_N$ and $f_N$ are time invariant, meaning

$$f_N(V_1, t_1 + T, V_2, t_2 + T, \ldots) = f_N(V_1, t_1, V_2, t_2, \ldots) \tag{2.30}$$

and one can define the auto-covariance

$$R(s) = \langle u'(t)u'(t+s) \rangle \tag{2.31}$$

or in normalized form respectively, the auto-correlation coefficient

$$\rho(s) = \frac{R(s)}{u'(t)^2}. \tag{2.32}$$

The auto-correlation coefficient is expected to drop off to zero for large time separations. The integral time scale, defined as

$$\bar{\tau} = \int_0^\infty \rho(s)ds \tag{2.33}$$

is a measure of how fast the process decorrelates with respect to the time. Apart from time-dependency turbulent flows are described as three-dimensional vector fields, meaning that statistical quantities generally depend on their spatial position. The one-point, one-time PDF of a random variable $u$ would then be

$$f(V, \mathbf{x}, t) = \frac{\partial F(V, \mathbf{x}, t)}{\partial V} \tag{2.34}$$

where $F(V, \mathbf{x}, t) = P\{u(\mathbf{x}, t) < V\}$ is the one-time, one-point CDF. In case of statistical homogeneity the statistics are invariant to a shift in space yielding for a N-point PDF:

$$f_N(V, \mathbf{x}_1 + \mathbf{X}, \mathbf{x}_2 + \mathbf{X}, \ldots) = f_N(V, \mathbf{x}_1, \mathbf{x}_2, \ldots). \tag{2.35}$$

Since the PDF is unknown in practice one can obtain mean quantities by averaging over a number of samples. With a sufficient high number of samples the average then approaches the value of the probabilistic average. In the present case of a fully developed open channel flow, including homogeneity with respect to both stream- and spanwise directions, one can take advantage of the statistical stationarity and homogeneity by averaging over both time and two-dimensional space. Henceforth, the angle brackets shall indicate an average as follows:

$$\langle u(y) \rangle = \frac{1}{L_x L_z \Delta t} \int_{t=t_0}^{t_0 + \Delta t} \int_{z=0}^{L_z} \int_{x=0}^{L_x} u(x, y, z, t) dx dz dt. \tag{2.36}$$

The two-point, one-time auto-covariance, also referred to as the two-point correlation, of a random field is given by

$$R_{ij}(\mathbf{r}, \mathbf{x}, t) = \langle u_i'(\mathbf{x}, t)u_j'(\mathbf{x} + \mathbf{r}, t) \rangle \tag{2.37}$$

where $\mathbf{r}$ is the separation vector between the two points. In case of statistical stationarity the dependence on $t$ and statistical homogeneity the one on $\mathbf{x}$ vanishes.

## 2.5. Wave Number Spectra

Having spatially homogeneous turbulence the flow data can be expressed in Fourier space representation. The Fourier transform of the two-point velocity covariance $R_{ij}$ yields

$$\Phi_{ij}(\boldsymbol{\kappa}, t) = \frac{1}{(2\pi)^3} \int_{-\infty}^{\infty} e^{-I\boldsymbol{\kappa}\cdot\mathbf{r}} R_{ij}(\mathbf{r}, t) d\mathbf{r}, \tag{2.38}$$

and shall be denoted as the velocity spectrum tensor with $\boldsymbol{\kappa} = (\kappa_1, \kappa_2, \kappa_3)$ the wave number vector, whose components $\kappa_i = 2\pi/\lambda_i$ are related to the wavelengths of the Fourier mode $\lambda_i$, and $I = \sqrt{-1}$. The inverse transform of velocity spectrum tensor yields

$$R_{ij}(\mathbf{r}, t) = \int_{-\infty}^{\infty} e^{I\boldsymbol{\kappa}\cdot\mathbf{r}} \Phi_{ij}(\boldsymbol{\kappa}, t) d\boldsymbol{\kappa}. \tag{2.39}$$

By setting $\mathbf{r}$=0 and using statistical stationarity and homogeneity again one can obtain an expression for one-dimensional wave number spectra with respect to either stream- or spanwise direction as follows

$$R_{ii}(0) = \langle u_i' u_i' \rangle = \int_{\kappa_x=0}^{\infty} \Phi_{ii}(\kappa_x) d\kappa_x = \int_{\kappa_z=0}^{\infty} \Phi_{ii}(\kappa_z) d\kappa_z. \tag{2.40}$$

Analogously one can derive expressions for so-called premultiplied one-dimensional wavelength spectra:

$$R_{ii}(0) = \langle u_i' u_i' \rangle = \int_{\log \lambda_x=0}^{\infty} \kappa_x \Phi_{ii}(\kappa_x) d\log \lambda_x = \int_{\log \lambda_z=0}^{\infty} \kappa_z \Phi_{ii}(\kappa_z) d\log \lambda_z \tag{2.41}$$

where $\lambda_x = 2\pi/\kappa_x$ is the stream- and $\lambda_z = 2\pi/\kappa_z$ the spanwise wavelength.

# 3. Methodology

In this chapter the methodology will be explained in terms of the numerical approach that is made in order to solve the afore derived wall normal velocity/vorticity formulation of the Navier-Stokes equations.

## 3.1. Numerical Methodology

In order to solve the flow equations 2.9a and 2.9b a spectral method, such as used by Kim, Moin, et al. (1987) for spatial discretization is applied where the flow variables are expanded in terms of Fourier Series in the homogeneous stream- and spanwise directions and Chebychev polynomials in wall normal direction.

The flow domain contains $N_x \times Ny \times N_z$ grid points. According to the spectral method there is equidistant grid spacing in the homogeneous directions:

$$x_i = i\frac{L_x}{N_x} \quad \forall \quad 0 \leq i \leq N_x - 1, \qquad z_k = k\frac{L_z}{N_z} \quad \forall \quad 0 \leq k \leq N_z - 1 \qquad (3.1)$$

while the grid points in wall normal direction are ordered by Chebychev-Gauss-Lobatto (CGL) points:

$$y_j = cos\left(\frac{\pi j}{N_y}\right) \quad \forall \quad 0 \leq j \leq N_y. \qquad (3.2)$$

In terms of time evolution an Euler implicit scheme for the viscous terms and third order low storage Runge-Kutta scheme for the non-linear terms is used, yielding:

$$\frac{\omega_y^{n+k/3} - \omega_y^n}{\Delta t \alpha_k} - \frac{1}{Re}\nabla^2(\omega_y^{n+k/3}) = h_\omega^{n+(k-1)/3} \qquad (3.3a)$$

$$\frac{(\nabla^2 v)^{n+k/3}}{\Delta t \alpha_k} - \frac{1}{Re}\nabla^2((\nabla^2 v)^{n+k/3}) = h_v^{n+(k-1)/3} \qquad (3.3b)$$

where k=1,2,3 are the Runge-Kutta steps with corresponding coefficients

$$\alpha_k = \left\{\frac{1}{3}, \frac{1}{2}, 1\right\}.$$

After being Fourier transformed with respect to the homogeneous directions, equation 3.3 is solved wavenumberwise using a Chebychev collocation method. Fast-Fourier-Transforms with de-aliasing according to the 2/3 rule are used (Uhlmann, 2000c). The method is in so far pseudo-spectral as the non-linear terms are calculated in physical space at each collocation point. The expansion of the problem variables $\phi = \{\omega_y, \nabla^2 v\}$ into spectral space in the periodic directions yields:

$$\phi(x,y,z) = \sum_{l=0}^{N_x} \sum_{m=0}^{N_z} \hat{\phi}_{lm}(y) \cdot e^{Il\alpha x} \cdot e^{Im\beta z} \tag{3.4}$$

where

$$l\alpha = \kappa_x, \ \alpha = 2\pi/L_x, \ l = 0,1,2..N_x \quad m\beta = \kappa_z, \ \beta = 2\pi/L_z, \ m = 0,1,2..N_z \tag{3.5}$$

and $I = \sqrt{-1}$ the imaginary number. The spatial derivatives of the problem variables in spectral space read as follows

$$\frac{\partial \phi(x,y,z)}{\partial x} = \sum_{l=0}^{N_x} \sum_{m=0}^{N_z} \hat{\phi}_{lm}(y) \cdot Il\alpha \cdot e^{Il\alpha x} \cdot e^{Im\beta z} \tag{3.6a}$$

$$\frac{\partial \phi(x,y,z)}{\partial z} = \sum_{l=0}^{N_x} \sum_{m=0}^{N_z} \hat{\phi}_{lm}(y) \cdot Im\beta \cdot e^{Il\alpha x} \cdot e^{Im\beta z} \tag{3.6b}$$

$$\frac{\partial \phi(x,y,z)}{\partial y} = \sum_{l=0}^{N_x} \sum_{m=0}^{N_z} \frac{d\hat{\phi}_{lm}(y)}{dy} \cdot e^{Il\alpha x} \cdot e^{Im\beta z} \ . \tag{3.6c}$$

In order to compute the convective terms all velocity and vorticity components have to be extracted from the problem variables first. This is done by the following relations for the Fourier coefficients, which are derived in A.3:

$$\left. \begin{aligned} \hat{u} &= \ I\frac{\kappa_x \frac{d\hat{v}}{dy} - \kappa_z \hat{\omega}_y}{\kappa_x^2 + \kappa_z^2} \\ \hat{w} &= \ I\frac{\kappa_z \frac{d\hat{v}}{dy} + \kappa_x \hat{\omega}_y}{\kappa_x^2 + \kappa_z^2} \\ \hat{\omega}_x &= \ I\frac{\kappa_z \hat{\phi} - \kappa_x \frac{d\hat{\omega}_y}{dy}}{\kappa_x^2 + \kappa_z^2} \\ \hat{\omega}_z &= \ -I\frac{\kappa_x \hat{\phi} + \kappa_z \frac{d\hat{\omega}_y}{dy}}{\kappa_x^2 + \kappa_z^2} \end{aligned} \right\} \ \forall \{\kappa_x, \kappa_z\}/\{\kappa_x = 0 \cap \kappa_z = 0\} \ . \tag{3.7}$$

For the constant mode ($\kappa_x = 0 \cap \kappa_z = 0$) the governing equations can be solved directly. In addition to the expansion in terms of truncated Fourier series with respect to the stream- and spanwise directions, the flow variables are expanded in terms of Chebyshev polynomials with respect to the wall-normal direction, receiving for the Fourier coefficients

$$\hat{\phi}_{lm}(y) = \sum_{q=0}^{N_y} a_q T_q(y), \quad (l,m) \neq (0,0) \tag{3.8}$$

where

$$T_q(y) = \cos(q \arccos(y)) \tag{3.9}$$

is a Chebyshev polynomial of the first kind, which reads in a Chebyshev-Gauss-Lobatto grid, such as defined in equation 3.2,

$$T_q(y_j) = \cos\left(q\frac{\pi j}{N_y}\right) \ .$$ (3.10)

The Fourier coefficients shall now be defined as

$$\hat{\phi}_{lmj} = \hat{\phi}_{lm}(y_j), \quad j = 0, 1, ..., N_y$$ (3.11)

with the $p$th derivative

$$\hat{\phi}_{lmj}^{(p)} = \left(\frac{d^p\hat{\phi}_{lm}(y)}{dy^p}\right)_{y=y_j} = \sum_{q=0}^{N_y} a_q^{(p)} T_q(y_j),$$ (3.12)

which can according to Canuto et al. (2006) be efficiently computed by using the following recursion relation:

$$2qa_q^{(p-1)} = c_{q-1}a_{q-1}^{(p)} - a_{q+1}^{(p)},$$ (3.13)

where

$$c_q = \begin{cases} 0, & q < 0 \ or \ q > N_y \\ 2, & q = 0 \\ 1, & \text{otherwise.} \end{cases}$$ (3.14)

Inserting equation 3.4 into equations 3.3a and 3.3b leads to the following expressions for each pair of modes $\kappa_x, \kappa_z$:

$$\left(\frac{d^2\hat{\phi}_{lm}^{(n+k/3)}(y)}{dy^2}\right)_{y=y_j} - \left(\kappa_x^2 + \kappa_z^2 + \frac{Re}{\alpha_k\Delta t}\right)\hat{\phi}_{lmj}^{(n+k/3)} = -\frac{Re}{\alpha_k\Delta t}\hat{\phi}_{lmj}^{(n)} - Re\cdot\hat{h}_{\hat{\phi}_{lmj}}^{(n+(k-1)/3)}$$ (3.15)

where

$$j = 0, 1, ..., N_y, \quad \phi = \{\omega_y, \nabla^2 v\} \quad \text{and } k = 1, 2, 3 \ .$$

Additionally a second order equation for $v$ has to be solved:

$$\left(\frac{d^2\hat{v}_{lm}^{(n+k/3)}(y)}{dy^2}\right)_{y=y_j} - \left(\kappa_x^2 + \kappa_z^2\right)\hat{v}_{lmj}^{(n+k/3)} = \nabla^2\hat{v}_{lmj}^{(n+k/3)}$$ (3.16)

The latter expressions will be solved using a Fortran code, which was provided by Professor Javier Jimenez and Professor Alfredo Pinelli of ETSI Aeronauticos, Universidad Politécnica, Madrid and further developed by Professor Markus Uhlmann, IfH, Karlsruhe Institut for Technology. The code imposes a constant mass flow rate and the timestep is adjusted to maintain a constant Courant-Friedrich-Lewy (CFL) number of 0.5.

## 3.2. Initial Conditions and Time Evolution

The initial flow fields for the present cases were obtained from an existing instantaneous flow field of a fully developed turbulent open channel flow with lower resolution, which has then been interpolated, by means of Chebyshev polynomials, to the number of grid points needed. Data from the smaller field was copied into the lower wave numbers before filling up the remaining Fourier coefficients with zeroes, as described in Uhlmann (2000a). The original flow field had been obtained from an initially laminar Poiseulle profile, where the most unstable two-dimensional eigenmode, obtained from a linear stability analysis, together with small perturbations had been added (Uhlmann, 2000b). The governing equations with the latter field as initial condition were then integrated in time until a statistically steady state was reached.

By changing the number of grid points, the box size or the Reynolds number for a given fully developed flow field, an initial transient phase of instationarity occurs before the flow reaches statistically stationary state again. The latter is monitored by the time evolution of the friction Reynolds number $Re_\tau$. Once the statistically stationary state is obtained, statistics are generated by averaging the instantaneous realizations of the flow field in time as well as in stream- and spanwise directions. The convergence of the statistics are indicated by a linear profile of the averaged total shear stress (cf. figure 4.9). In order to receive converged statistics the lower Reynolds number cases, which contain less grid points, have to be integrated over longer time period compared to the higher ones due to their lower amount of stream- and spanwise spatial samples. Time evolution series of the friction Reynolds number for the different cases of this work (cf. table 4.2) are shown in Figure 3.1 having the time dimensionless in bulk time units $t^* = t\frac{u_b}{h}$, where $u_b$ is the bulk velocity

$$u_b = \frac{1}{h}\int_{y=0}^{h}\langle u\rangle dy. \tag{3.17}$$

Note that the different Reynolds number flows are obtained by adjusting the kinematic viscosity $\nu$. The MPI-based computations were carried out on the ForHLR1 of Steinbuch Centre for Computing, Karlsruhe, using up to 160 processors in parallel as shown in table 3.1.

| case | $n_{proc}$ | $t_{exec}$ | $n_{tot}$ | $\Delta t_{exec}$ | $\Delta T u_b/h$ |
|------|-----------|-----------|----------|-------------------|------------------|
| *oc*200 | 20 | 50h | 300000 | 0.6s | 2582 |
| *oc*400 | 40 | 128h | 200000 | 2.3s | 868 |
| *oc*600 | 160 | 285h | 97500 | 10.5s | 273 |

Table 3.1.: Computational parameters for the present open channel flow simulation. $n_{proc}$: number of processors; $t_{exec}$: total execution time (wall time); $n_{tot}$: number of time steps; $\Delta t_{exec}$: execution time per time step; $\Delta T u_b/h$: total physical time in bulk time units

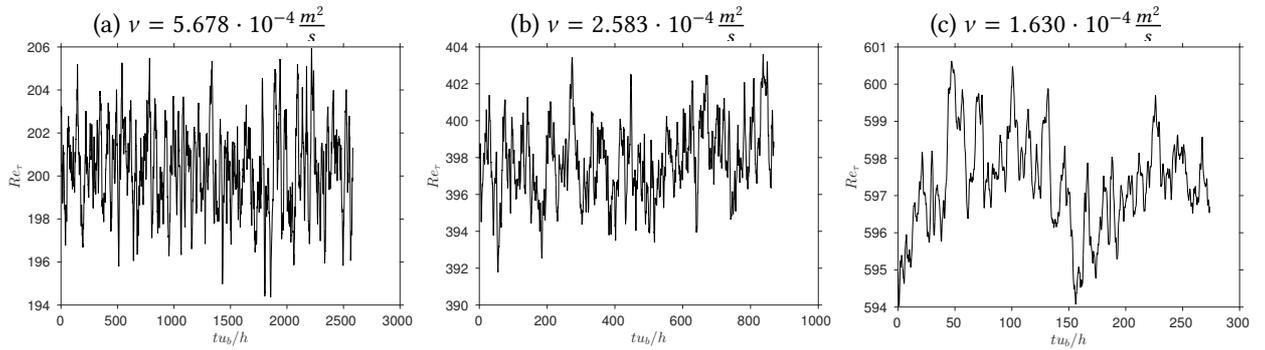

Figure 3.1.: Time evolution of the friction Reynolds number $Re_\tau$ for the time period where statistics are accumulated for different flow cases.

In addition to the aforementioned turbulent statistics, so-called spectral statistics are generated. Therefore, a number of horizontal planes, where the flow variables are only averaged in time without applying spatial averaging, were chosen. Spectral statistics are needed in order to determine two-point correlations and energy spectra, which become important when analysing different flow scales (cf. sections 4.1 and 4.3). Post-processing on statistics have mainly been computed using MATLAB, whereas instantaneous flow fields were analysed using the latter as well as Paraview.

# 4. Results

Results are reported in the upcoming chapter. First domain size and grid spacing are evaluated using two-point velocity correlations before fully developed turbulent statistics are analysed with special emphasis on the Reynolds stresses and root mean square vorticity fluctuations, which become highly anisotropic when approaching the free surface. Next, the coherent large-scale or very large-scale motions respectively are analysed with the aid of spectral statistics. Finally, instantaneous realizations of flow fields are visualized in order to show surface-near vortical motions that contribute to the different behaviour in open channel flows with respect to closed channel ones.

## 4.1. Domain Size and Grid Spacing

In order to determine whether the computational domain is large enough to contain the largest turbulent scale, two-point correlations, such as defined by equation 2.37, have been computed in the homogeneous directions at different distances from the wall. A sufficiently large domain would result in two-point correlations that fall off to zero at high separations. Additionally, in order to resolve the smallest turbulent scales, the grid resolution needs to be sufficiently high.

The computational domain size and grid spacing of the first open channel simulations of this work have been estimated with results from Moser et al. (1999) closed channel simulations and led to configurations as shown in table 4.1.

| case | $Re_\tau$ | $L_x$ | $L_z$ | $\alpha$ | $\beta$ | $N_x \times N_y \times N_z$ | $\Delta x^+$ | $\Delta z^+$ | $\Delta y^+_{max}$ |
|------|-----------|-------|-------|----------|---------|------------------------------|--------------|--------------|---------------------|
| $oc200'$ | 200 | $4\pi$h | $2\pi$h | 0.25 | 0.5 | $192 \times 129 \times 128$ | 13.2 | 6.6 | 2.5 |
| $oc400'$ | 400 | $4\pi$h | $2\pi$h | 0.25 | 0.5 | $512 \times 193 \times 384$ | 9.9 | 6.6 | 3.3 |
| $oc600'$ | 600 | $4\pi$h | $2\pi$h | 0.25 | 0.5 | $768 \times 257 \times 768$ | 9.7 | 4.9 | 3.6 |

Table 4.1.: Initial cases for open channel simulation, h=2

Red coloured values for the cases $oc200'$ and $oc400'$ in table 4.1 were larger than commonly-accepted resolutions of spectral methods in wall-bounded flows ($\Delta x^+ < 10$, $\Delta y^+_{max} < 4$, $\Delta z^+ < 5$, as a rule of thumb). Furthermore, two-point correlations of case $oc600'$ exhibited high values for the streamwise velocity fluctuations at high separations at vertical positions near the free surface with respect to both homogeneous directions. The latter is a first indicator that flow structures in an open channel flow might be larger than in closed channel one when approaching the surface. The number of grid points and the domain size have then successively been adapted until configurations such as shown in table 4.2 were obtained.

| case | $Re_\tau$ | $L_x$ | $L_z$ | $\alpha$ | $\beta$ | $N_x \times N_y \times N_z$ | $\Delta x^+$ | $\Delta z^+$ | $\Delta y^+_{max}$ |
|------|-----------|-------|-------|----------|---------|------------------------------|--------------|--------------|---------------------|
| $oc200$ | 200.60 | $4\pi$h | $2\pi$h | 0.25 | 0.5 | $256 \times 129 \times 256$ | 9.9 | 4.9 | 2.5 |
| $oc400$ | 398.12 | $4\pi$h | $2\pi$h | 0.25 | 0.5 | $512 \times 193 \times 512$ | 9.8 | 4.9 | 3.2 |
| $oc600$ | 597.46 | $8\pi$h | $4\pi$h | 0.125 | 0.25 | $1536 \times 257 \times 1536$ | 9.8 | 4.9 | 3.7 |

Table 4.2.: Final cases for open channel simulation, h=2

Henceforth, the results of the latter simulations will be discussed. Figures 4.1, 4.2 and 4.3 show the two-point correlations for all cases at different vertical positions. The streamwise velocity fluctuations of the highest Reynolds number case are apparently not decorrelated well with respect to the stream- and spanwise directions at the channel top, as shown in Figure 4.3b, but rising the domain size even further would exceed the scope of this work and might be subject to further investigation.

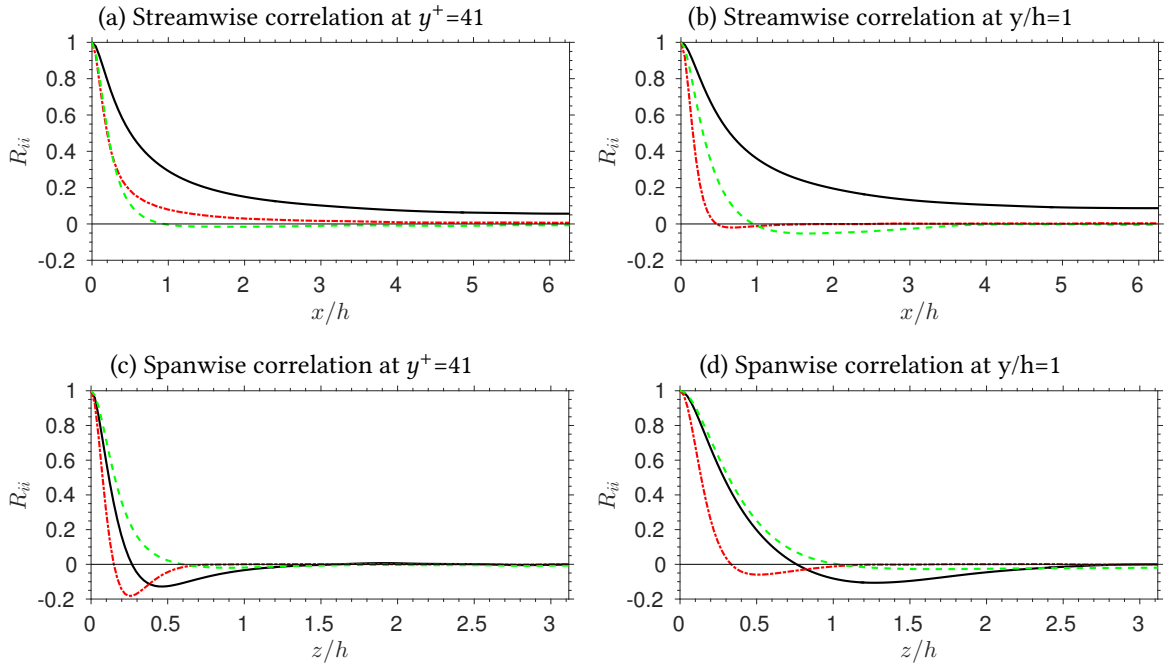

Figure 4.1.: Two-point correlations: ——, $R_{uu}$; ---, $R_{vv}$; - - -, $R_{ww}$. $Re_\tau$=200.

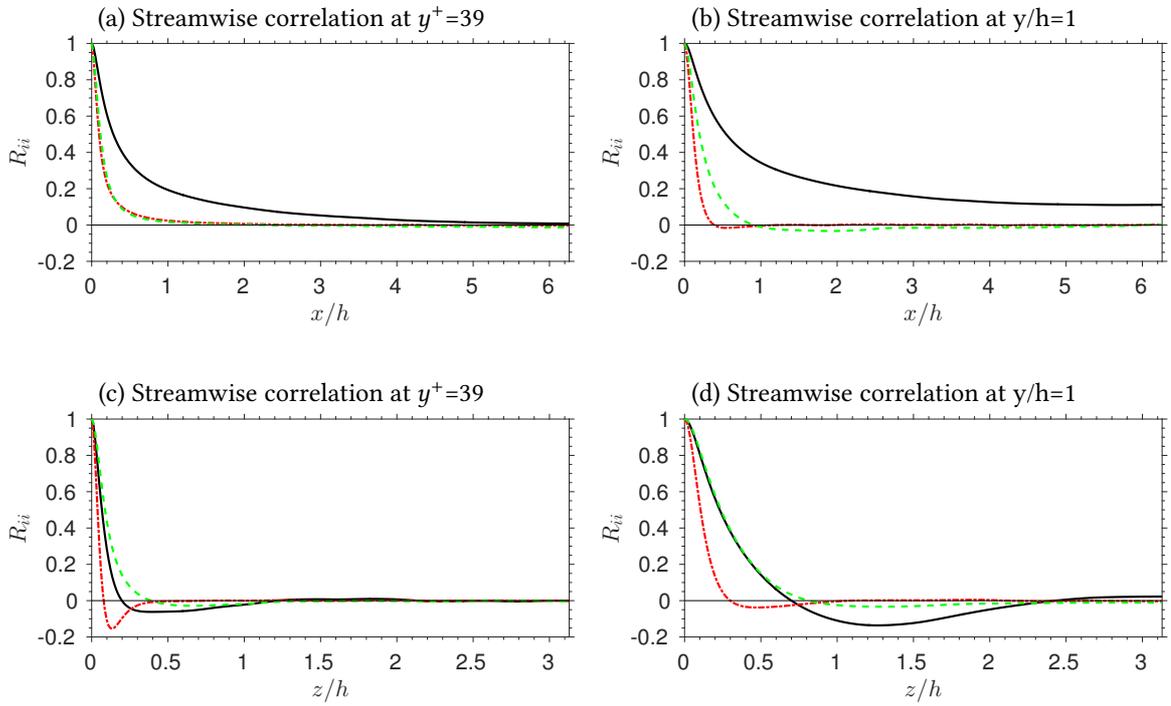

Figure 4.2.: Two-point correlations: — $R_{uu}$; --- $R_{vv}$; --- $R_{ww}$. $Re_\tau$=400.

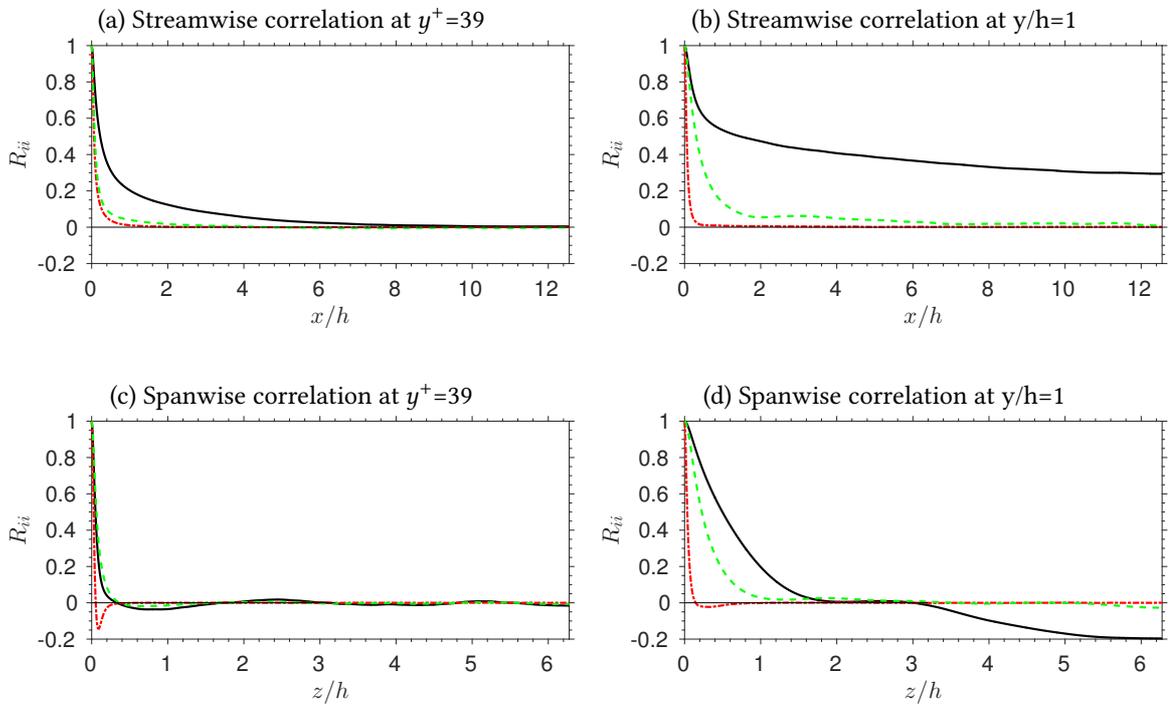

Figure 4.3.: Two-point correlations: — $R_{uu}$; --- $R_{vv}$; --- $R_{ww}$. $Re_\tau$=600.

## 4.2. Turbulent Statistics

After simulations have been carried out over sufficiently large time period in bulk time units (cf. table 3.1), space averaging in the homogeneous directions and time averaging were performed yielding the turbulent statistics. The statistics have then been compared to those received from closed channel simulation (Moser et al., 1999), cf. table 4.3.

| case | $Re_\tau$ | $L_x$ | $L_z$ | $\alpha$ | $\beta$ | $N_x \times N_y \times N_z$ | $\Delta x^+$ | $\Delta z^+$ | $\Delta y_c^+$ |
|------|-----------|-------|-------|----------|---------|------------------------------|--------------|--------------|----------------|
| cc180 | 178.13 | $4\pi$h | $\frac{4}{3}\pi$h | 0.5 | 1.5 | $192 \times 129 \times 192$ | 11.7 | 3.9 | 4.4 |
| cc395 | 392.24 | $2\pi$h | $\pi$h | 1 | 2 | $384 \times 193 \times 288$ | 6.4 | 4.3 | 6.5 |
| cc590 | 587.19 | $2\pi$h | $\pi$h | 1 | 2 | $576 \times 257 \times 576$ | 6.4 | 3.2 | 7.2 |

Table 4.3.: Cases for closed channel simulation from Moser et al. (1999), h=1

### 4.2.1. Mean Velocity Profile

The profile of the mean velocity in fully developed turbulent channel flows depends on only two non-dimensional parameters, which can be derived by dimensional analysis. The following ansatz for the mean velocity profile can be stated:

$$\frac{d\langle u \rangle}{y} = \frac{u_\tau}{y} \phi \left( \frac{y}{h}, y^+ \right), \tag{4.1}$$

with the function $\phi$ yet to be determined. Prandtl (1925) postulated that in the vicinity of the wall ($y/h \ll 1$) the mean velocity profile is just determined by the viscous scales and independent of $h$ and called this region the inner layer. Furthermore, one can show by Taylor expansion at $y$=0, together with the no-slip condition, the so-called linear velocity law

$$u^+ = y^+ \tag{4.2}$$

holds within the viscous sublayer ($y^+ < 5$). Since at high Reynolds numbers the outer part of the inner layer corresponds to large values of $y^+$, $\phi$ looses its dependency on the viscosity as well and becomes constant. The expression for the mean velocity derived by Von Kármán (1930) then looks as follows

$$u^+ = \frac{1}{\kappa} \ln(y^+) + B, \tag{4.3}$$

where $\kappa$ is called the von Kármán constant. This logarithmic law of the wall by von Kármán will further be denoted as log law. Nikuradse (1932) obtained the first experimental results ($\kappa = 0.4$ and $B$=5.5) in air pipe flows. Nezu (2005) summarized some results for the constants in the log law profiles of different flows, in particular: $\kappa$=0.41, $B$=5.0 by Coles (1968) and $\kappa$=0.41, B=5.2 by Brederode and Bradshaw (1974) in boundary layers; $\kappa$=0.41, $B$=5.17 in closed channel flows by Dean (1978) and $\kappa$=0.41, $B$=5.29 in open channel flows by Rodi and Asce (1986), reported over the Reynolds number range $439 \leq Re_\tau \leq 6139$. There is a known Reynolds number dependence in the log law constant $B$. Kim, Moin, et al. (1987) reported $B$=5.5 for the marginal Reynolds number case ($Re_\tau$=180) instead of $B$=5.0, used in Moin and Kim (1982) ($Re_\tau$=640), having $\kappa$=0.4, both were closed channel simulations. More recently, Abe, Kawamura, and Matsuo (2001) reported $\kappa$=0.41 and $B$=5.2 for $Re_\tau = 640$ for closed channel flow. Figure 4.4 shows the profiles of the mean streamwise velocity component, non-dimensionalized by the friction velocity $u_\tau$. Note that the curves are plotted together with those from closed channel simulations of Moser et al. (1999), which are indicated by the dotted blue lines that lie above the open channel profiles in the free-surface vicinity.

The mean velocity profiles collapse well with the ones computed by Kim, Moin, et al. (1987) for closed channel flows and there is only a little discrepancy in the vicinity of the free surface, where the closed channel slopes lie above the open channel ones. For the log law plotted in the figures 4.4b and 4.4c the constants $\kappa$=0.41 and $B$=5.2 were taken, which Pope (2000) proposed. For the marginal Reynolds number case the constant $B$ is adjusted to $B$=5.6. Unlike the closed channel profile, the open channel profile looks more than a straight line even above the logarithmic region, which matches with Keulegan (1938), where the log law has been assumed to describe the mean velocity profile until the channel top for open channel flows.

Open channel flow simulations of Handler et al. (1993) and Lam and Banerjee (1992) concluded that there is an open channel effect on the log law since they obtained either a smaller slope (2.4 instead of 2.5 for Handler et al. (1993)) or smaller constant $B$ (5.1 instead of 5.5 for Lam and Banerjee (1992)). However, such an effect can not be extracted from the data of this work. The exact values of the constants $\kappa$ and $B$ calculated with a linear regression function from open and closed channel simulation are shown in table 4.4 and match well. Note that the simulations of Handler et al. (1993) and Lam and Banerjee (1992) were stated in the very marginal turbulent regime ($Re_\tau = 134$ and 171 respectively) where the log layer region barely exists.

|        | oc200 | cc180 | oc400 | cc395 | oc600 | cc590 |
|--------|-------|-------|-------|-------|-------|-------|
| $\kappa$ | 0.374 | 0.375 | 0.406 | 0.405 | 0.408 | 0.406 |
| $B$    | 4.75  | 4.85  | 5.24  | 5.16  | 5.22  | 5.23  |

Table 4.4.: Constants of the log law $u^+ = \frac{1}{\kappa}y^+ + B$ computed by linear regression having $y^+ > 30$ and $y/h < 0.3$. oc: open channel, cc: closed channel data from Moser et al. (1999), number in the header is $Re_\tau$.

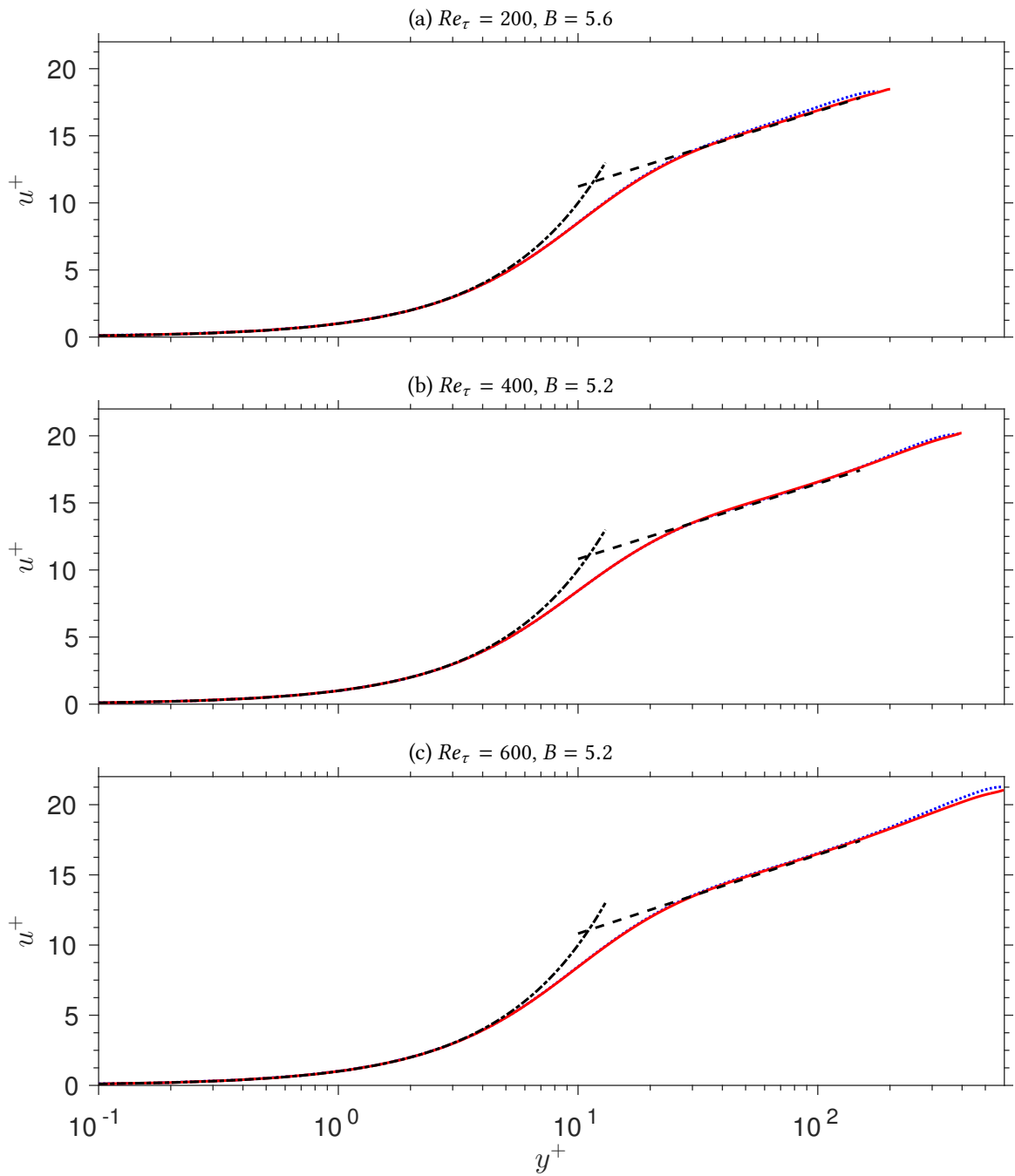

Figure 4.4.: Mean streamwise logarithmic velocity profiles: ——, open channel; ┄┄,Moser et al. (1999) closed channel; -··-, linear law of the wall; - -, log law for fully developed channel flows: $u^+ = \frac{1}{0.41} ln(y^+) + B$.

### 4.2.2. Reynolds Stresses

The Reynolds stress tensor for channel flows with two homogeneous directions contains four different components and looks as follows:

$$\mathbf{R} = \begin{bmatrix} \langle u'u' \rangle & \langle u'v' \rangle & 0 \\ \langle v'u' \rangle & \langle v'v' \rangle & 0 \\ 0 & 0 & \langle w'w' \rangle \end{bmatrix},$$
(4.4)

with $\langle v'u' \rangle = \langle u'v' \rangle$ due to its symmetry. The derivation of the stress tensor for channel flows with two homogeneous directions can be looked up in section A.5. The normal components of the Reynolds stress tensor contribute to the turbulent kinetic energy, defined as

$$k = \frac{1}{2}(\langle u'u' \rangle + \langle v'v' \rangle + \langle w'w' \rangle)$$
(4.5)

and their square roots, which are the root mean square values of the velocity fluctuations

$$u_{i,rms} = \sqrt{\langle u_i' u_i' \rangle}, \ i = 1, 2, 3,$$
(4.6)

are denoted as turbulent intensities. Figure 4.6 shows the open channel turbulent intensities normalized by $u_\tau$ compared to those obtained by closed channel simulation from Moser et al. (1999). Note that for the open channel configuration the channel top at $y/h=1$ corresponds to the channel centerline of the closed channel configuration. The data matches well in the near-wall region, where the components show asymptotic behaviour (cf. Pope (2000)):

$$\lim_{y \to 0} \langle u'u' \rangle = O(y^2)$$
$$\lim_{y \to 0} \langle v'v' \rangle = O(y^4)$$
$$\lim_{y \to 0} \langle w'w' \rangle = O(y^2)$$
$$\lim_{y \to 0} \langle u'v' \rangle = O(y^3).$$
(4.7)

Analogously asymptotics can be derived for the surface-near region, yielding

$$\lim_{y \to h} \langle u'u' \rangle = O(1)$$
$$\lim_{y \to h} \langle v'v' \rangle = O((h-y)^2)$$
$$\lim_{y \to h} \langle w'w' \rangle = O(1)$$
$$\lim_{y \to h} \langle u'v' \rangle = O(h-y).$$
(4.8)

Figure 4.5 shows the Reynolds stresses computed with the current data, both in wall- and surface-near regions, where they exhibit asymptotic behaviour such as indicated by the aforementioned equations.

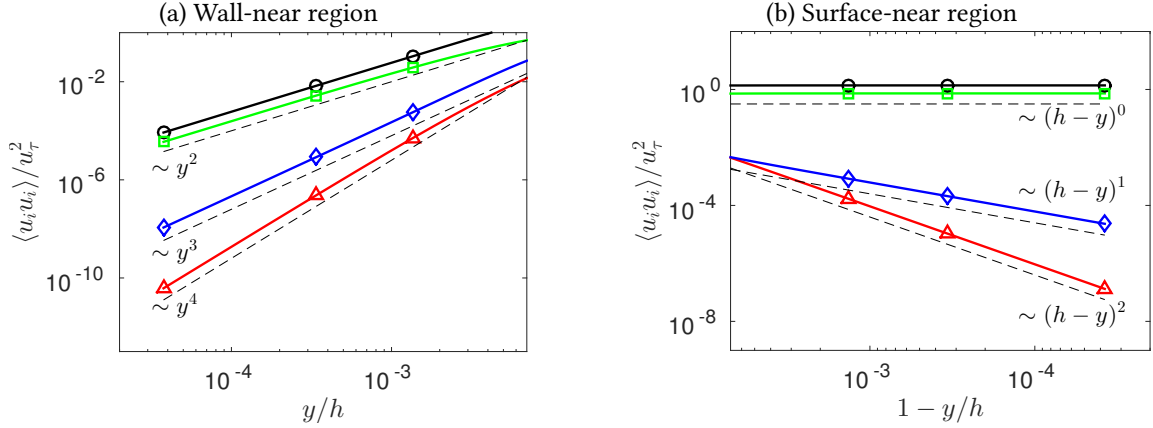

Figure 4.5.: Asymptotic behaviour of Reynold stress components. **o**, $\langle u'u'\rangle$; △, $\langle v'v'\rangle$; □, $\langle w'w'\rangle$; ◇, $\langle u'v'\rangle$. $Re_\tau = 600$.

In case of the marginal Reynolds number simulation ($Re_\tau = 200$) the curves of turbulent intensities do not collapse exactly with the ones obtained by closed channel simulation, which is caused by the slightly different Reynolds number ($Re_\tau = 180$ for closed channel). The streamwise component shows the well known peak in the buffer layer region ($y^+ \approx 15$). The maximum of the turbulent kinetic energy is a signature of the coherent turbulent structures. These streamwise elongated streaky structures are regions of high, either positive or negative, streamwise velocity fluctuations and will, according to Kline et al. (1967), henceforth be denoted as high and low velocity streaks.

Reaching the channel top, turbulent intensities show a different behaviour with respect to the closed channel case. Due to impermeability at the free surface the vertical component of the fluctuating velocity drops to zero, while its energy is transferred to the horizontal components. Most of the energy of the vertical component is transferred to the spanwise one, while the streamwise component increases much less. This is in agreement with results obtained from numerical studies of low Reynolds number turbulent open channel flow, such as Swean et al. (1991), Komori et al. (1993), Borue et al. (1995) or Nagaosa (1999). Handler et al. (1993) found the pressure-strain correlation term to be the key contributor to the intercomponent energy transfer below the free surface in open channel flow. Figure 4.7 shows the diagonal components of the pressure-strain term $\Pi_{ii} = \langle p'S'_{ii}\rangle$ in the transport equation for Reynolds stresses for $Re_\tau = 400$. Figures for the other Reynolds number cases show qualitatively similar results and can be looked up in the appendix (figure A.1). Since surface-normal motions are blocked by the free surface, the surface-normal pressure-strain component $\Pi_{22}$ decreases and becomes negative while approaching the free surface, meaning it acts as a sink in the transport equation of the surface-normal Reynolds stress component. While the streamwise component of the pressure-strain $\Pi_{11}$ shifts from negative values to slightly positive at the free surface, the spanwise component $\Pi_{33}$ remains positive in the near-surface boundary layer exhibiting values of higher magnitude. The latter effect is believed to be responsible for the dominant energy transfer from the vertical Reynolds stress component to the spanwise one.

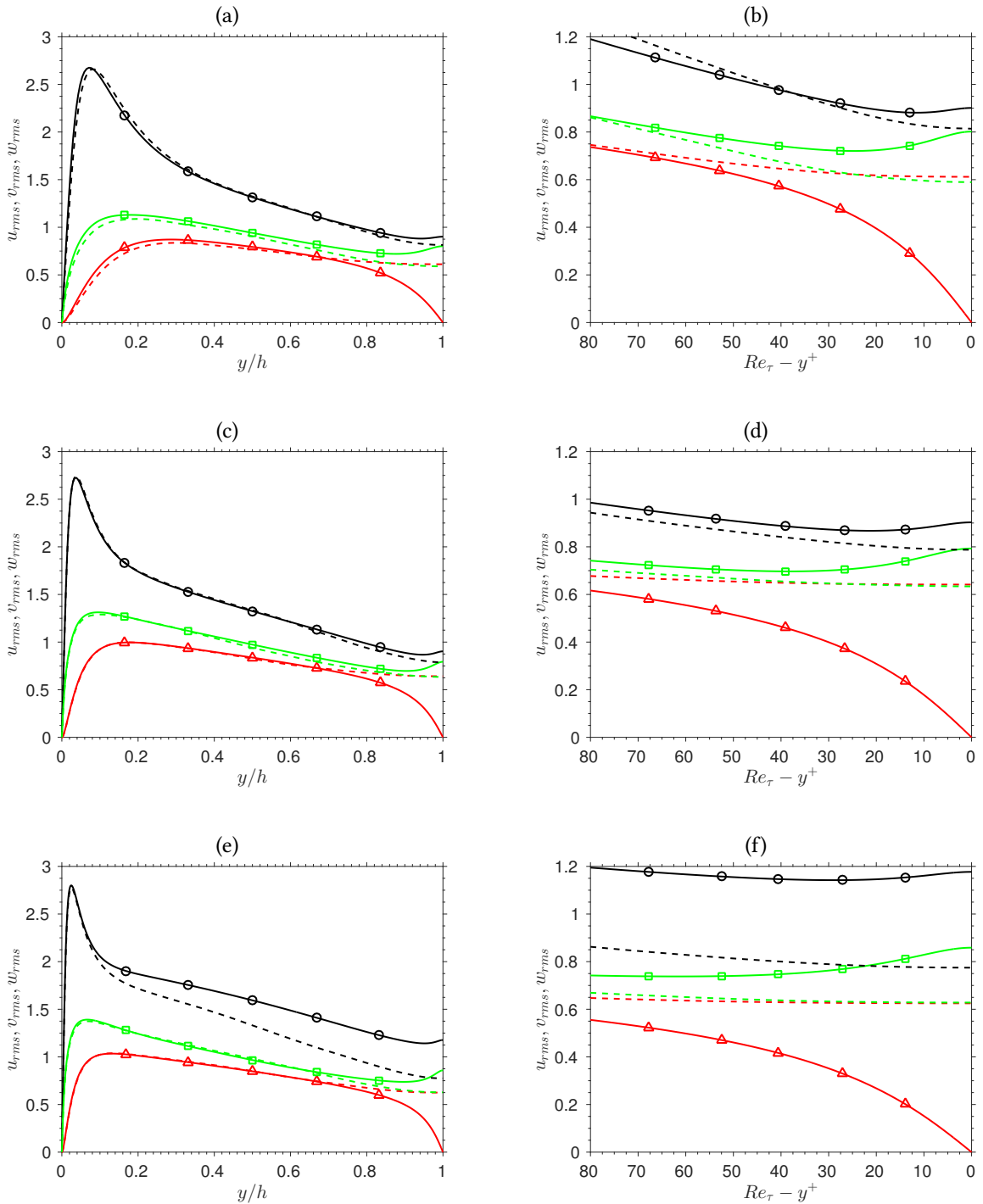

Figure 4.6.: Turbulent intensities: $\mathbf{o}$, $u_{rms}$; $\triangle$, $v_{rms}$; $\square$, $w_{rms}$; Dashed lines indicate Moser et al. (1999) data from the closed channel, normalized by $u_\tau$. The left column contains the profiles in outer units ($y/h$) and the right one in wall units ($y^+$). (a),(b): $Re_\tau = 200$; (c),(d): $Re_\tau = 400$; (e),(f): $Re_\tau = 600$.

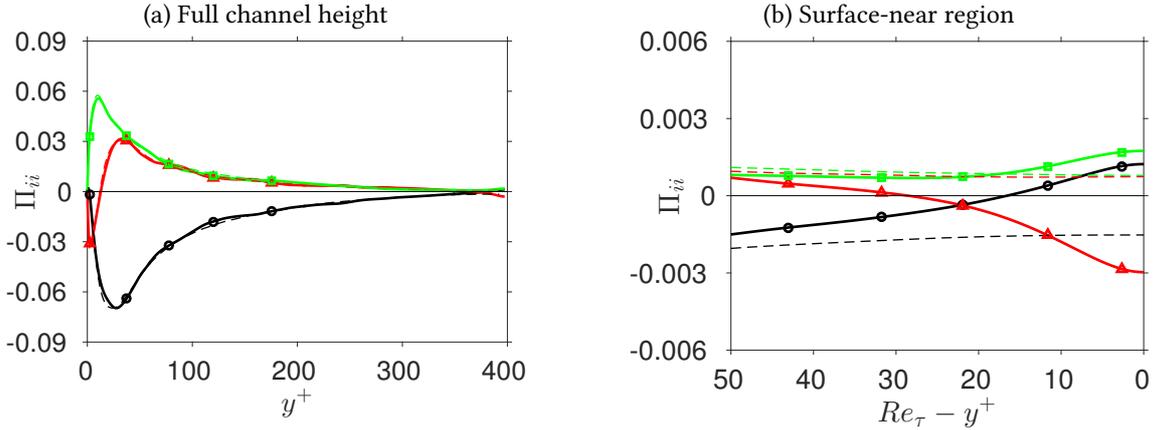

Figure 4.7.: Diagonal components of the pressure-strain term for $Re_\tau = 400$: **o**, $\Pi_{11}$; △, $\Pi_{22}$; □, $\Pi_{33}$. Dashed lines indicate closed channel data from Moser et al. (1999).

The center of turbulent closed channels exhibits almost isotropic flow behaviour in terms of Reynolds stresses and root mean square vorticity fluctuations, as will be discussed below. The region in open channel flows, where Reynolds stress profiles deviate from closed channel case by showing an anisotropic behaviour, will therefore be denoted as velocity anisotropy layer (Nagaosa, 1999). Analogously a vorticity anisotropy layer near the free surface exists, which will be discussed in section 4.2.4. Figure 4.6 shows the velocity anisotropy layer near the free surface, where the intercomponent energy transfer occurs, to exhibit a thickness of approximately $\delta_v \approx 0.3h$. Whether the layer thickness of $\delta_v^+ \approx 50$ scales in wall or in outer flow units ($\delta_v/h \approx 1/3$) could not be concluded in Nagaosa (1999), since a single Reynolds number case of $Re_\tau = 150$ was regarded. The additional data of different Reynolds numbers of this work revealed the velocity anisotropy layer to scale in bulk units. Furthermore, the aforementioned figure shows a second effect that becomes apparent for the highest Reynolds number case. Besides the intercomponent energy transfer in the near-surface boundary layer the curve for the streamwise turbulent intensities starts to separate at much lower vertical positions, leading to a higher level of total turbulent kinetic energy in the surface-near part of the channel. Comparison of our data with experimental open channel flow data in the same Reynolds number regime from Chen et al. (2014), such as shown in figure 4.8, reveals the same discrepancy. Test cases have been set up and will be investigated in future work in order to clarify whether the latter behaviour is a spurious phenomenon, which arises from insufficient grid resolution, marginal box size or both of it. Note that the reference data of Chen et al. (2014) is obtained from a flume flow with an aspect ratio of $L_z/h = 8.6$, meaning turbulent statistics might have been influenced by side wall effects. Nevertheless, the results obtained by the highest

Reynolds number case of the current work should be treated cautiously.

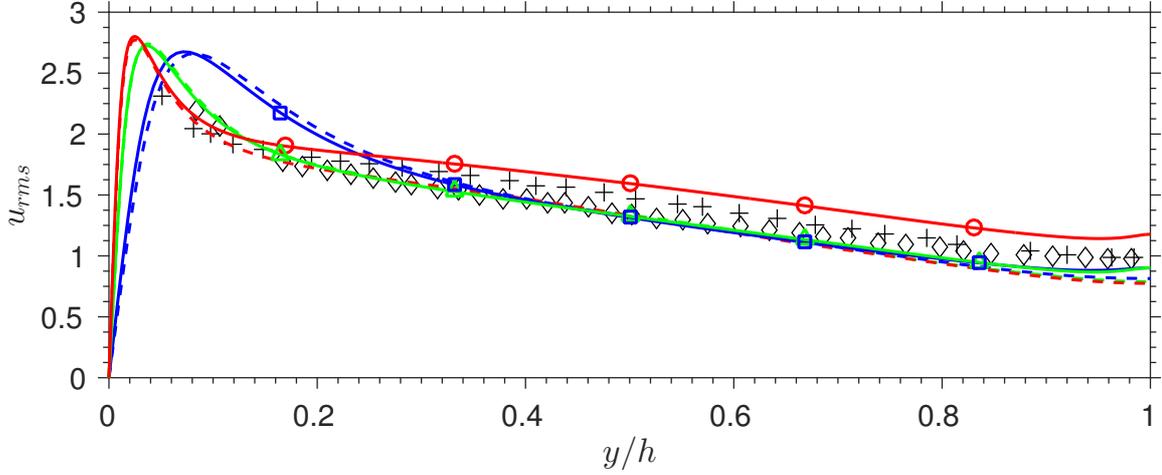

Figure 4.8.: Streamwise turbulent intensities $u_{rms}$: $\square$, $Re_\tau = 200$; $\triangle$, $Re_\tau = 400$; $\mathbf{o}$, $Re_\tau = 600$. Experimental open channel flow data from Chen et al. (2014): $\diamond$, $Re_\tau = 382$; $+$, $Re_\tau = 609$. Dashed lines indicate Moser et al. (1999) closed channel flow data.

The Reynolds shear stresses as shown in Figure 4.9 collapse very well with those from closed channel simulation apart from the above mentioned discrepancy for the marginal Reynolds number case. The linear profile of the total shear stress indicates that the flow is statistically stationary, which is the case in the here considered fully developed channel flow and the number of samples in space and time averaging is sufficient to receive reliable statistics. As the Reynolds number increases the Reynolds stresses become more dominant since the wall region becomes thinner in outer flow units.

### 4.2.3. Lumley Triangles

In order to determine the anisotropy of a turbulent flow the Reynolds stress anisotropy tensor **b** can be computed from the Reynolds stress tensor:

$$b_{ij} = \frac{\langle u_i' u_j' \rangle}{2k} - \frac{\delta_{ij}}{3} \qquad (4.9)$$

with

$$\delta_{ij} = \begin{cases} 1, \ i = j \\ 0, \ i \neq j \end{cases}$$

the Kronecker delta. In order to plot the Lumley triangle as it is described in Lumley (1978) it is necessary to compute the invariants of the anisotropy tensor as follows

$$\eta = \sqrt{\frac{1}{6} b_{ij} b_{ji}} \qquad (4.10a)$$

$$\xi = \left( \frac{1}{6} b_{ij} b_{jk} b_{ki} \right)^{\frac{1}{3}}, \qquad (4.10b)$$

where Einstein summation convention implies summation over repeated indices. Figure 4.10 shows the Lumley triangles computed with the recent simulation data and the closed channel data from Kim, Moin, et al. (1987) for comparison. The invariants values have to stay within the triangle for all flows. The origin of the coordinate systems corresponds to isotropic flow, the diagonals to axisymmetric state with one dominant diagonal component (second quadrant) or one weak diagonal component (first quadrant) respectively. The point at $\eta = \xi = 1/3$ corresponds to one component turbulence and the line between the latter point and $\eta = 1/6, \xi = -1/6$ is the two component region. In case of an open channel, the flow state changes from axisymmetric with one dominant component towards axisymmetric with one weak component, as approaching the free surface. On the surface itself the open channel contains a pure two-component flow, because the vertical Reynolds stress component becomes damped to zero due to impermeability at the top. Closed channel flow in contrary approaches the isotropic state, as reaching the channel centerline, since the limiting behaviour of the wall is no more present. A good description of the characteristic shapes of the structures related to the Reynolds stress tensor can be found in Simonsen and Krogstad (2005).

### 4.2.4. Vorticity

The root mean square of the vorticity fluctuations shown in Figure 4.11 are normalized as follows

$$\omega_{i,rms}^+ = \sqrt{\omega_i' \omega_i'} \frac{\nu}{u_\tau^2}. \tag{4.11}$$

In the vicinity of the wall the vorticity profiles match well with the closed channel ones with again only little discrepancy in the lowest Reynolds number case, where the compared Reynolds numbers differ slightly. When reaching the free surface, vorticities show different behaviour compared to the closed channel centerline. In the latter case the components of the root mean square vorticity fluctuations show almost purely isotropic behaviour, meaning that they are of the same size. For the open channel this is obviously not the case. Due to impermeability and shear free boundary condition at the surface (equation 2.11), the vorticity vector simplifies as follows

$$\boldsymbol{\omega}(x, y = h, z) = \nabla \times \mathbf{u}(x, y = h, z) = \begin{pmatrix} \frac{\partial w}{\partial y} - \frac{\partial v}{\partial z} \\ \frac{\partial u}{\partial z} - \frac{\partial w}{\partial x} \\ \frac{\partial v}{\partial x} - \frac{\partial u}{\partial y} \end{pmatrix}_{y=h} = \begin{pmatrix} 0 \\ \omega_{y,top} \\ 0 \end{pmatrix}. \tag{4.12}$$

Surface parallel vorticities vanish at the top leading to perpendicular vortices at the surface. As Nagaosa (1999) points out, the anisotropic vorticity layer, where surface parallel components of the root mean square vorticity fluctuation decrease, is about ten wall units thick and much thinner than the anisotropic velocity layer with 50 wall units for $Re_\tau = 150$, which is in good agreement with the data of this study. The additional different Reynolds number information of this work furthermore confirms the thickness of the vorticity anisotropy layer $\delta_\omega \approx 10\delta_\nu$ to be independent of the Reynolds number.

The velocity anisotropy layer in contrary scales in outer flow units, as mentioned above ($\delta_v \approx 0.3h$). The resulting connection of vortices to the surface is a known phenomenon of open channel flows and such vortices will henceforth be denoted as surface-attached vortices (cf. Handler et al. (1993), Pan and Banerjee (1995), Nagaosa and Saito (1997)). The existence of surface-attached vortices together with the impermeability condition lead to quasi-two-dimensional turbulent flow beneath the free surface, which has been discussed for example by Kumar et al. (1998). The loss of isotropy and shifting towards two-dimensional flow can as well be seen in the Lumley triangles of open channel flows, cf. figure 4.10, and is one of the main differences compared to closed channel flows.

According to figure 4.11 there seems to be a Reynolds number dependence for the root mean square of the fluctuating vorticities normalized in wall units, which is more apparent at the free surface than at the wall. Since the same behaviour can be extracted from closed channel flow data, it is not believed to be a free surface effect. Note that normalization with outer flow units did not turn out in better results.

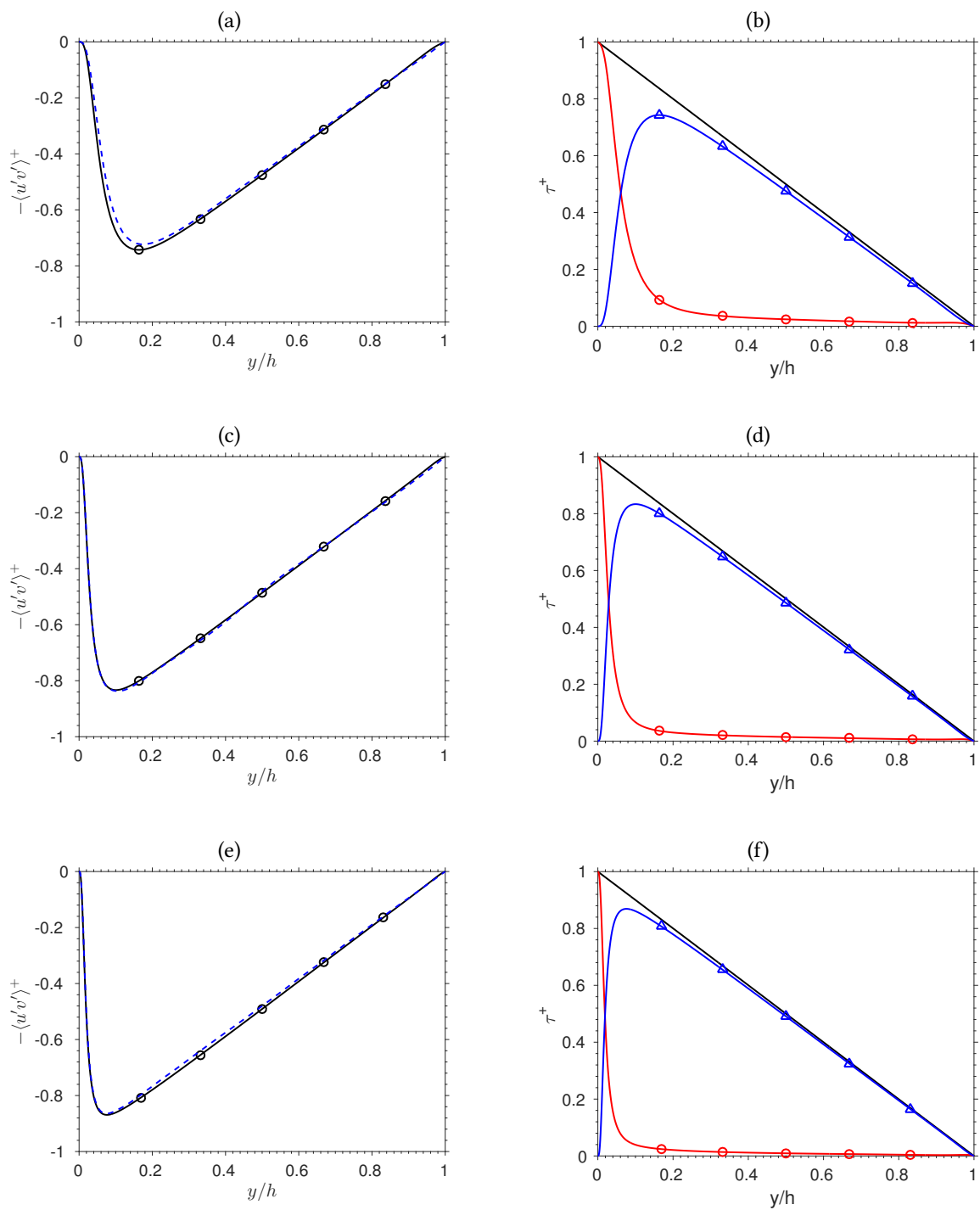

Figure 4.9.: Reynolds shear stress normalized by the wall shear velocity: (a),(c),(e): **o**, open channel simulation data; **- -**, data from Moser et al. (1999). (b),(d),(f): Total shear stress from open channel simulation (——) and its decomposition into Reynolds (△) and viscous (○) part: $\frac{\tau_{tot}}{u_\tau^2} = \nu \frac{\partial \langle u \rangle}{\partial y} \frac{1}{u_\tau^2} + \frac{\langle u'v' \rangle}{u_\tau^2}$. (a),(b): $Re_\tau = 200$; (c),(d): $Re_\tau = 400$; (e),(f): $Re_\tau = 600$.

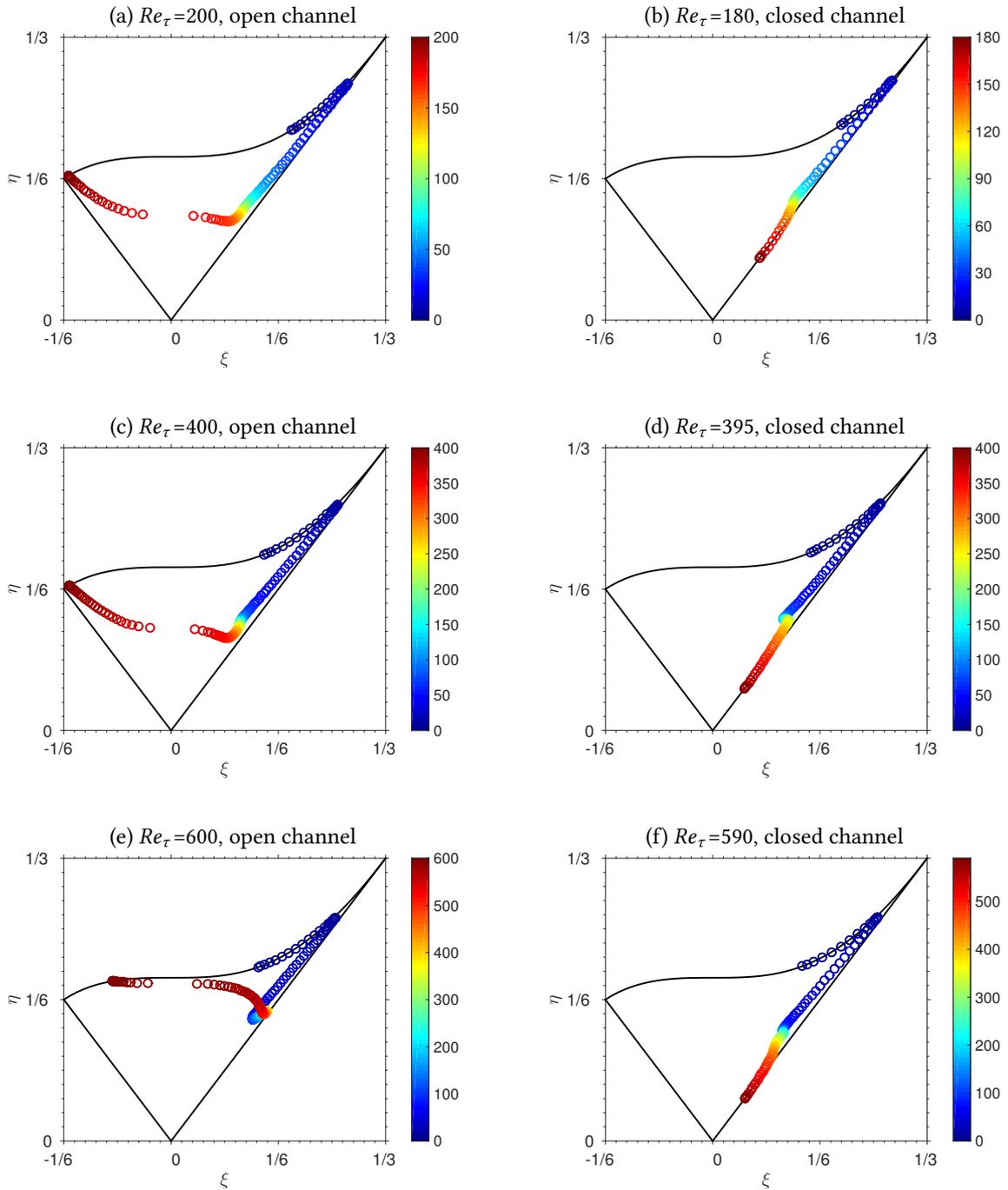

Figure 4.10.: Lumley triangles for all cases of the open channel simulation compared those of the closed channel configuration by Moser et al. (1999). The vertical position of the anisotropy invariants in wall units is indicated by the colorbar.

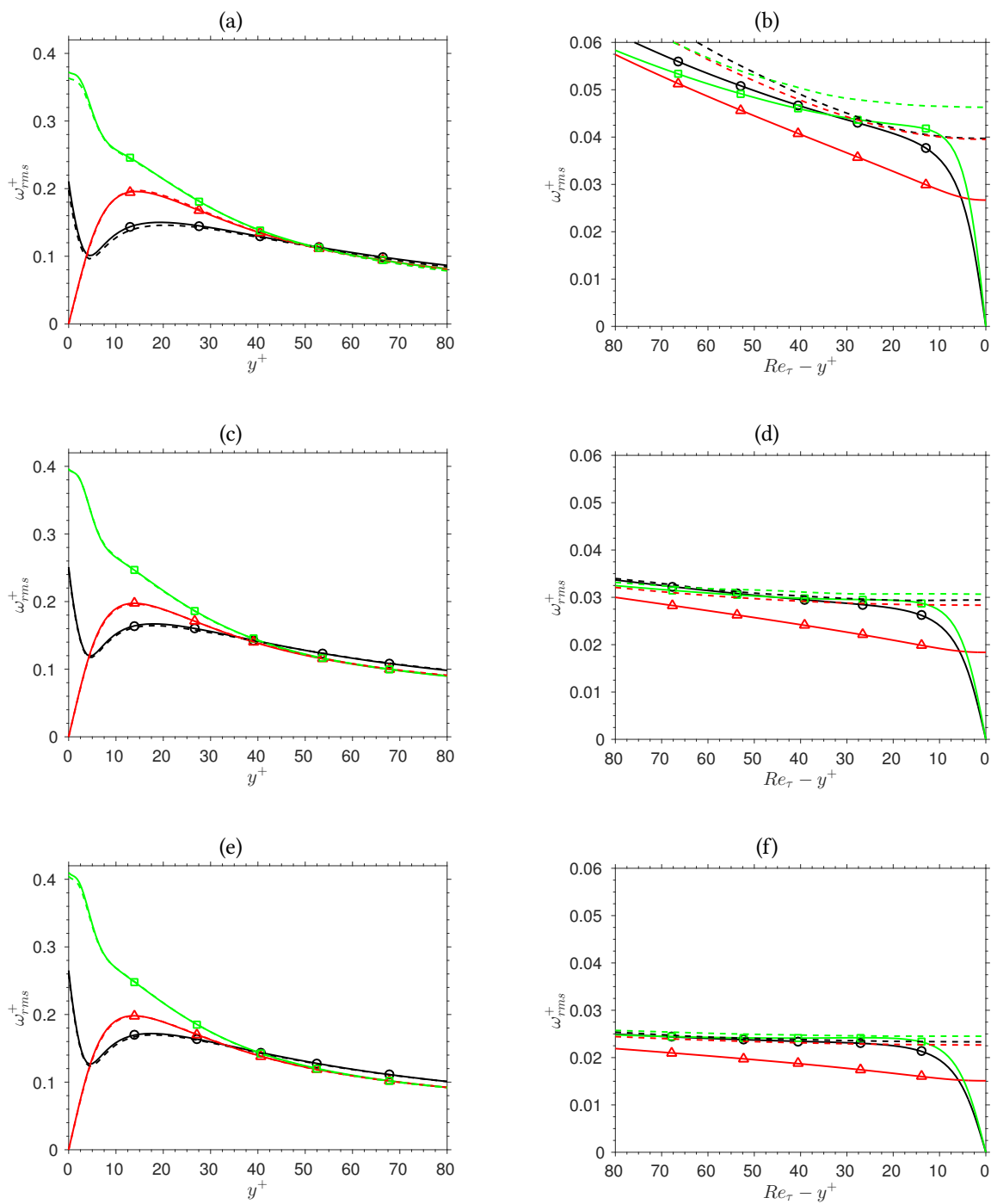

Figure 4.11.: Root mean square of the vorticities, normalized by $\nu/u_\tau^2$: ○, $\omega_{x,rms}^+$; △, $\omega_{y,rms}^+$; □, $\omega_{z,rms}^+$; Dashed lines indicate Moser et al. (1999) data from the closed channel. The left column contains the profiles near the wall whereas the right column shows the surface near ones. (a),(b): $Re_\tau = 200$; (c),(d): $Re_\tau = 400$; (e),(f): $Re_\tau = 600$.

### 4.2.5. Turbulent Kinetic Energy Equation Budget

Subtracting the average from the instantaneous momentum equation yields a transport equation for the fluctuating velocity, which then, after being multiplied with the velocity fluctuation and averaged, results in a transport equation for the turbulent kinetic energy $k$ (cf. equation 4.5):

$$\frac{\partial k}{\partial t} + \langle u_j \rangle \frac{\partial k}{\partial x_j} + \frac{1}{\rho} \frac{\partial \langle u_j' p' \rangle}{\partial x_j} + \frac{1}{2} \frac{\partial \langle u_i' u_i' u_j' \rangle}{\partial x_j} - 2\nu \frac{\partial \langle u_j' S_{ij}' \rangle}{\partial x_j} = -\langle u_i' u_j' \rangle \frac{\partial \langle u_i \rangle}{\partial x_j} - 2\nu \langle S_{ij}' S_{ij}' \rangle \quad (4.13)$$

where

$$S_{ij}' = \frac{1}{2} \left( \frac{\partial u_i'}{\partial x_j} + \frac{\partial u_j'}{\partial x_i} \right) \quad (4.14)$$

is the rate-of-strain tensor of the fluctuating velocity. The term $-\langle u_i' u_j' \rangle \frac{\partial \langle u_i \rangle}{\partial x_j}$ is also referred to as the production term of turbulent kinetic energy $\mathcal{P}$ and generally positive. The term $2\nu \langle S_{ij}' S_{ij}' \rangle$ is denoted as the viscous dissipation term $\varepsilon$ and also generally positive, but acting as a sink of turbulent kinetic energy due to a negative sign in equation 4.13. For the present case of fully developed plane channel flow the equation can be simplified to

$$-\underbrace{\nu \frac{d^2 k}{dy^2}}_{(i)} + \underbrace{\frac{1}{\rho} \frac{d \langle v' p' \rangle}{dy}}_{(ii)} + \underbrace{\frac{1}{2} \frac{d \langle v' u_j' u_j' \rangle}{dy}}_{(iii)} = \mathcal{P} - \varepsilon, \quad (4.15)$$

where the terms correspond to: viscous diffusion (*i*), pressure transport (*ii*), turbulent convection (*iii*), production $\mathcal{P}$ and dissipation $\varepsilon$. Figure 4.12 shows the terms of the turbulent kinetic energy equation of the present open channel simulation. Note that the angled brackets in case of terms (*ii*), (*iii*) and $\varepsilon$ indicate averaging over the homogeneous directions as well as over several realizations of the flow field (30 for $Re_\tau = 200$, two for $Re_\tau = 400$ and one for $Re_\tau = 600$), which leads to the fact that statistics for the latter terms (especially the higher order turbulent convection term) may not be fully converged. Nevertheless, the data matches well with results obtained from closed channel simulation by Moser et al. (1999) in the vicinity of the wall. When approaching the free surface, in contrary, the viscous diffusion is dropping with respect to the closed channel, whereas turbulent and pressure transport terms are switching their signs, leading to a either positive (pressure transport) or negative (turbulent transport) value at the free surface, instead of asymptotically approaching a zero value as they do for the closed channel case. Furthermore can be observed that, while the profiles scale fairly well in viscous units in the wall-near region, there is some discrepancy between the different Reynolds number profiles in the surface-near layer. While the production term stays about the same as in closed channel flows, the dissipation appears to be lower, exhibiting a local peak at $Re_\tau - y^+ \approx 8$ in the free surface boundary layer.

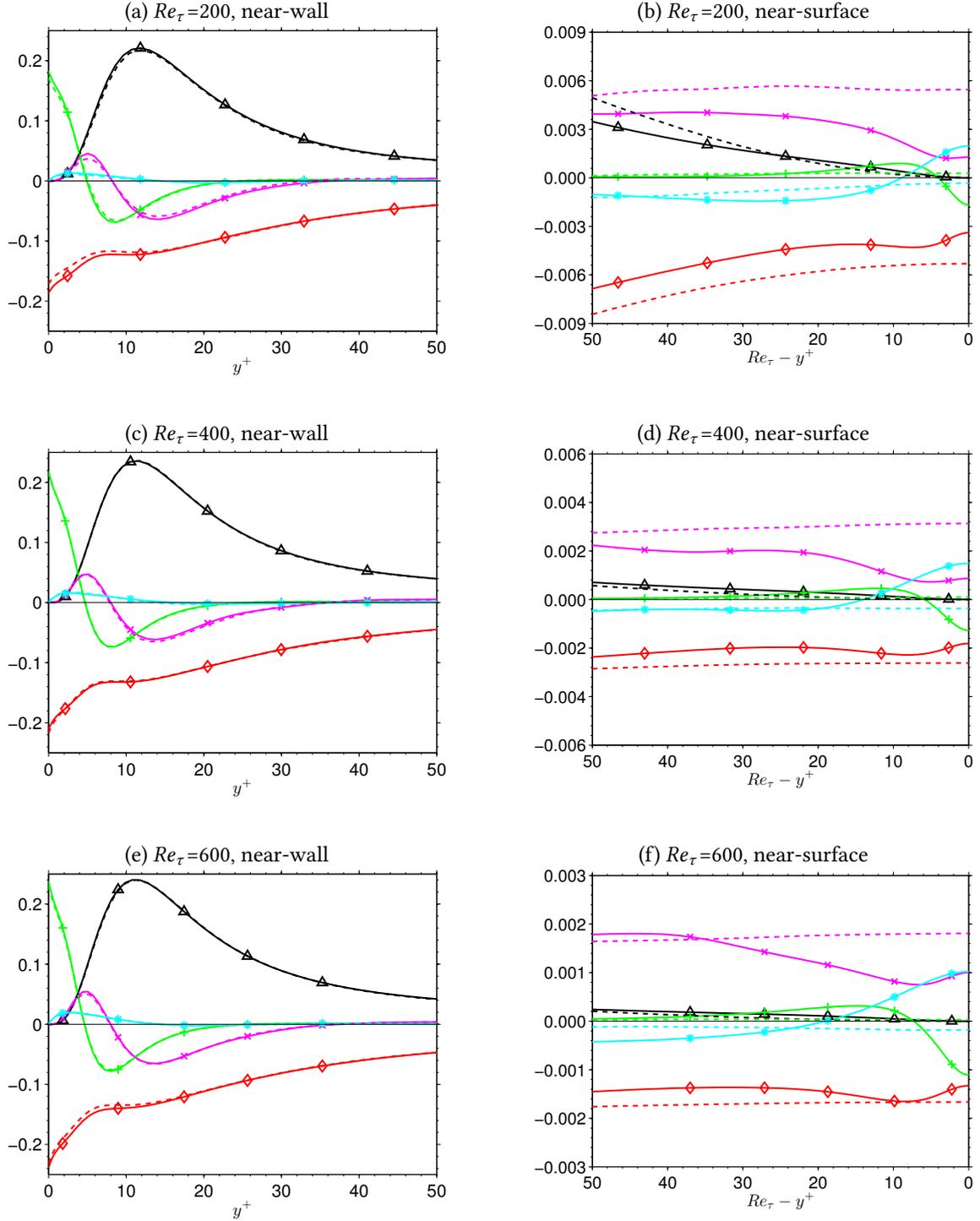

Figure 4.12.: Turbulent kinetic energy budget for open channel flow:

$\triangle$, production term $\mathcal{P} = -\langle u'v'\rangle \frac{d\langle u\rangle}{dy}$; $\diamond$, dissipation term $-\varepsilon = -2\nu\langle S'_{ij}S'_{ij}\rangle$;

+, viscous diffusion term $\nu\frac{d^2 k}{dy}$; ×, turbulent convection term $\frac{1}{2}\frac{d}{dy}\langle v'u'_i u'_i\rangle$;

*, pressure transport term $\frac{1}{\rho}\frac{d\langle v'p'\rangle}{dy}$. Terms are normalized by viscous scales. Dashed lines indicate values from corresponding closed channel cases of Moser et al. (1999). (a,c,e): near-wall profiles in wall units; (b,d,f): Near-surface profiles, the x-axis shows the distance from the free surface or the channel centerline respectively in wall units ($Re_\tau - y^+$).

The latter phenomenon leads to a discrepancy in the ratio profile of turbulent kinetic energy production to dissipation, such as shown in figure 4.13. While the wall-near region shows the same behaviour for both configurations including the well known peak at $y^+ \approx 12$, the ratio exhibits higher values in an open channel flow when approaching the free surface than in the closed channel case. The latter effect is compensated in the core region above the logarithmic layer of the open channel, since the overall integral of the production to dissipation ratio has to equal unity due to statistically stationary flow. Note that the phenomenon becomes more apparent in the lower Reynolds number cases, because the core region becomes smaller when the logarithmic region spans more of the flow domain. In order to elucidate the physical mechanism that acts on the dissipation term in the free-surface vicinity of turbulent open channel flow, further investigations, which would exceed the scope of this work, will be necessary.

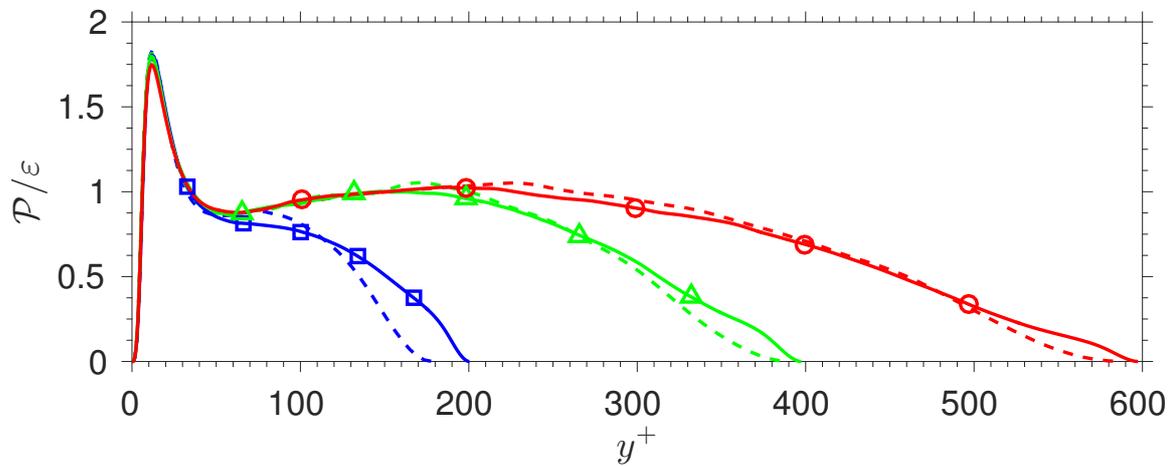

Figure 4.13.: Profiles of the ratio of production to dissipation $\mathcal{P}/\varepsilon$ in open channel flow: □, $Re_\tau = 200$; △, $Re_\tau = 400$; ○, $Re_\tau = 600$; Dashed lines indicate Moser et al. (1999) data from the closed channel.

## 4.3. Coherent Structure Analysis

One emphasis of this thesis is the analysis of large and very large coherent structures in open channel flows. Note that, since the term large-scale motions (LSM) most commonly corresponds to bulges and packets of hairpin vortices in boundary layers having a mean extension of $\lambda_x \times \lambda_z = (2-3)\delta \times (1-1.5)\delta$, the even longer structures appearing in channels and pipes will further be denoted as very large-scale motions (VLSM), referring to Kim and Adrian (1999). In order to determine the size of VLSM it is common practice to analyse the averaged premultiplied energy spectra of velocity fluctuations. The location of the energy peak in the spectrum corresponds thereby to the wavelength of the energy-containing motions (Perry et al., 1986).

Large-scale analysis for closed channel flows using premultiplied spectra can amongst others be found in Abe, Kawamura, and Choi (2004), Del Álamo and Jiménez (2003) and Jiménez (1998). The latter investigation was done on the data of Kim, Moin, et al. (1987) and Moser et al. (1999), the reference data for this work. The spanwise premultiplied spectra of the streamwise velocity fluctuations showed peaks at $\lambda_z^+ \approx 100$ near the wall, corresponding to the well known spanwise spacing of the near-wall velocity streaks, cf. Kline et al. (1967). As moving away from the wall the spectral peaks were shifted to longer wavelengths and stayed at the second numerical wavelength near the center, which might be a sign that they were constrained by the numerical box, as the author considers. The question how the behaviour in a wider box would be, was yet to be cleared. However, the second peak, which appeared above $y/h \approx 0.5$ corresponded to $\lambda_z/h \approx 2$. The streamwise spectra had most of their energy at very long wavelengths and were clearly constrained by the numerical box. The more recent work of Del Álamo and Jiménez (2003) analysed VLSM in very large boxes ($L_x = 12\pi h$ for $Re_\tau = 180$ and $L_x = 8\pi h$ for $Re_\tau = 550$) and found the streamwise extension of the VLSM to be at least $\lambda_x \geq 5h$, while confirming the spanwise extension of $\lambda_z/h \approx 2$. Abe, Kawamura, and Choi (2004) found the wavelength of the spanwise extension of VLSM in the upper channel half to be around $1.3 \leq \lambda_z/h \leq 1.6$ and the streamwise at least $\lambda_x > 3h$ in a closed channel simulation having Reynolds number similar to Moser et al. (1999) but larger numerical boxes ($L_x \times L_z = 4\pi h \times 2\pi h$).

Note that streamwise spectra for intermediate to high Reynolds numbers exhibit a peak at short wavelengths near the wall before showing a $\phi_{uu} \sim \kappa^{-1}$ range when moving away from the wall and then, after a certain point, showing one short wavelengths peak again. Perry et al. (1986) found the $\kappa^{-1}$ range in the region $y^+ > 140$ and $y/R < 0.3$ to extend between $\lambda_x/R \approx 5$ and $\lambda_x/R \approx 15$ for $1600 \leq Re_\tau \leq 3900$ in smooth pipes, where $R$ is the pipe radius. For $y/R > 0.3$ the short wavelength end of the range then settles at $\lambda_x/R \approx 3$ until the range collapses to one peak above $y/R \approx 0.6$. According to Kim and Adrian (1999) the power spectra within the $k^{-1}$ range can be decomposed into two modes, which scale with inner and outer units respectively. By extracting either the low wave number end of the $\kappa^{-1}$ range or rather the single peak, if there is one, it is possible to estimate the maximum streamwise extension of the VLSM with respect to the streamwise direction. Using this method the latter studies, which was an experimental investigation on annular pipe flows within a Reynolds number range $1058 \leq Re_\tau \leq 3175$ observed two different phenomena within the VLSM. The structures related to the streamwise velocity fluctuation feature a streamwise extension of the order of $2R$, with $R$ the pipe radius, spanning much of the region from the wall to the pipe centerline. Then the length of the VLSM extend up to $14R$ concentrated around the logarithmic region.

Monty et al. (2009) compared experimental data of turbulent boundary layer, closed channel and pipe flows of similar Reynolds numbers ($Re_\tau \approx 3000$) and measurement resolutions. Their comparison revealed that all mentioned flow configurations exhibit the same streamwise scales in terms of LSM ($\lambda_x \approx (2-3)h$). VLSM ($14 < \lambda_x/h < 20$) were only found in closed channel and pipe flows respectively. The near-wall peak ($y^+ \approx 15, \lambda_x^+ \approx 1000$) is found in all three cases and by moving away from the wall the bimodal spectrum evolves, having a high wavelength peak at $\lambda_x \approx 6h$ at $y \approx 0.06h$. In the boundary layer these streamwise length-scales are referred to as superstructures (Hutchins and Marusic, 2007). The high wavelength peak for boundary layers then moves to smaller wavelengths for $y > 0.06h$ and the spectrum finally shows a single peak at the wavelength related to LSM for $y > 0.3h$. For closed channel and pipe flows the high wavelength peak shifts to even longer wavelengths corresponding to VLSM ($14 < \lambda_x/h < 20$), while the low wavelength peak gets settled at $\lambda_x \approx 3h$. Monty et al. (2009) found the VLSM to grow as far as $y \approx 0.7h$ but pointed out that it is difficult to quantify their wavelength beyond $y \approx 0.33h$, since they grow with wall distance while their energy decays. Furthermore the peak related to LSM becomes more dominant in the outer flow region, which might be the reason why the long streamwise extension of VLSM was not captured in the investigation Kim and Adrian (1999) in the outer flow region.

There remains a lack of data corresponding to open channel flow configurations for Reynolds numbers higher than marginal ones, which should be partly filled by this investigation. The study of Handler et al. (1993) found an increase by a factor of two with respect to the spanwise extension, when comparing scales of the streamwise velocity fluctuations near the free surface of an open with closed channel in the marginal turbulent regime ($Re_\tau = 134, 125$ respectively), having a domain size of $L_x \times L_z = 4\pi h \times 3\pi/2h$.

Figure 4.14 shows the energy spectra of the velocity fluctuations of the highest Reynolds number case ($Re_\tau = 600$) near the wall ($y^+ \approx 5$) compared to the ones from Moser et al. (1999) in wall units. Note that the spectra of this work are shifted towards low wave numbers compared to the closed channel ones, since the computational domain is much larger with respect to the stream- and spanwise directions. The profiles collapse fairly well especially in the regime of high wave numbers, whereas there is a difference at low wave numbers, particularly for the spanwise spectrum of the streamwise velocity fluctuations.

Spanwise wave number energy spectra in linear scales for the stream- and spanwise velocity fluctuations near the wall are compared to the ones of Moser et al. (1999) in figure 4.15. The data of the $Re_\tau = 600$ case of this work exhibits a large local peak at a low wave number, which is less prominent in the reference data. This behaviour has been reported by Abe, Kawamura, and Choi (2004) for closed channel flows as well, where it was associated with VLSM in the outer layer that have not been captured by Moser et al.

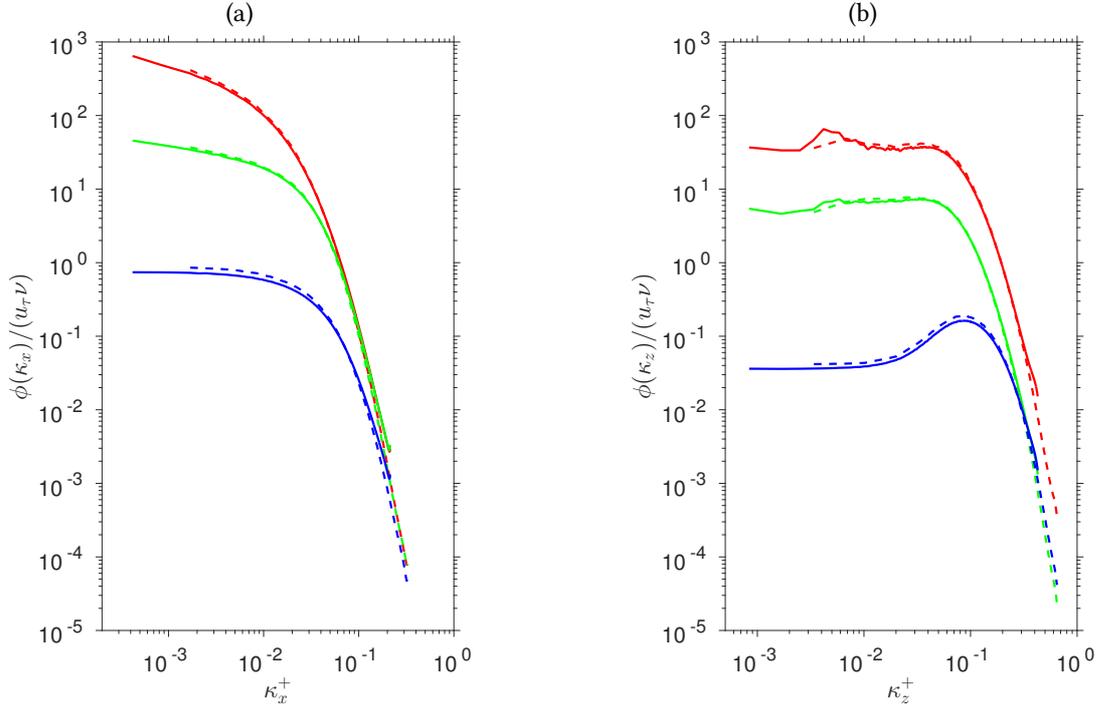

Figure 4.14.: One-dimensional wave number energy spectra of velocity fluctuations for $Re_\tau = 600$ at $y^+ = 5.04$: (a) streamwise spectra, (b) spanwise spectra. ——, $\phi_{uu}/(u_\tau \nu)$; ——, $\phi_{vv}/(u_\tau \nu)$; ——, $\phi_{ww}/(u_\tau \nu)$. Dashed lines indicate Moser et al. (1999) closed channel data for $Re_\tau = 590$ at $y^+ = 5.34$.

(1999) due to the smaller computational domain. The peak value of the energy density normalized with wall units is 0.11 compared to approximately 0.09 from Abe, Kawamura, and Choi (2004). This might be an indicator that the contribution of the outer layer VLSM to the inner layer energy spectrum is higher in open channel flows than in closed ones. But it could as well be caused by the still marginal domain of the latter study, which might have been insufficient with respect to capturing all the large-scale motions. Nevertheless, the effects of the outer layer VLSM on the mean flow variables near the wall are negligible small, cf. figure 4.6. Since the local peak does not appear in the lower Reynolds number cases the penetration of VLSM into the wall-near layer is believed to be an effect of high Reynolds numbers.

The scales of the LSM and VLSM respectively for turbulent open channel flow will be estimated by analysis of one-dimensional premultiplied energy spectra of streamwise velocity fluctuation in the upcoming sections with respect to spanwise scales and streamwise scales separately.

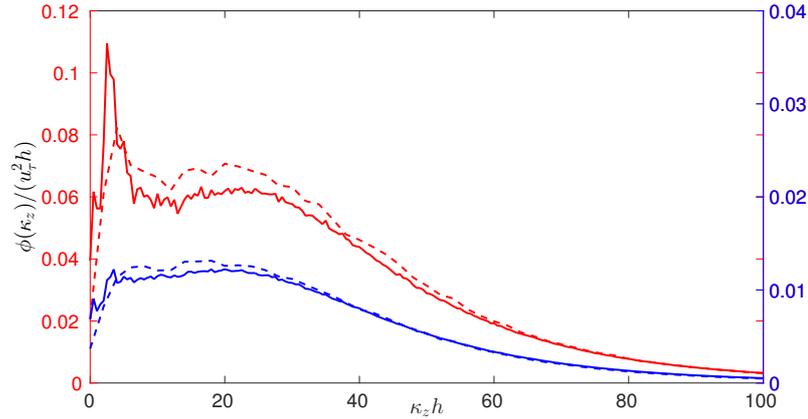

Figure 4.15.: One-dimensional spanwise wave number energy spectra of the stream- and spanwise velocity fluctuations for $Re_\tau = 600$ at $y^+ = 5.04$: —, $\phi_{uu}(\kappa_z)/(u_\tau^2 h)$; —, $\phi_{ww}(\kappa_z)/(u_\tau^2 h)$. Dashed lines indicate Moser et al. (1999) closed channel data for $Re_\tau = 590$ at $y^+ = 5.34$.

### 4.3.1. Spanwise Scales

A simulation with $Re_\tau = 600$ and box size of $L_x \times L_z = 5\pi h \times 2\pi h$ (cf. table 4.5) resulted in spanwise spectra close to the channel surface, which did not exhibit a peak at all, as. figure 4.16 indicates. The spanwise extension of the numerical box was apparently too small to contain the VLSM and the box width was extended to $L_z = 4\pi h$, which resulted in spanwise premultiplied energy spectra that exhibited peaks up to the free surface.

| case | $Re_\tau$ | $L_x$ | $L_z$ | $\alpha$ | $\beta$ | $N_x \times N_y \times N_z$ | $\Delta x^+$ | $\Delta z^+$ | $\Delta y_{max}^+$ |
|------|-----------|-------|-------|----------|---------|------------------------------|--------------|--------------|---------------------|
| oc600" | 600 | 5πh | 2πh | 0.2 | 0.5 | $1024 \times 257 \times 1024$ | 9.1 | 3.6 | 3.6 |

Table 4.5.: Intermediate case for open channel simulation, h=2

Figure 4.17 shows the spanwise premultiplied spectra of the streamwise velocity fluctuation for all Reynolds number cases of the current work in the $(y,z)$-plane. The spectra have been normalised by their peak value of each $y$-location. The colormap of the energy contour reaches from 0.8 times the local maximum (dark blue) to local maximum value (dark red). The black-and-white lines indicate 0.9 times the local maximum of the corresponding data from closed channel simulation of Moser et al. (1999). For the near-wall spectra the well known peak at $\lambda_z^+ \approx 100$ appears for all cases, corresponding to the average spacing of the near-wall streaks. Moving away from the wall leads to a shift of the peak position towards longer wavelengths. With increasing Reynolds numbers the near-wall and outer layer peaks become more separated having a $\kappa^{-1}$ range in between at $0.2 \lesssim y/h \lesssim 0.4$. Complete spanwise one-dimensional premultiplied energy spectra at different $y$ locations are shown in the appendix of this work in figure A.7.

Near the wall closed and open channel flow data collapse well. In the surface-near region the peak wavelength in the spanwise premultiplied spectra is increased by a factor of two for open compared to closed channel flow within the Reynolds number range of this work. This is in agreement with the investigation by Handler et al. (1993) for marginal Reynolds number flow ($Re_\tau = 134$). While the growth of the spanwise extension of large-scaled motions is completed above the log layer for the highest Reynolds number case, it takes almost the full channel height for the structures to grow in the lower Reynolds number cases exhibiting their final width just below the free surface.

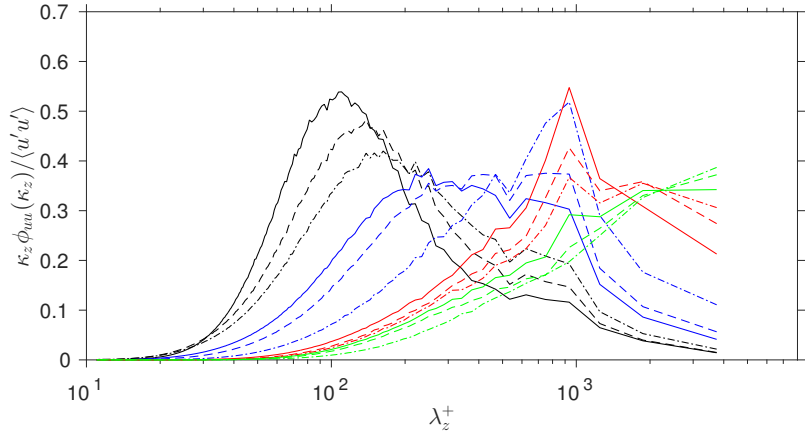

Figure 4.16.: One-dimensional premultiplied spanwise power spectra $\kappa_z \phi_{uu}(\kappa_z)/\langle u'u'\rangle$ as a function of $\lambda_z^+$ for the "oc600" case at different $y^+$ positions: —, $y^+$=5; - -, $y^+$=19; -·-, $y^+$=30; —, $y^+$=59; - -, $y^+$=80; -·-, $y^+$=151; —, $y^+$=298; - -, $y^+$=405; -·-, $y^+$=454; —, $y^+$=503; - -, $y^+$=541; -·-, $y^+$=595.

The contour lines in figure 4.18 indicate 0.9 times the maximum value of the spanwise one-dimensional premultiplied energy spectra for the different Reynolds number cases. The figure shows the near-wall peaks to collapse in wall units at a wavelength of $\lambda_z^+ \approx 100$, whereas the ones at the free surface collapse in bulk units at a wavelength of $\lambda_z/h = \pi$. The latter value is double the wavelength obtained from the closed channel data of Moser et al. (1999) or Abe, Kawamura, and Choi (2004), who reported $\lambda_z/h \approx 1.3 - 1.6$ for a similar range of Reynolds numbers. Note that due to the usage of discrete wave modes in numerical simulation the true spacing of the VLSM in physical open channel flow should be at least within the values $3/4\pi \leq \lambda_z/h \leq 3/2\pi$.

Likewise in closed channel flow the spanwise wavelength of the most energy containing motions grows almost linearly within the logarithmic wall layer between $y^+ = 30$ and $y/h$=0.3. In this region of self-similarity the spanwise spacing of the high energetic motions follows approximately the law of

$$\lambda_x = 5y, \; y^+ > 30 \; and \; y/h < 0.3 \tag{4.16}$$

indicated by the dotted black line in figure 4.18. At the edge of the similarity region the spanwise extension of the most energy containing motions reaches a value of $\lambda_z/h \approx 1.3 - 1.6$ for $Re_\tau = 200, 400$, before growing again in the vicinity of the free surface ($\lambda/h > 0.8$), finally reaching a value of $\pi$. In case of closed channel flow, in contrary, the value stays at 1.3-1.6 up to the channel centerline. For the highest Reynolds number case the influence of the free surface becomes more dominant throughout most of the flow domain and the growth of the maximum wavelength is finished at the end of the similarity region where its value already reached the open channel surface value of $\pi$.

Since the spanwise spacing of VLSM was found to be $\lambda_z = \pi h$ in bulk units independent of the Reynolds number it has yet to be clarified why a box with spanwise extension of $L_z = 2\pi$ was insufficiently large in terms of capturing VLSM in the vicinity of the free surface in the case of $Re_\tau = 600$. This behaviour will be investigated in future work.

### 4.3.2. Streamwise Scales

For the streamwise scales the limitation of the numerical box is even more apparent, especially for the highest Reynolds number case, as shown in figure 4.3b, and therefore the results should be treated with caution. Figure 4.19 shows the premultiplied energy spectra of the streamwise velocity fluctuations normalized the same way as in figure 4.17. Again the black-and-white lines indicate 0.9 times the local energy maximum at each $y$-position for closed channel data (Moser et al., 1999). The complete stremwise one-dimensional premultiplied energy spectra at different $y$ locations can be found in the appendix in figure A.8.

For the lower Reynolds cases the profiles of the streamwise premultiplied spectra match well with the ones from closed channel simulation (Moser et al., 1999) throughout most of the flow domain. Near the free surface they tend to slightly larger values before dropping back to closed channel values at the free surface itself. For the highest Reynolds number case a $\kappa^{-1}$ ranges seem to establish above around $y^+ \approx 60$ (cf. figure A.8). The collapse of the wavelength of the energetic peak in the upper channel half with the size of the largest numerical wavelength indicates the numerical box to be too small to fully capture the streamwise energetic scales for $Re_\tau = 600$. The energetic scales for $Re_\tau = 200, 400$ shown in the aforementioned figure correspond rather to LSM than to VLSM, meaning that turbulent coherent structures with very large streamwise extension evolve with higher Reynolds numbers. The latter would be in agreement with Kim and Adrian (1999), who found the longest streamwise extension of most energetic motions corresponding to streamwise velocity fluctuation to raise between Reynolds numbers of $Re_\tau = 1058$ and $Re_\tau = 1987$ from $\lambda_x \approx 10R$ to $\lambda_x \approx 14R$ before reaching an asymptotic state.

When comparing the streamwise extension of the most energy containing motions for different Reynolds numbers the scaling of these motions become apparent. As for the spanwise scales the spacing of the streamwise ones collapses in wall units in the vicinity of the wall exhibiting a value of $\lambda_x^+ \approx 1000$, which is in good agreement with Moser et al. (1999) and the literature corresponding to wall-bounded turbulent flows. The spacing of the outer flow high energetic motions collapses fairly well in wall units, having values around $\lambda_x \approx (3-4)h$ throughout most of the flow domain before dropping to values of $\lambda_x \approx (2-3)h$. Streamwise similarity can not be found in this regime of Reynolds numbers, since the scale separation between the streamwise extension of near-wall structures and those in outer layer is too small. It is likely that a self-similar region will establish in case of higher Reynolds numbers, which for closed channel flows of $Re_\tau = 2000$ has been proven by Jiménez (2012).

In order to separate the scales between the highly anisotropic near-wall structures and the VLSM one can decompose the premultiplied energy spectra into two modes as Del Álamo and Jiménez (2003) described. Figure 4.21 shows stream- and spanwise premultiplied spectra, which have been obtained by integrating the two-dimensional velocity spectrum tensor of streamwise velocity fluctuations $\phi_{uu}(\kappa_x, \kappa_z)$ only over the wave numbers above (below) a certain barrier with the respect to one direction in order to obtain the one-dimensional spectrum with respect to the other direction. The streamwise spectrum in figure 4.21a for example has been obtained by integrating over all spanwise modes, where $\lambda_z/h < 0.75$. The spectra have been normalized with viscous units.

Structures related to modes where $\lambda_x^+ \approx 1000$ and $\lambda_z^+ \approx 100$ are highly concentrated near the wall at $y^+ \approx 15$, as figures 4.21(c,d) indicates. Those are signatures of the well known low-speed streaks appearing in all wall-bounded turbulent flows and scaling in wall units. The premultiplied spectra of structures where $\lambda_x > 5h$ and $\lambda_z > 0.75h$, such as shown in the figures 4.21(a,b), are signatures of very large-scale motions. Most of their energy appears to be at wavelengths much larger than $\lambda_x = 5h$ but unfortunately it is not possible to estimate their true streamwise extension from the current data, which the position of the maximum energy at the largest wavelengths indicates. Nevertheless, since Del Álamo and Jiménez (2003) were able to capture the VLSM in terms of decaying premultiplied energy spectra at the largest wavelengths within the same box size, as the red lines in figure 4.21 indicate, VLSM of open channel flow appear to be longer than the ones from closed channel flow at this Reynolds number. Figure 4.21a also shows that VLSM penetrate into the buffer layer, which has already been indicated in figure 4.15.

Recapitulating we found large-scale structures, which exhibit a spanwise extension of $\lambda_z = \pi$ at the free surface for all Reynolds number cases of this thesis, an increase by a factor of two compared to the structures at closed channel centerlines. If the scale-separation is sufficiently high ($Re_\tau = 600$) the linear growth of these structures within the similarity region ($y^+ > 30$, $y/h < 0.3$) became apparent.

For the streamwise scales of VLSM the picture was less clear. Low Reynolds number flows ($Re_\tau = 200, 400$) contained large-scale structures with a size of $\lambda_x \approx (3-4)h$ throughout most of the flow domain with a drop down to $\lambda_x \approx (2-3)h$ at the channel surface, which are similar results compared to closed channel flows. For the highest Reynolds number case ($Re_\tau = 600$) very large-scale motions seemed to appear, but could not be captured well, since they interfered with the box size. A simulation of the same Reynolds number having a larger numerical box will be subject to further investigation. None of the cases exhibited a similarity region corresponding to the streamwise extension of large-scale motions, since the scale separation at these Reynolds numbers is very little.

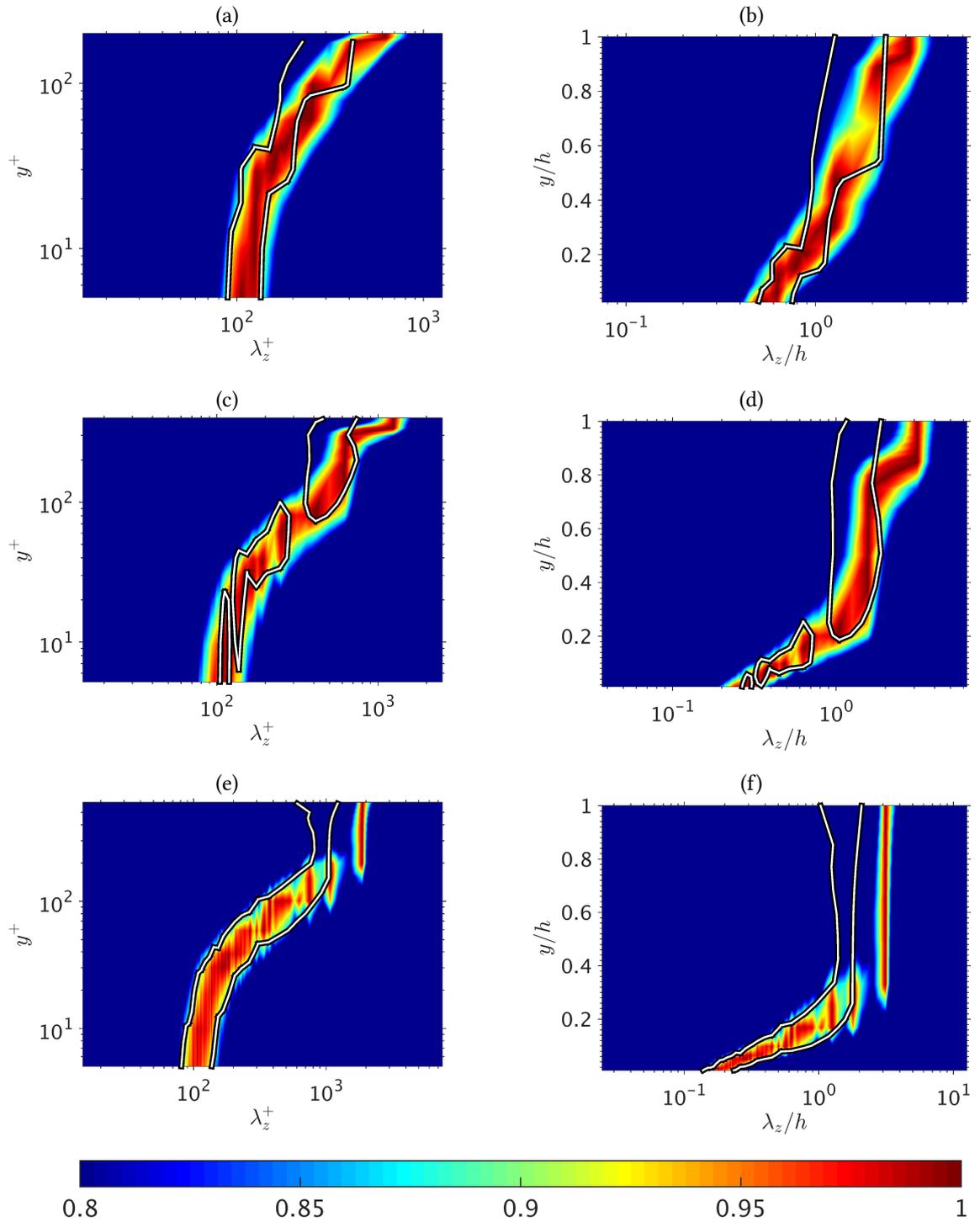

Figure 4.17.: Contour maps of premultiplied one-dimensional spanwise energy spectra of streamwise velocity fluctuations $\kappa_z \phi_{uu}(\kappa_z)$ as a function of the spanwise wavelength $\lambda_z$ and the distance from the wall $y$. The spectra are normalized with the local maximum of each y-plane. Black-and-white lines indicate 0.9 times the local energy maximum at each $y$-position for closed channel data (Moser et al., 1999). (a,b): $Re_\tau = 200$, (c,d): $Re_\tau = 400$, (e,f): $Re_\tau = 600$; (a,c,e): wall units, (b,e,f): outer flow units.

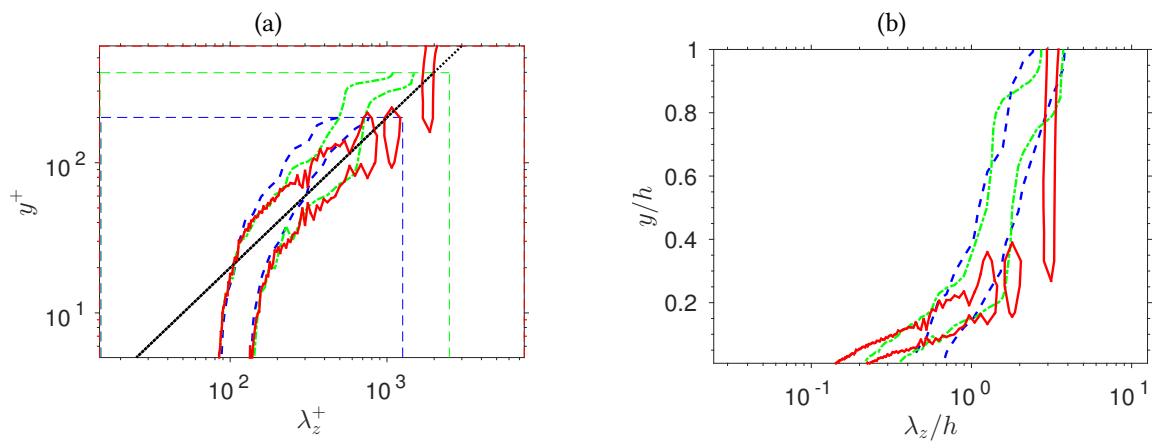

Figure 4.18.: Contour lines of premultiplied one-dimensional spanwise energy spectra of streamwise velocity fluctuations $\kappa_z \phi_{uu}(\kappa_z)$ of open channel simulations as a function of the spanwise wavelength $\lambda_z$ and the distance from the wall $y$. The contour lines indicate values of 0.9 times the local maximum. $\cdot\text{-}\cdot$, $Re_\tau = 200$; $\text{-}\text{-}\text{-}$, $Re_\tau = 400$; $\text{——}$, $Re_\tau = 600$. Dashed thin lines indicate the corresponding box size.

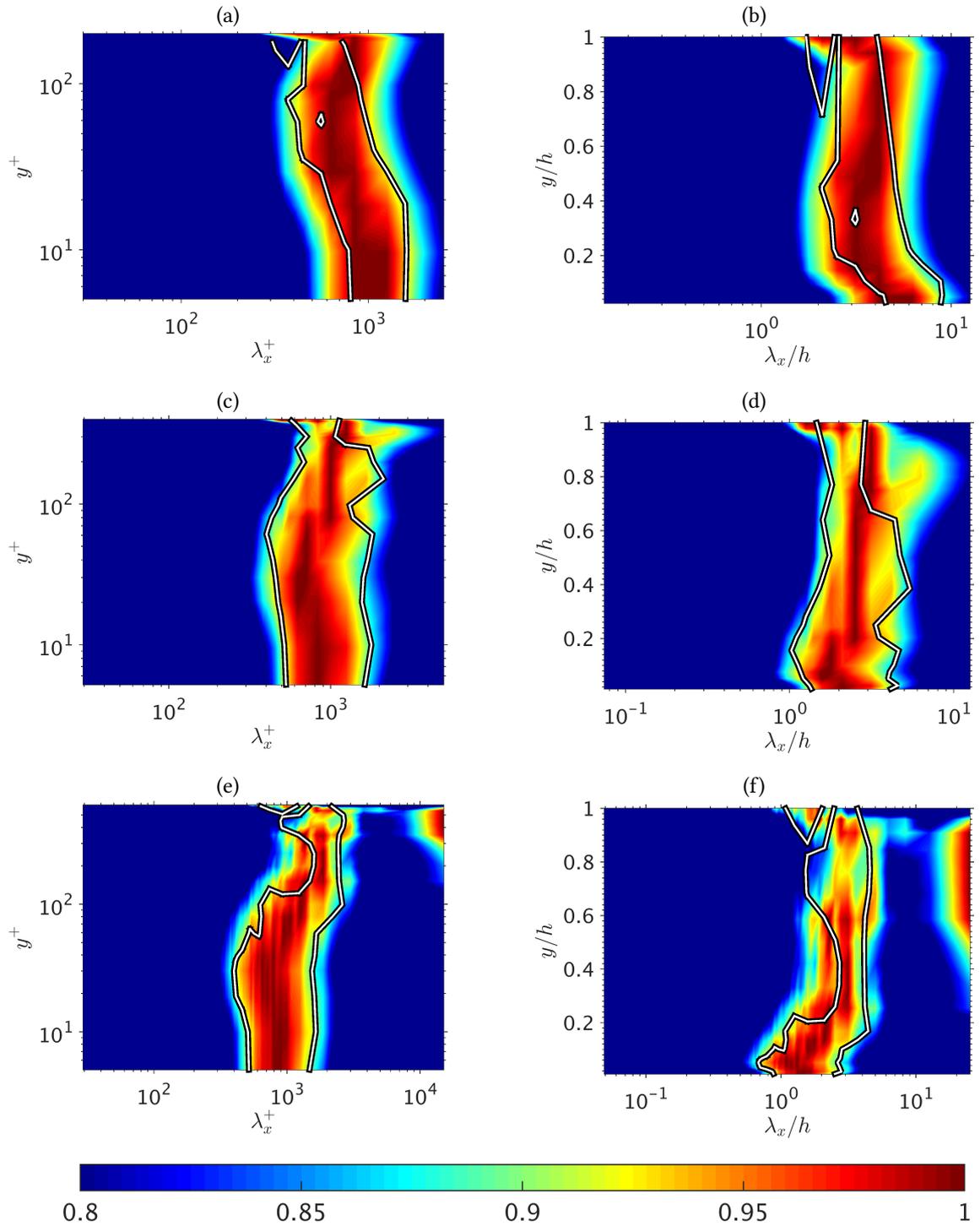

Figure 4.19.: Contour maps of premultiplied one-dimensional streamwise energy spectra of streamwise velocity fluctuations $\kappa_x \phi_{uu}(\kappa_x)$ as a function of the streamwise wavelength $\lambda_z$ and the distance from the wall y. The spectra are normalized with the local maximum of each y-plane. Black-and-white lines indicate closed channel data (Moser et al., 1999). (a,b): $Re_\tau = 200$, (c,d): $Re_\tau = 400$, (e,f): $Re_\tau = 600$; (a,c,e): wall units, (b,e,f): outer flow units.

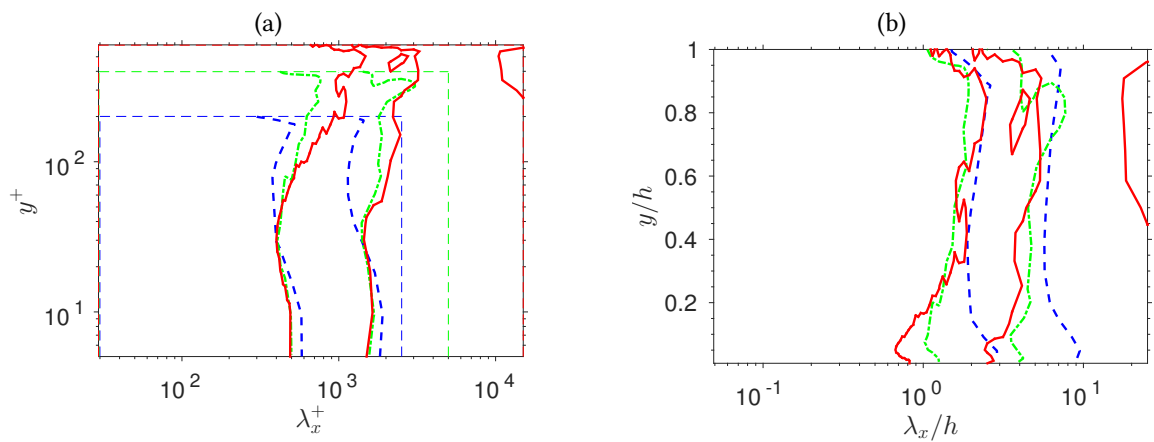

Figure 4.20.: Contour lines of premultiplied one-dimensional spanwise energy spectra of streamwise velocity fluctuations $\kappa_z \phi_{uu}(\kappa_z)$ of open channel simulations as a function of the spanwise wavelength $\lambda_z$ and the distance from the wall $y$. The contour lines indicate values of 0.9 times the local maximum. - -, $Re_\tau = 200$; - - -, $Re_\tau = 400$; ——, $Re_\tau = 600$. Dashed thin lines indicate the corresponding box size.

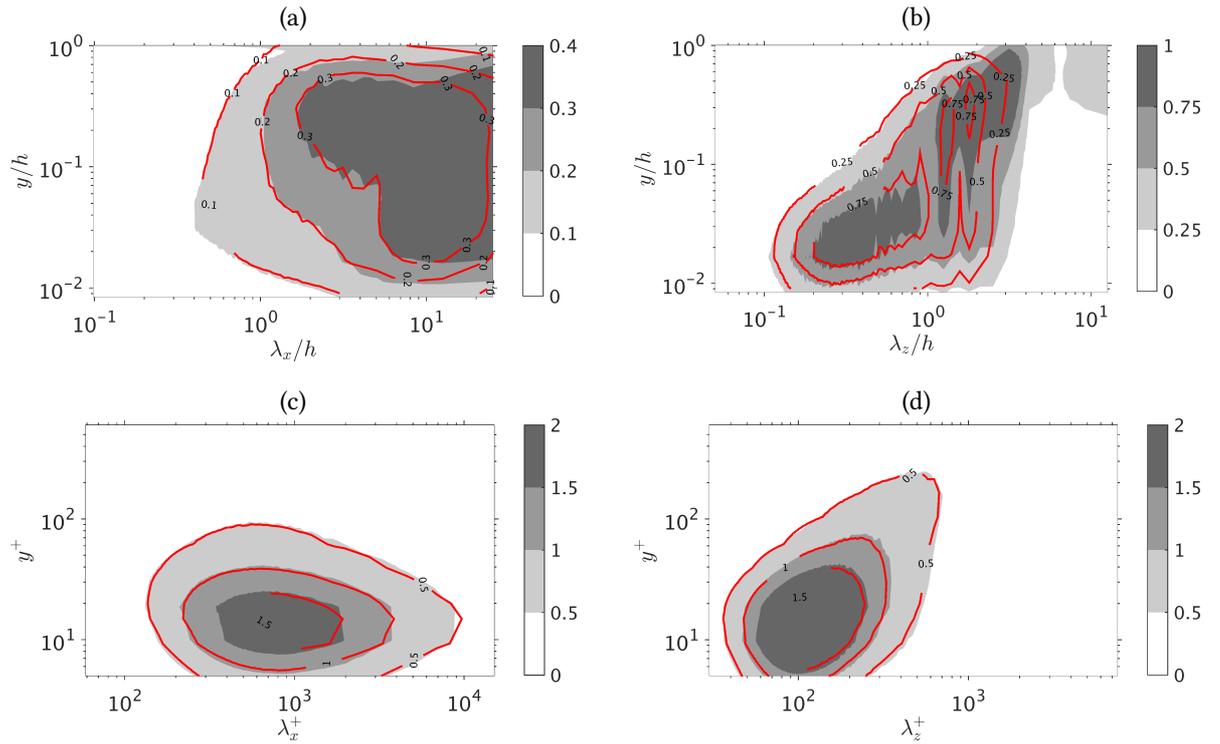

Figure 4.21.: Contour maps of decomposed premultiplied one-dimensional energy spectra of streamwise velocity fluctuations as a function of wavelength and wall distance for $Re_\tau = 600$. (a,c): Streamwise spectra $\kappa_x \phi_{uu}(\kappa_x)/u_\tau^2$ as a function of the streamwise wavelength $\lambda_x$, (b,d) spanwise spectra $\kappa_z \phi_{uu}(\kappa_z)/u_\tau^2$ as a function of the spanwise wavelength $\lambda_z$. (a,b): outer flow units, (c,d): wall units. (a) Spanwise modes with $\lambda_z > 0.75h$; (b) streamwise modes with $\lambda_x > 5h$; (c) spanwise modes with $\lambda_z < 0.75h$; (d) streamwise modes with $\lambda_x < 5h$. Red lines indicate closed channel data from Del Álamo and Jiménez (2003).

## 4.4. Instantaneous Flow Fields

In order to visualize the VLSM, isocontours of the instantaneous streamwise velocity fluctuations normalized by $u_\tau$ are shown in figure 4.22 for the lowest and highest Reynolds number case respectively. Since the intensity of the structures near the wall is much larger than in the outer flow, the contours in the lower channel half have been removed to clarify the picture. The instantaneous snapshots confirm the results obtained from spectra analysis. Figure 4.22 exhibits two streamwise elongated structures with the same velocity fluctuation value in the $Re_\tau = 200$ channel having $L_z = 2\pi$ and four in the $Re_\tau = 600$ one with $L_z = 4\pi$, leading to an instantaneous spanwise spacing of $\pi$ and matching the average spanwise spacing well. The streamwise extension of the instantaneous structures is in the order of the box length $L_x$, meaning that much longer numerical boxes are necessary for capturing all of the turbulent large-scales. A wider range of length scales of the higher Reynolds number structure can furthermore be extracted from figure 4.22, which is not surprising, since the global dimension of the structure scales with $h$ and higher Reynolds numbers cover a wider range of scales.

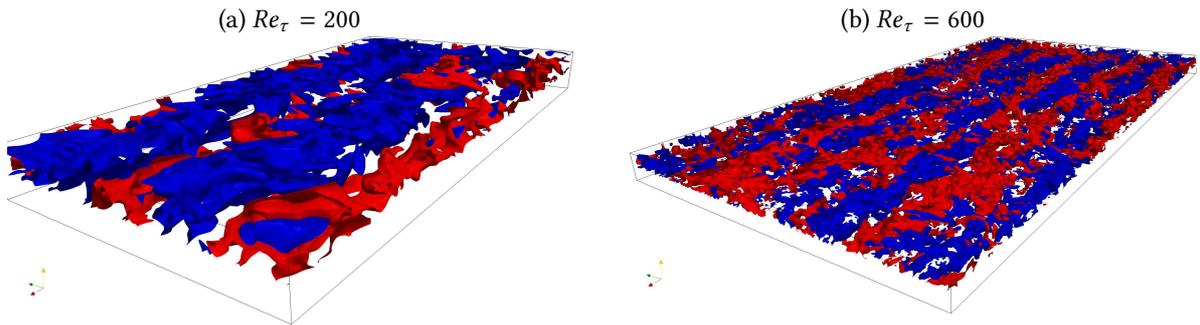

(a) $Re_\tau = 200$        (b) $Re_\tau = 600$

Figure 4.22.: Isocontours of the instantaneous streamwise velocity fluctuations nomalized by $u_\tau$ of the lowest and highest Reynolds number case respectively. Low velocity surfaces are blue ($u'/u_\tau$=-1), high velocity ones red ($u'/u_\tau$=1). The lower channel half has been hidden due to better visibility.

According to Nagaosa (1999) two types of coherent vortical structures exist below the free surface of an open channel flow. One elongated near-horizontal, quasi-streamwise vortex parallel to the free surface and the surface-attached vortex perpendicular to it. The latter studies also verified the observations of Kumar et al. (1998) and Pan and Banerjee (1995) that surface-attached vortices are long living and referred it to a small dissipation rate of the surface-normal vorticity.

The evolution of quasi-streamwise vortices from near-wall towards the free surface is well described in Nagaosa (1999) and shall further be elucidated through instantaneous flow field visualizations. According to Schoppa and Hussain (1997) quasi-streamwise vortices are generated nearby a lifted low-speed streak. The lift up of low-speed streaks is a dominant mechanism of near-wall turbulence and corresponding to a bursting event, which is the major Reynolds stress producing event. Furthermore the quasi-streamwise vortices move towards the free surface accompanying the lifted streak. The vortex pair then produces high intensities of vertical velocity fluctuations when interacting with the free surface, so-called splats ($v' > 0$) and anti-splats ($v' < 0$). Surface-attached vortices are produced by the connection of quasi-streamwise vortices with the free surface. According to Nagaosa (1999) the quasi-streamwise vortices are mainly responsible for splats and antisplats whereas the surface-attached ones play a less significant role. Perot and Moin (1995) and Walker et al. (1996) described splats and antisplats to be the two types of motion that determine the intercomponent energy transfer at the free surface, cf. figures 4.6 and 4.11.

In order to identify vortex tubes a Q-criterion has been computed. Hunt et al. (1988) defines a vortex as connected fluid with a positive second invariant of the velocity gradient tensor $\nabla \mathbf{u}$. According to Jeong and Hussain (1995) the second invariant Q is defined as

$$Q = \frac{1}{2}(\|\Omega\|^2 - \|\mathbf{S}\|^2) = \frac{1}{2}\left(\frac{\partial u_i}{\partial x_i}\frac{\partial u_i}{\partial x_i} - \frac{\partial u_i}{\partial x_j}\frac{\partial u_j}{\partial x_i}\right) = -\frac{1}{2}\frac{\partial u_i}{\partial x_j}\frac{\partial u_j}{\partial x_i} \qquad (4.17)$$

where $\|\mathbf{S}\| = \sqrt{tr(\mathbf{S}\mathbf{S}^T)}$, $\|\Omega\| = \sqrt{tr(\Omega\Omega^T)}$. $S_{ij} = \frac{1}{2}\left(\frac{\partial u_i}{\partial x_j} + \frac{\partial u_j}{\partial x_i}\right)$ and $\Omega_{i,j} = \frac{1}{2}\left(\frac{\partial u_i}{\partial x_j} - \frac{\partial u_j}{\partial x_i}\right)$ are the symmetric and antisymmetric components of $\nabla \mathbf{u}$ and known as the rate-of-strain tensor and the vorticity tensor respectively. Q defines vortices as areas where the vorticity magnitude is greater than the magnitude of rate-of-strain. The criterion also adds the condition on the pressure inside the vortex to be lower than ambient pressure, cf. Holmén (2012).

Figure 4.23a shows a pair of lifted quasi-streamwise vortices from the $Re_\tau = 200$ case beneath the free surface accompanied by splats and anti-splats. High positive (negative) intensities of vertical velocity fluctuations in the angled vertical plane beneath the free surface correspond here to splats (antisplats). The quasi-streamwise vortices are accompanied by a lifted low-speed streak originating from the buffer layer, as figure 4.23b indicates. The ejection event responsible for the lift-up of the low-speed streak together with the (anti-)splats induced by the quasi-streamwise vortices is responsible for pressure fluctuations at the free surface, which are represented by an elevated contour surface in figure 4.23b.

Surface-attached vortices identified by the Q-criterion are shown in Figure 4.24, where surrounding vortices with the same Q have been removed for better visibility. The figure contains a part of the full channel indicated by the square in figure 4.25. The instantaneous picture shows a surface-attached vortex, possessing a tail reaching down into the wall region, as well as one that already got rid of it.

A comparison of the pressure distribution at the free surface with the vertical vorticity component can be extracted from figure 4.26. Since surface parallel vortices become damped to zero at the free surface, as mentioned above (cf. equation 4.12), only the vertical vorticity component remains. Areas with high intensity of vertical vorticity correspond to

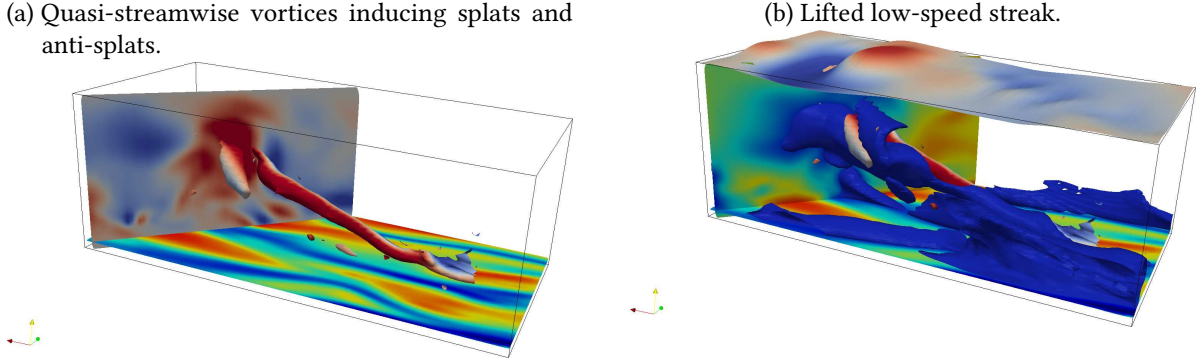

(a) Quasi-streamwise vortices inducing splats and anti-splats.

(b) Lifted low-speed streak.

Figure 4.23.: (a): Isocontours show quasi-streamwise vortices identified by Q-criterion having $Qh^2/u_\tau^2 = 494.5$. Their surface is covered with contours of the verctial velocity fluctuation, reaching from dark blue ($v'/u_\tau = -1.76$) to dark red ($v'/u_\tau = 1.76$). The angled vertical plane shows the distribution of the vertical velocity fluctuation as well, using the same colormap. Direction of the main flow is from right to left. The lower horizontal plane shows the streamwise velocity fluctuation as contours, reaching from dark blue ($u'/u_\tau = -5.27$) to dark red ($u'/u_\tau = 5.27$). (b) shows the same vortices as (a) together with isocontours of a lifted low-speed streak having $u'/u_\tau = -3.52$. The angled vertical plane shows the distribution of the streamwise velocity flucutation here using the same colormap as the contours of the lower horizontal plane. The upper plane shows contours of the instantaneous pressure distribution as an elecated surface having a colormap reaching from $p/(\rho u_\tau^2) = -3.09$ (dark blue) to $p/(\rho u_\tau^2) = 3.09$ (dark red). $Re_\tau = 200$. Box size $b_x \times b_y \times b_z = 0.2L_x \times L_y \times 0.2L_z$, corresponding to the region marked with a square and diagonal in figure 4.26.

surface-attached vortices, such as pictured in figure 4.24. The very local negative peaks in the pressure distribution correspond hereby with high positive or negative intensities of the vertical vorticity respectively as the arrows indicate. This implies that these negative pressure peaks indicate the position of surface-attached vortices, since the pressure in the vortex core is lower than ambient pressure, as mentioned before.

Figure 4.27 shows the vertical component of the vorticity at the free surface on a semi-transparent plane from above. Beneath the three-dimensional VLSM are plotted the same way as in figure 4.22. The figure indicates that the preferential location of high intensive surface-normal vorticity appears to be in the region between the structures corresponding to positive and negative fluctuations respectively. Furthermore the instantaneous pictures indicate the spacing of the vertical vorticity component to scale in wall units, which would agree with its aforementioned relation to the surface of VLSM. Nevertheless the statistical proof has yet to be provided.

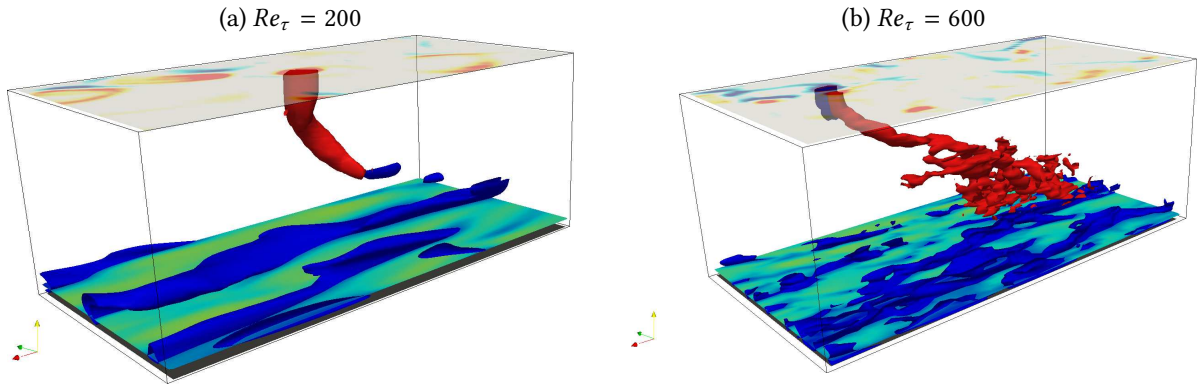

Figure 4.24.: Isocontours show surface-attached vortices identified by Q-criterion. Direction of the main flow is coming out of the paper. The lower channel part contains low velocity streaks shown as isocontours. The channel top is covered by a semi-transparent contourplot of vertical vorticity $\omega_y$, cf. figure 4.25. (a) $Re_\tau = 200$, box size $b_x \times b_y \times b_z = 0.2L_x \times L_y \times 0.2L_z$ with vortices having $Qh^2/u_\tau^2 = 185.4$ and low velocity streaks of $u'/u_\tau = -1.76$. (b) $Re_\tau = 600$, box size $b_x \times b_y \times b_z = 0.1L_x \times L_y \times 0.1L_z$ with $Qh^2/u_\tau^2 = 253.4$ and low velocity streaks of $u'/u_\tau = -2.06$.

In summary VLSM could be captured within instantaneous realizations of flow fields. The two dominant vortex families of free surface flows, near-surface quasi-streamwise vortices and surface-attached vortices, could be tracked having the latter preferentially located in the region between VLSM, corresponding to positive streamwise velocity fluctuations and negative ones respectively. The vortices are assumed to be mainly responsible for the pressure distribution at the free surface since their location coincidences with either very large negative peaks in the pressure distribution in case of surface-attached vortices or more extensive negative or positive regions of pressure fluctuation in case of quasi-streamwise vortices, which induce splats and anti-splats that impact on the free surface. The latter events are known to be mainly responsible for intercomponent energy transfer (Nagaosa, 1999).

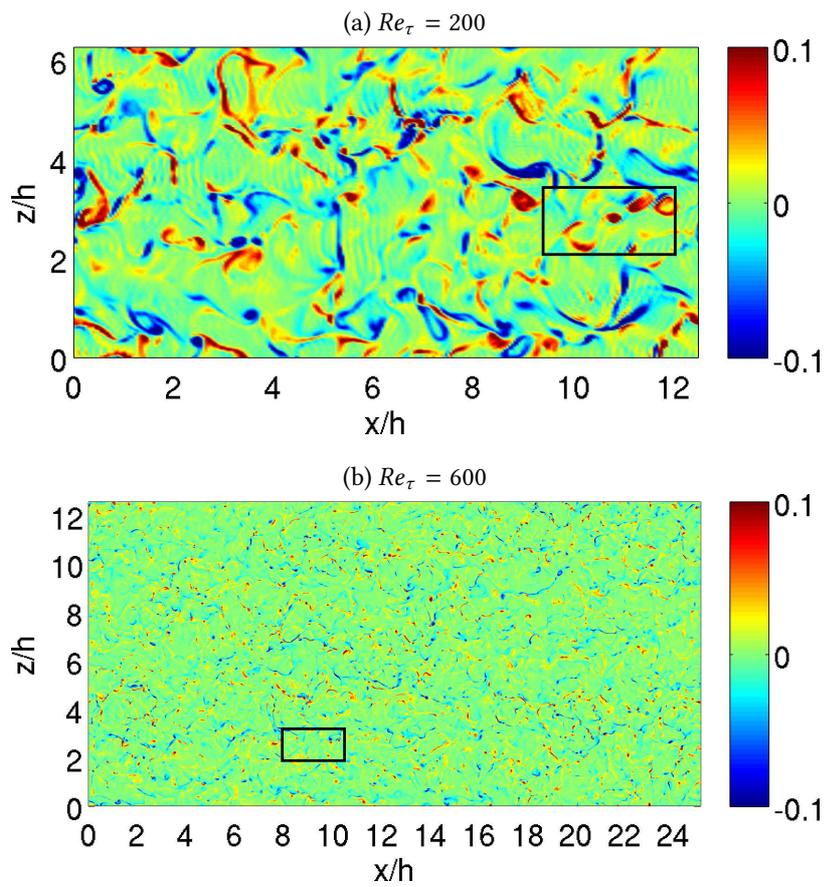

Figure 4.25.: Visualization of the normalized instantaneous y-component of vorticity ($\omega_y \nu / u_\tau^2$) at the channel top, (x,z) plane. The square indicates the box used in figure 4.24.

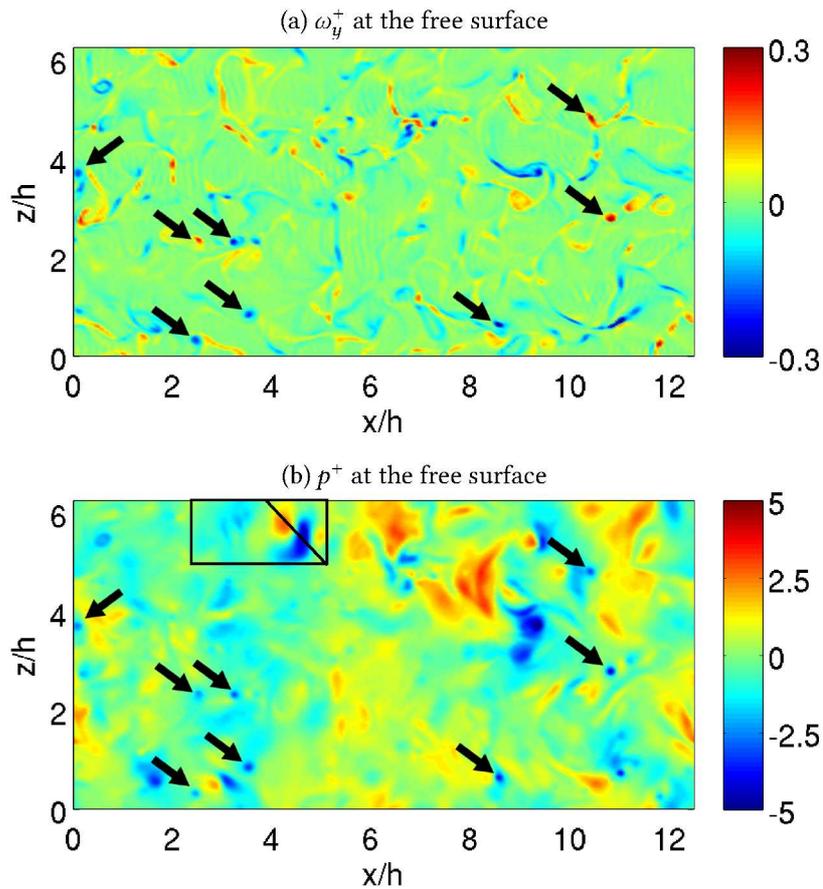

Figure 4.26.: (a): Normalized instantaneous y-component of vorticity $(\omega_y v / u_\tau^2)$ at the free surface. (b): Normalized instantaneous pressure distribution $(p/(\rho u_\tau^2))$ at the free surface. Arrows show the position of very local negative peaks w.r.t. the pressure and rather positiv or negative ones w.r.t. the vertical vorticity. The square indicates the box used in figure 4.23. $Re_\tau = 200$.

(a) $Re_\tau = 200$ at the free surface

(b) $Re_\tau = 400$ at the free surface

(c) $Re_\tau = 600$ at the free surface

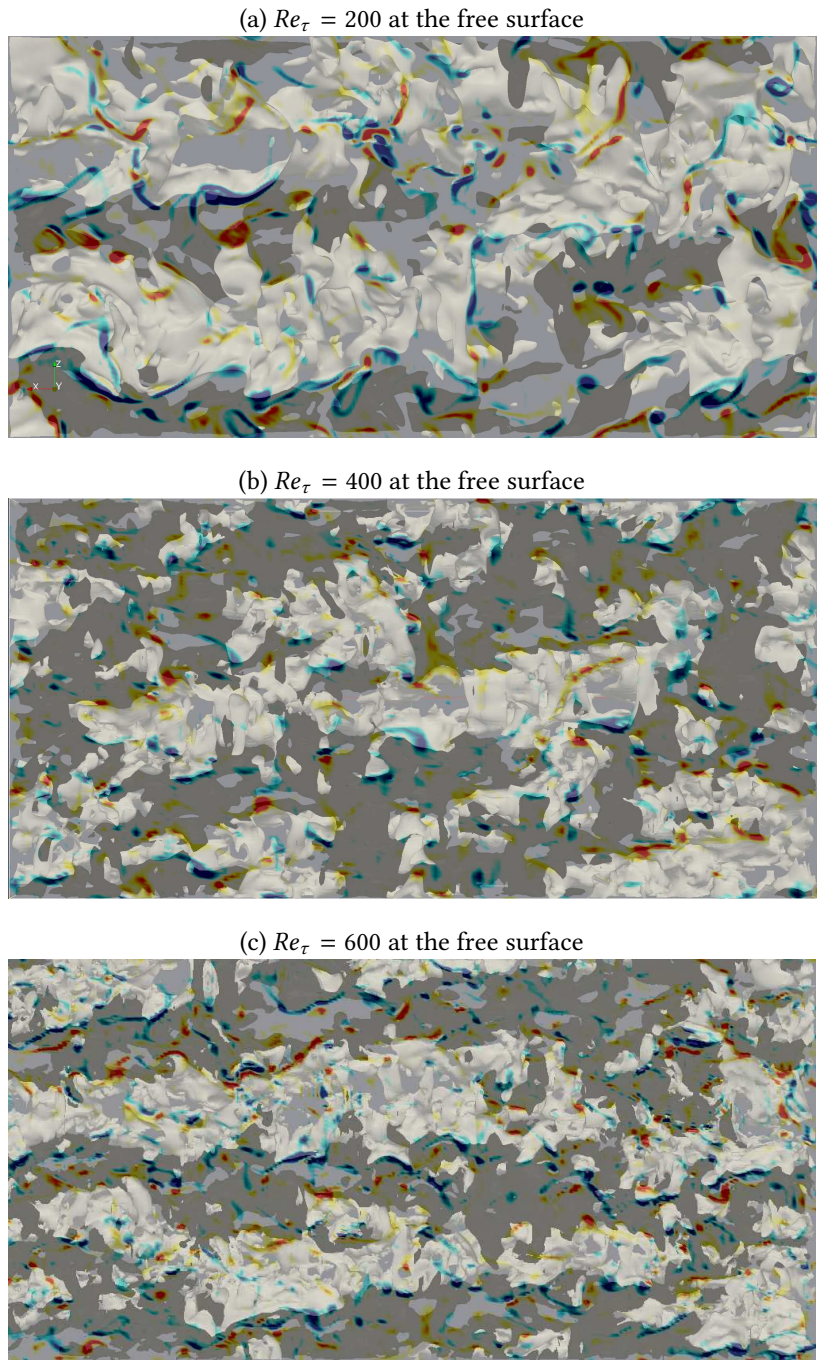

Figure 4.27.: Isocontours of the instantaneous streamwise velocity fluctuations of the upper channel half (black conours indicate structures of positive velocity fluctuations, white of low velocity ones) superimposed by a semi-transparent plane of the vertical vorticity component at the channel top. The view on the channel is from above. The flowdirection is from right to left. For the $Re_\tau = 600$ case a box of the size $b_x \times b_z = 0.5L_x \times 0.5L_z$ is shown for compareability.

# 5. Conclusion

The outcome of this work will be summarized and future work will be determined in the concluding chapter.

## 5.1. Summary and Conclusion

Direct numerical simulations of open channel flow have been carried out for friction Reynolds numbers of $Re_\tau = 200, 400, 600$. While for the lower Reynolds number cases all turbulent scales of interest could be spatially resolved well ($L_x \times L_z = 4\pi h \times 2\pi h$, $\Delta x^+ \times \Delta z^+ \approx 10 \times 5$), the highest case revealed problems possibly at both ends of the range of scales. A grid spacing of $\Delta z^+ \approx 5$ in the spanwise direction might have been too coarse to resolve the smallest turbulent scales at $Re_\tau = 600$ and therefore responsible for the deviation of the streamwise Reynolds stress profile compared to experimental results, which is currently subject to further investigations. Furthermore, a streamwise box extension of $L_x = 8\pi h$ appeared to be insufficient in order to capture the largest turbulent structures of the flow, which led to the appearance of spurious peaks in the streamwise one-dimensional premultipled energy spectra $\kappa \phi_{uu}$ in the vicinity of the free surface at wavelengths corresponding to the box size. The same phenomenon occurred in the spanwise direction somewhat weaker, where the box size had been extended to $L_z = 4\pi h$ and premultipled energy spectra were still exhibiting large amounts of energy at wavelengths of $\lambda_z = L_z$. The same can be observed from the two-point velocity correlations with respect to stream- and spanwise directions, which did not decay to zero within the current domain size. Therefore the results obtained from the highest Reynolds number case have to be treated extremely cautiously and further simulations involving higher resolutions and larger boxes are necessary in order to validate the current results.

An influence of the free surface on the mean velocity profile over the logarithmic region does not present within the data of this work. The log law of wall-bounded turbulence appeared to be also valid in the case of open channel flows, and a comparison of our data with the one from Moser et al. (1999) revealed no significant disagreement, including the low Reynolds number effect to act the same way on the log law constant $B$ as it does for closed channel flows. Beyond the logarithmic region, open channel mean velocity profiles differ from closed channel ones by following the log law even up to the free surface, which confirms the well known feature of open channel flow that was first discovered by Keulegan (1938).

Turbulence statistics revealed the existence of dual-layer structure of the free-surface boundary layer, which exhibits strong anisotropy both in Reynolds stresses and vorticity.

While the anisotropy layer of the Reynolds stresses scaled in the outer flow units, $\delta_v \approx 0.3h$, the anisotropy layer of the root mean square vorticity fluctuations was found to scale in wall units, having $\delta_\omega \approx 10\delta_v$. Previous investigations on free surface turbulence, such as Handler et al. (1993), suggested the pressure-strain term in the transport equations of the Reynolds stress components to be the key contributor to the intercomponent energy transfer. The same was found to be true for the different cases of this investigation, where the pressure-strain term becomes dominant within a thin layer near the free surface ($\delta_\Pi \approx 0.1h$) corresponding to the most anisotropic region in the vicinity of the free surface.

Production to dissipation ratio in the turbulent kinetic energy budget equation appeared to be higher beneath the free surface of an open channel flow than in the corresponding region below the centerline of closed channel flows. This is caused by lower dissipation in case of the open channel flow, whereas production of turbulent kinetic energy stays about the same order for both types of flow. The lower dissipation is balanced by the transport terms of viscous diffusion, turbulent convection and pressure transport, which exhibit striking different behaviour in the vicinity of the free surface than in the closed channel case.

An analysis of one-dimensional premultipled energy spectra of streamwise velocity fluctuations revealed large scale coherent structures as a form of the most energetic wavelength, which exhibit a spanwise extension of $\lambda_z = \pi$ at the free surface independent of the Reynolds number. This means an increase by a factor of two compared to the structures at closed channel centerlines. If the scale-separation is sufficiently high ($Re_\tau = 600$) the linear growth of these structures within the similarity region ($y^+ > 30$, $y/h < 0.3$) became apparent.

Regarding the streamwise direction, the lower Reynolds number simulations showed large-scale structures with a size of $\lambda_x \approx (3-4)h$ throughout most of the flow domain with a drop down to $\lambda_x \approx (2-3)h$ at the channel surface, which are similar results compared to closed channel flows. For the highest Reynolds number case ($Re_\tau = 600$) very large-scale motions seemed to appear, but could not be captured well, since they interfered with the box size as mentioned above. Nevertheless, since VLSM were captured in closed channel flow simulations with similar box size they are believed to be longer in open channel flow ($\lambda_x > 10h$). Due to insufficient scale separation, a similarity region with respect to the streamwise extension of large scale motions could not be established at these Reynolds numbers.

Very large-scale motions could be captured within visualizations of instantaneous flow field realizations in terms of isocontours of streamwise velocity fluctuations. In terms of vortical structures the two dominant species near the free surface, surface-attached vortices and near-surface quasi-streamwise vortices, could be tracked. The evolution of the latter structures, as it is described in Nagaosa (1999), could be extracted from instantaneous visualizations and their impact on the pressure distribution on the free surface has been acknowledged. Furthermore instantaneous plots of the wall-normal vorticity component together with isocontours of the streamwise velocity fluctuations beneath the surface suggested, that the preferential location of surface-attached vortices is in between the VLSM in terms of spanwise spacing. As the range of scales of the VLSM rises with the Reynolds number the size of the vortical structures in the vicinity of the free surface appears to scale in wall units, which statistically has yet to be quantified.

## 5.2. Outlook

As two-point correlations and spectral data of the $Re_\tau = 600$ case showed, the computational box of this work was too small to capture most energy containing streamwise VLSM in the vicinity of the free surface, cf. figures 4.3b and 4.19. The spanwise box extension appeared to be marginal at least, as figure 4.3d indicates. Furthermore, a discrepancy occurring in the averaged streamwise Reynolds stress profiles indicated the small scales to possibly be not fully resolved. Therefore, test cases involving higher spatial resolutions have been set up and are yet to be evaluated. Finally, a direct numerical simulation of the latter Reynolds number case involving a numerical box with both larger stream- and spanwise extension and sufficiently high grid resolution need to be performed in the future in order to determine the complete extension of VLSM in open channel flows and to validate the statistical data of this work with respect to the corresponding Reynolds number case.

Furthermore the current work mainly focussed on the VLSM corresponding to the streamwise velocity fluctuation, since they contribute the most to the kinetic energy budget. Investigations on the other velocity components could be interesting as well, especially those structures corresponding to the spanwise velocity fluctuation are expected to exhibit a large increase in streamwise extension when compared to closed channel flow structures (Handler et al., 1993). Structures related to $\langle v'v' \rangle$ or $\langle u'v' \rangle$ are, on the other hand, less likely to be significant, since their intensities decrease progressively to zero towards the free surface.

Although computations have not been cheap in terms of computational resources, the current simulations are still located somewhere between the marginal and the intermediate Reynolds number regime. With additional computational power, direct numerical simulations of significantly higher Reynolds number open channel flows, such as performed by Lee and Moser (2015) for closed channel flow, can be carried out. Such simulations are necessary to obtain clear separation of turbulent scales. Open channel VLSM can be then compared to the ones in the canonical wall-bounded flows. According to Kim and Adrian (1999) VLSM reach an asymptotic state in terms of streamwise extension for Reynolds numbers of $Re_\tau > 2000$, which makes higher Reynolds number simulations of open channel flows desirable. General conclusions for high Reynolds number flows, such as used in numerous civil engineering applications, could then be drawn. Note that Monty et al. (2009) showed that VLSM behave similarly in closed channel and pipe flows, exhibiting a streamwise extension of $\lambda_x = (14 - 20)h$ at friction Reynolds numbers of $Re_\tau \approx 3000$. According to the data of this work it is not unlikely that corresponding structures in open channel flows of the same Reynolds number would exhibit longer extension, which is yet to be confirmed.

In order to determine robust results corresponding to the spacing of vorticity and pressure fluctuations the latter values should be treated statistically and their spectra should be analysed such as done for the velocity fluctuations in this work. Furthermore, the alignment between the wall-normal vorticity component at the free surface and the VLSM has as well yet to be quantified statistically.

# A. Appendix

## A.1. Fourth-order Transport Equation for Vertical Velocity

We write the non-dimensional momentum equation 2.8b in the following form:

$$\frac{\partial u_i}{\partial t} = -\frac{\partial p}{\partial x_i} + H_i + \frac{1}{Re}\frac{\partial^2 u_i}{\partial x_j \partial x_j}, \ i = 1,2,3, \tag{A.1}$$

where $H_i = -u_j \frac{\partial u_i}{\partial x_j}$ is the convective term. Differentiating the streamwise momentum equation with respect to x, vertical with respect to y and spanwise with respect to z, yields

$$\frac{\partial}{\partial x}\left(\frac{\partial u}{\partial t}\right) = -\frac{\partial^2 p}{\partial x} + \frac{\partial H_1}{\partial x} + \frac{1}{Re}\frac{\partial}{\partial x}\left(\frac{\partial^2 u}{\partial x_j \partial x_j}\right), \tag{A.2a}$$

$$\frac{\partial}{\partial y}\left(\frac{\partial v}{\partial t}\right) = -\frac{\partial^2 p}{\partial y} + \frac{\partial H_2}{\partial y} + \frac{1}{Re}\frac{\partial}{\partial y}\left(\frac{\partial^2 v}{\partial x_j \partial x_j}\right), \tag{A.2b}$$

$$\frac{\partial}{\partial z}\left(\frac{\partial w}{\partial t}\right) = -\frac{\partial^2 p}{\partial z} + \frac{\partial H_3}{\partial z} + \frac{1}{Re}\frac{\partial}{\partial z}\left(\frac{\partial^2 w}{\partial x_j \partial x_j}\right). \tag{A.2c}$$

Taking now the sum of A.2a, A.2b and A.2c we get an expression for the Laplacian of the pressure:

$$\frac{\partial}{\partial t}\underbrace{\left(\frac{\partial u}{\partial x} + \frac{\partial v}{\partial y} + \frac{\partial w}{\partial z}\right)}_{\nabla \mathbf{u}=0} = -\nabla^2 p + \left(\frac{\partial H_1}{\partial x} + \frac{\partial H_2}{\partial y} + \frac{\partial H_3}{\partial z}\right) + \frac{1}{Re}\frac{\partial^2}{\partial x_j \partial x_j}\underbrace{\left(\frac{\partial u}{\partial x} + \frac{\partial v}{\partial y} + \frac{\partial w}{\partial z}\right)}_{\nabla \mathbf{u}=0}$$
$$\tag{A.3}$$

$$\Leftrightarrow \nabla^2 p = \left(\frac{\partial H_1}{\partial x} + \frac{\partial H_2}{\partial y} + \frac{\partial H_3}{\partial z}\right) \tag{A.4}$$

Now we take the Laplacian of the y-component of the Navier-Stokes equations.

$$\nabla^2\left(\frac{\partial v}{\partial t}\right) = -\nabla^2\left(\frac{\partial p}{\partial y}\right) + \nabla^2 H_2 + \frac{1}{Re}\nabla^4 v$$

$$\Leftrightarrow \frac{\partial v}{\partial t}\nabla^2 v = -\frac{\partial}{\partial y}\left(\nabla^2 p\right) + \nabla^2 H_2 + \frac{1}{Re}\nabla^4 v$$

$$\overset{(A.4)}{=} -\frac{\partial}{\partial y}\left(\frac{\partial H_1}{\partial x} + \frac{\partial H_2}{\partial y} + \frac{\partial H_3}{\partial z}\right) + \left(\frac{\partial^2}{\partial x^2} + \frac{\partial^2}{\partial y^2} + \frac{\partial^2}{\partial z^2}\right)H_2 + \frac{1}{Re}\nabla^4$$

$$= \underbrace{-\frac{\partial}{\partial y}\left(\frac{\partial H_1}{\partial x} + \frac{\partial H_3}{\partial z}\right) + \left(\frac{\partial^2}{\partial x^2} + \frac{\partial^2}{\partial z^2}\right)H_2}_{h_v} + \frac{1}{Re}\nabla^4 \tag{A.5}$$

## A.2. Vertical Vorticity Equation

Let us define vorticity as

$$\boldsymbol{\omega} = (\nabla \times \mathbf{u}) = \begin{pmatrix} \frac{\partial w}{\partial y} - \frac{\partial v}{\partial z} \\ \frac{\partial u}{\partial z} - \frac{\partial w}{\partial x} \\ \frac{\partial v}{\partial x} - \frac{\partial u}{\partial y} \end{pmatrix} = \begin{pmatrix} \omega_x \\ \omega_y \\ \omega_z \end{pmatrix}. \tag{A.6}$$

In order to receive the vorticity transport equation we apply the curl operator to the momentum equation 2.6b.

$$\nabla \times \left( \frac{\partial \mathbf{u}}{\partial t} \right) + \nabla \times ((\mathbf{u} \cdot \nabla)\mathbf{u}) = -\nabla \times (\nabla p) + \frac{1}{Re} \nabla \times (\nabla^2 \mathbf{u}) \tag{A.7}$$

Using the definition of vorticity and vector identity A.21 the equation simplifies as follows:

$$\frac{\partial \boldsymbol{\omega}}{\partial t} + \nabla \times ((\mathbf{u} \cdot \nabla)\mathbf{u}) = \frac{1}{Re} \nabla^2 \boldsymbol{\omega} \tag{A.8}$$

With A.22 the second term of the equation above can be reformulated as:

$$\nabla \times ((\mathbf{u} \cdot \nabla)\mathbf{u}) = \frac{1}{2} \underbrace{\nabla \times (\nabla(\mathbf{u} \cdot \mathbf{u}))}_{=0, \text{ see A.21}} - \nabla \times (\mathbf{u} \times \underbrace{(\nabla \times \mathbf{u})}_{=\boldsymbol{\omega}})) \tag{A.9}$$

Then by using vector identity A.23 the term can transformed further to:

$$\nabla \times ((\mathbf{u} \cdot \nabla)\mathbf{u}) = -\nabla \times (\mathbf{u} \times \boldsymbol{\omega}) = -\mathbf{u} \underbrace{(\nabla \cdot \boldsymbol{\omega})}_{\nabla \times (\nabla \mathbf{u})=0} + \boldsymbol{\omega} \underbrace{(\nabla \cdot \mathbf{u})}_{\nabla \mathbf{u}=0} - (\boldsymbol{\omega} \cdot \nabla)\mathbf{u} + (\mathbf{u} \cdot \nabla)\boldsymbol{\omega} \tag{A.10}$$

We substitute the term back into equation A.8 and yield the vorticity transport equation.

$$\frac{\partial \boldsymbol{\omega}}{\partial t} + (\mathbf{u} \cdot \nabla)\boldsymbol{\omega} = (\boldsymbol{\omega} \cdot \nabla)\mathbf{u} + \frac{1}{Re} \nabla^2 \boldsymbol{\omega} \tag{A.11}$$

It's y-component

$$\frac{\partial \omega_y}{\partial t} + (\mathbf{u} \cdot \nabla)\omega_y = (\boldsymbol{\omega} \cdot \nabla)v + \frac{1}{Re} \nabla^2 \omega_y \tag{A.12}$$

can be rewritten to

$$\frac{\partial \omega_y}{\partial t} = \underbrace{\left( \frac{\partial H_1}{\partial z} - \frac{\partial H_3}{\partial x} \right)}_{h_\omega} + \nabla^2 \omega_y \tag{A.13}$$

where $H_1 = u_j \frac{\partial u}{\partial x_j}$ and $H_3 = u_j \frac{\partial w}{\partial x_j}$.

## A.3. Derivation of Fourier Coefficients of Velocity and Vorticity Components

The definition for the wall normal vorticity ($\omega_y = \frac{\partial u}{\partial z} - \frac{\partial w}{\partial x}$) leads to the Fourier coefficients:

$$\hat{\omega}_y = I\kappa_z \hat{u} - I\kappa_x \hat{w} \tag{A.14}$$

Using continuity equation in spectral space

$$I\kappa_x \hat{u} + \frac{d\hat{v}}{dy} + I\kappa_z \hat{w} = 0$$

$$\Leftrightarrow \quad I\hat{w} = -I\frac{\kappa_x}{\kappa_z}\hat{u} - \frac{1}{\kappa_z}\frac{d\hat{v}}{dy} \tag{A.15}$$

on equation A.14 gives us

$$\hat{\omega}_y = I\kappa_z \hat{u} + I\frac{\kappa_x^2}{\kappa_z}\hat{u} + \frac{\kappa_x}{\kappa_z}\frac{d\hat{v}}{dy}$$

$$= I(\kappa_z + \frac{\kappa_x^2}{\kappa_z})\hat{u} + \frac{\kappa_x}{\kappa_z}\frac{d\hat{v}}{dy} \ . \tag{A.16}$$

We then multiply both sides of the equation with $I\kappa_z$ to receive

$$I\kappa_z\hat{\omega}_y = -(\kappa_z^2 + \kappa_x^2)\hat{u} + I\kappa_x \frac{d\hat{v}}{dy}$$

$$\Leftrightarrow \quad \hat{u} = I\frac{\kappa_x \frac{d\hat{v}}{dy} - \kappa_z\hat{\omega}_y}{\kappa_x^2 + \kappa_z^2} \ . \tag{A.17}$$

For $\omega_z = (\frac{\partial v}{\partial x} - \frac{\partial u}{\partial y})$ we get in spectral space

$$\hat{\omega}_z = I\kappa_x \hat{v} - \frac{d\hat{u}}{dy}$$

$$\overset{(A.17)}{=} I\kappa_x \hat{v} - I\frac{\kappa_x \frac{d^2\hat{v}}{dy^2} - \kappa_z \frac{d\hat{\omega}_y}{dy}}{\kappa_x^2 + \kappa_z^2}$$

$$= I\frac{\kappa_x}{\kappa_x^2 + \kappa_z^2}\left(\underbrace{\kappa_x^2\hat{v} + \kappa_z^2\hat{v} - \frac{d^2\hat{v}}{dy^2}}_{=-\hat{\varphi}} - \frac{\kappa_z}{\kappa_x}\frac{d\hat{\omega}_y}{dy}\right)$$

$$\hat{\omega}_z = -I\frac{\kappa_x\hat{\varphi} + \kappa_z\frac{d\hat{\omega}_y}{dy}}{\kappa_x^2 + \kappa_z^2} \tag{A.18}$$

Substituting $\hat{u}$ into equation A.14 leads to

$$\hat{\omega}_y = -\kappa_z \frac{\kappa_x \frac{d\hat{v}}{dy} - \kappa_z\hat{\omega}_y}{\kappa_x^2 + \kappa_z^2} - I\kappa_x \hat{w}$$

$$\Leftrightarrow \quad \hat{w} = I\frac{\kappa_z\frac{d\hat{v}}{dy} + \kappa_x\hat{\omega}_y}{\kappa_x^2 + \kappa_z^2} \tag{A.19}$$

Using the relation ($\phi = \frac{\partial \omega_z}{\partial x} - \frac{\partial \omega_x}{\partial y}$) in spectral space gives

$$\hat{\phi} = I\kappa_x \hat{\omega}_z - I\kappa_z \hat{\omega}_x$$

$$\Leftrightarrow \quad \hat{\omega}_x = \frac{\kappa_x}{\kappa_z}\hat{\omega}_z - \frac{1}{I\kappa_z}\hat{\phi}$$

$$\overset{(A.18)}{=} I\frac{-\kappa_x^2\hat{\phi} - \kappa_x\kappa_z\frac{d\hat{\omega}_y}{dy} + (\kappa_x^2 + \kappa_z^2)\hat{\phi}}{\kappa_z(\kappa_x^2 + \kappa_z^2)}$$

$$\hat{\omega}_x = I\frac{\kappa_z\hat{\phi} - \kappa_x\frac{d\hat{\omega}_y}{dy}}{\kappa_x^2 +^2} \tag{A.20}$$

## A.4. Vector Identities

For any scalar $\phi$ the curl of a gradient is equals to zero.

$$\nabla \times (\nabla \phi) = 0 \tag{A.21}$$

$$\frac{1}{2}(\mathbf{a} \cdot \mathbf{a}) = (\mathbf{a} \cdot \nabla)\mathbf{a} + \mathbf{a} \times (\nabla \times \mathbf{a}) \tag{A.22}$$

$$\nabla \times (\mathbf{a} \times \mathbf{b}) = \mathbf{a}(\nabla \cdot \mathbf{b}) - \mathbf{b}(\nabla \cdot \mathbf{a}) + (\mathbf{b} \cdot \nabla)\mathbf{a} - (\mathbf{a} \cdot \nabla)\mathbf{b} \tag{A.23}$$

## A.5. Reynolds Stress Tensor for Plane Channel Flow

The Reynolds stress tensor in general looks as follows

$$\mathbf{R} = \begin{bmatrix} \langle u'u' \rangle & \langle u'v' \rangle & \langle u'w' \rangle \\ \langle v'u' \rangle & \langle v'v' \rangle & \langle v'w' \rangle \\ \langle w'u' \rangle & \langle w'v' \rangle & \langle w'w' \rangle \end{bmatrix} \tag{A.24}$$

For a plane channel flow with statistical homogeneity w.r.t. the stream- and spanwise direction the following condition for the velocity PDF $f(V_1, V_2, V_3; x, y, z)$ holds (cf. equation 2.35):

$$\frac{\partial f}{\partial x} = \frac{\partial f}{\partial z} = 0 \tag{A.25}$$

Furthermore the flow is statistically invariant under reflections of the z coordinate axis implying

$$f(V_1, V_2, V_3; x, y, z) = f(V_1, V_2, -V_3; x, y, -z) \tag{A.26}$$

Computing the mean spanwise velocity component at $z = 0$ using equation 2.21 yields

$$\langle w \rangle (x, y, z = 0) = \int_{-\infty}^{\infty} V_3 f(V_1, V_2, V_3, x, y, z = 0) dV_3 \tag{A.27}$$

$$= \int_{-\infty}^{\infty} -V_3 \underbrace{f(V_1, V_2, -V_3, x, y, z = 0)}_{\overset{(A.26)}{=} f(V_1, V_2, V_3; x, y, z=0)} dV_3 \tag{A.28}$$

Adding $\int_{-\infty}^{\infty} V_3 f(V_1, V_2, V_3, x, y, z = 0) dV_3$ on both sides of the equation yields

$$2 \int_{-\infty}^{\infty} V_3 f(V_1, V_2, V_3, x, y, z = 0) dV_3 = 0 \tag{A.29}$$

$$\Leftrightarrow \langle w \rangle(x, y, z = 0) = 0 \tag{A.30}$$

Finally applying statistical homogeneity condition (equation A.25) leads to:

$$\langle w \rangle(x, y, z) = 0 \tag{A.31}$$

The correlation term $\langle u'w' \rangle$ at $z = 0$ yields

$$\langle u'w' \rangle(x, y, z = 0) = \int_{-\infty}^{\infty} \int_{-\infty}^{\infty} (V_1 - \langle u \rangle)(V_3 - \langle w \rangle) f(V_1, V_2, V_3, x, y, z = 0) dV_1 dV_3 \tag{A.32}$$

$$\overset{(A.31)}{=} \int_{-\infty}^{\infty} \int_{-\infty}^{\infty} (V_1 - \langle u \rangle) V_3 f(V_1, V_2, V_3, x, y, z = 0) dV_1 dV_3 \tag{A.33}$$

$$= \int_{-\infty}^{\infty} \int_{-\infty}^{\infty} (V_1 - \langle u \rangle)(-V_3) \underbrace{f(V_1, V_2, -V_3, x, y, z = 0)}_{\overset{(A.26)}{=} f(V_1, V_2, V_3; x, y, z = 0)} dV_1 dV_3 \tag{A.34}$$

Now $\int_{-\infty}^{\infty} \int_{-\infty}^{\infty} (V_1 - \langle u \rangle) V_3 f(V_1, V_2, V_3, x, y, z = 0) dV_1 dV_3$ on both sides of the equation is added:

$$2 \int_{-\infty}^{\infty} \int_{-\infty}^{\infty} (V_1 - \langle u \rangle) V_3 f(V_1, V_2, V_3, x, y, z = 0) dV_1 dV_3 = 0 \tag{A.35}$$

$$\Leftrightarrow \langle u'w' \rangle(x, y, z = 0) = 0 \tag{A.36}$$

Using equation A.25 again yields

$$\langle u'w' \rangle(x, y, z) = 0. \tag{A.37}$$

The same can be shown for the term $\langle v'w' \rangle$. Furthermore $\langle v'u' \rangle = \langle u'v' \rangle$, $\langle w'u' \rangle = \langle u'w' \rangle$ and $\langle w'v' \rangle = \langle v'w' \rangle$ due to symmetry, so that the Reynolds stress tensor simplifies to

$$\mathbf{R} = \begin{bmatrix} \langle u'u' \rangle & \langle u'v' \rangle & 0 \\ \langle u'v' \rangle & \langle v'v' \rangle & 0 \\ 0 & 0 & \langle w'w' \rangle \end{bmatrix} \tag{A.38}$$

## A.6. Figures of Flow Statistics

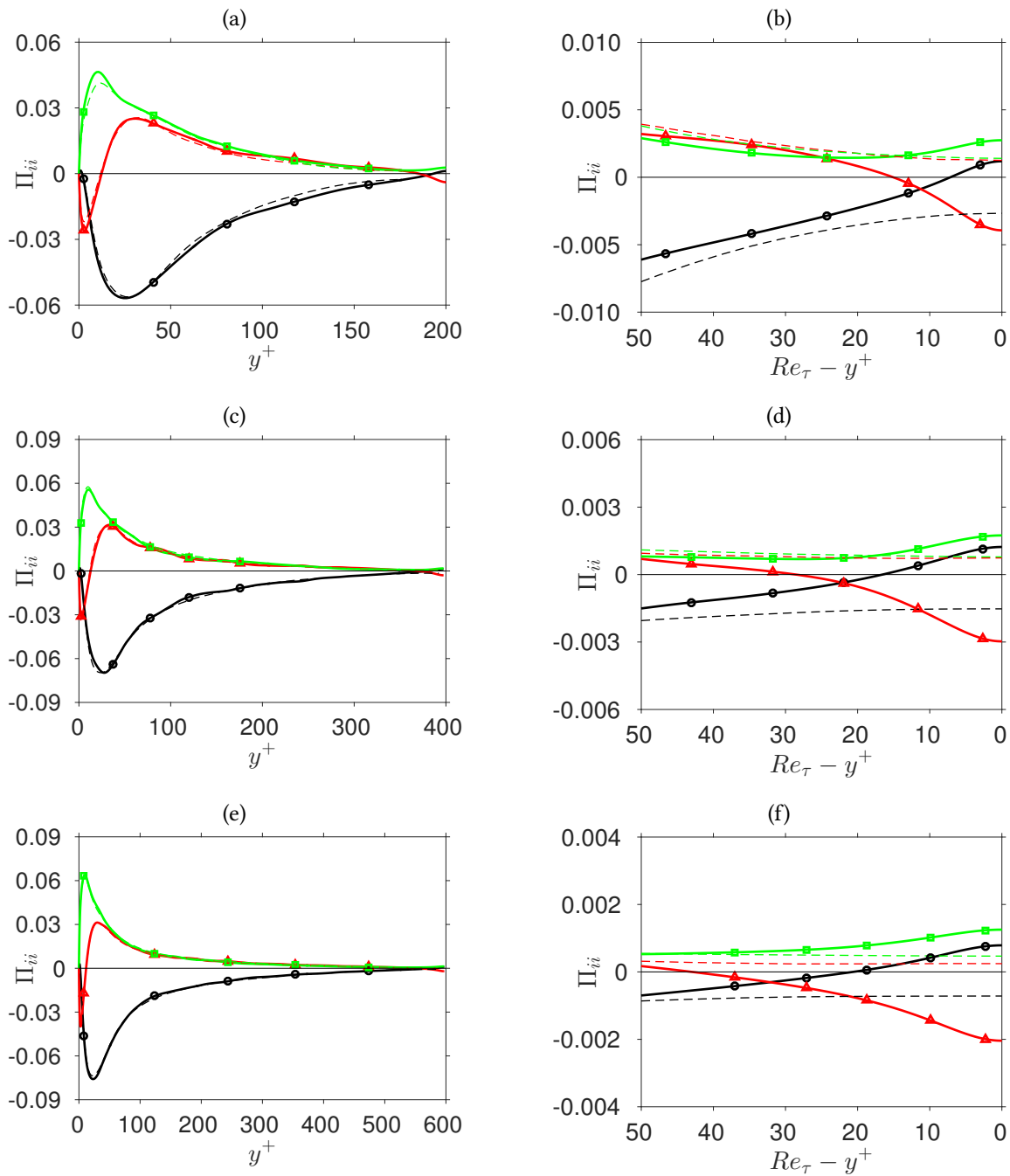

Figure A.1.: Diagonal components of the pressure-strain term: **o**, $\Pi_{11}$; $\triangle$, $\Pi_{22}$; $\square$, $\Pi_{33}$. Left column contains data of the full channel height wheras the right column shows surface near data. (a),(b): $Re_\tau = 200$; (c),(d): $Re_\tau = 400$; (e),(f): $Re_\tau = 600$. Dashed lines indicate closed channel data from Moser et al. (1999).

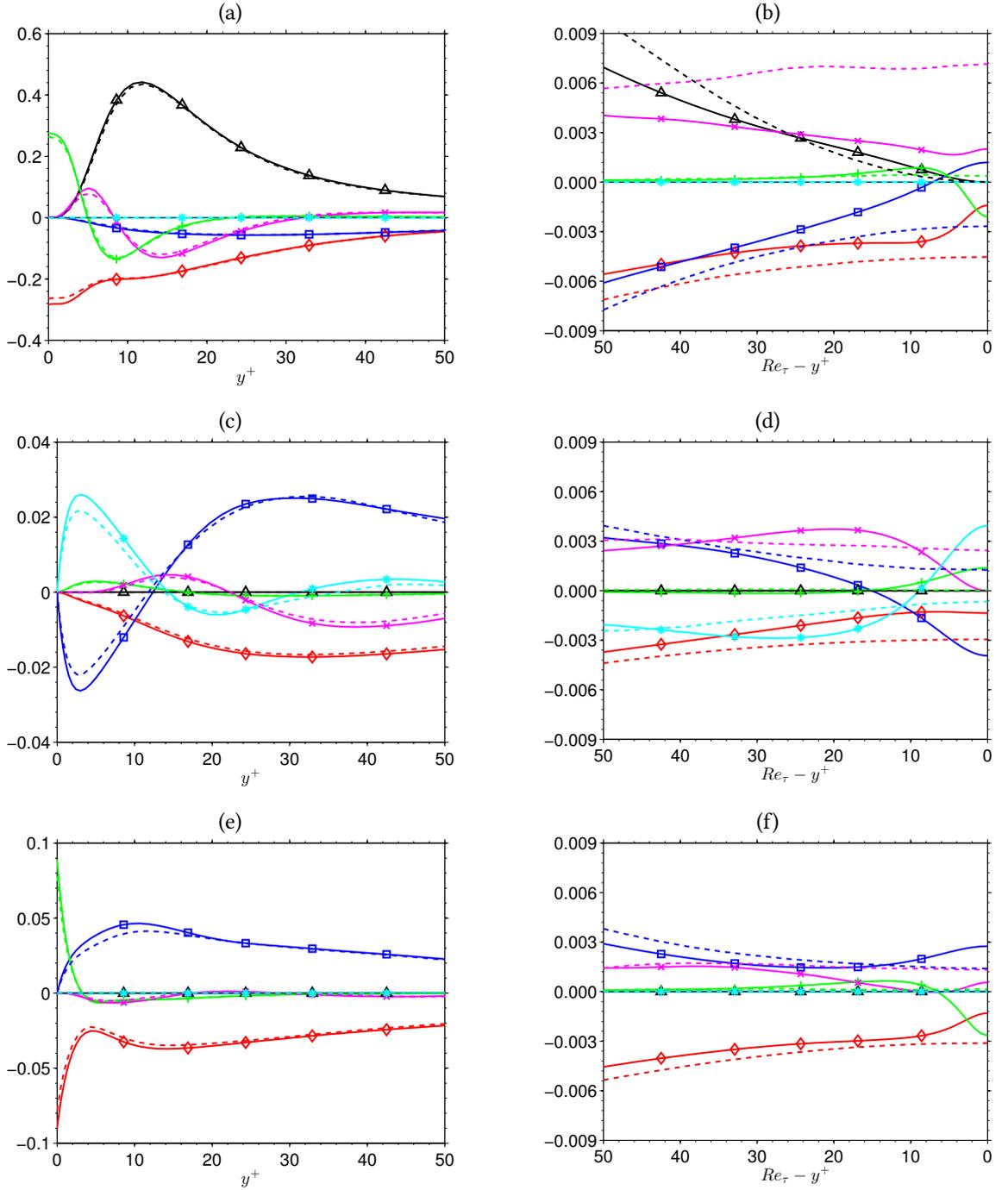

Figure A.2.: Budget terms for normal Reynolds stress transport equations for $Re_\tau = 200$: $\triangle$, production term $\mathcal{P}_{ii} = -2\langle u_i' v' \rangle \frac{d\langle u_i \rangle}{dy}$; $\diamond$, dissipation term $-\varepsilon_{ii} = -2\nu \langle \frac{\partial u_i}{\partial x_k} \frac{\partial u_i}{\partial x_k} \rangle$; $+$, viscous diffusion term $\nu \frac{d^2 \langle u_i' u_i' \rangle}{dy}$; $\times$, turbulent convection term $\frac{d}{dy} \langle v' u_i' u_i' \rangle$; $*$, pressure diffusion term $\frac{2}{\rho} \frac{d \langle u_i' p' \rangle}{dx_i}$; $\square$, pressure-strain term $\frac{2}{\rho} \langle p' \frac{\partial u_i'}{\partial x_i} \rangle$. Terms are normalized by viscous scales. Dashed lines indicate values from corresponding closed channel cases of Moser et al. (1999). (a,c,e): Near-wall profiles in wall units; (b,d,f): Near-surface profiles, the x-axis shows the distance from the free surface or the channel centerline respectively in wall units ($Re_\tau - y^+$).

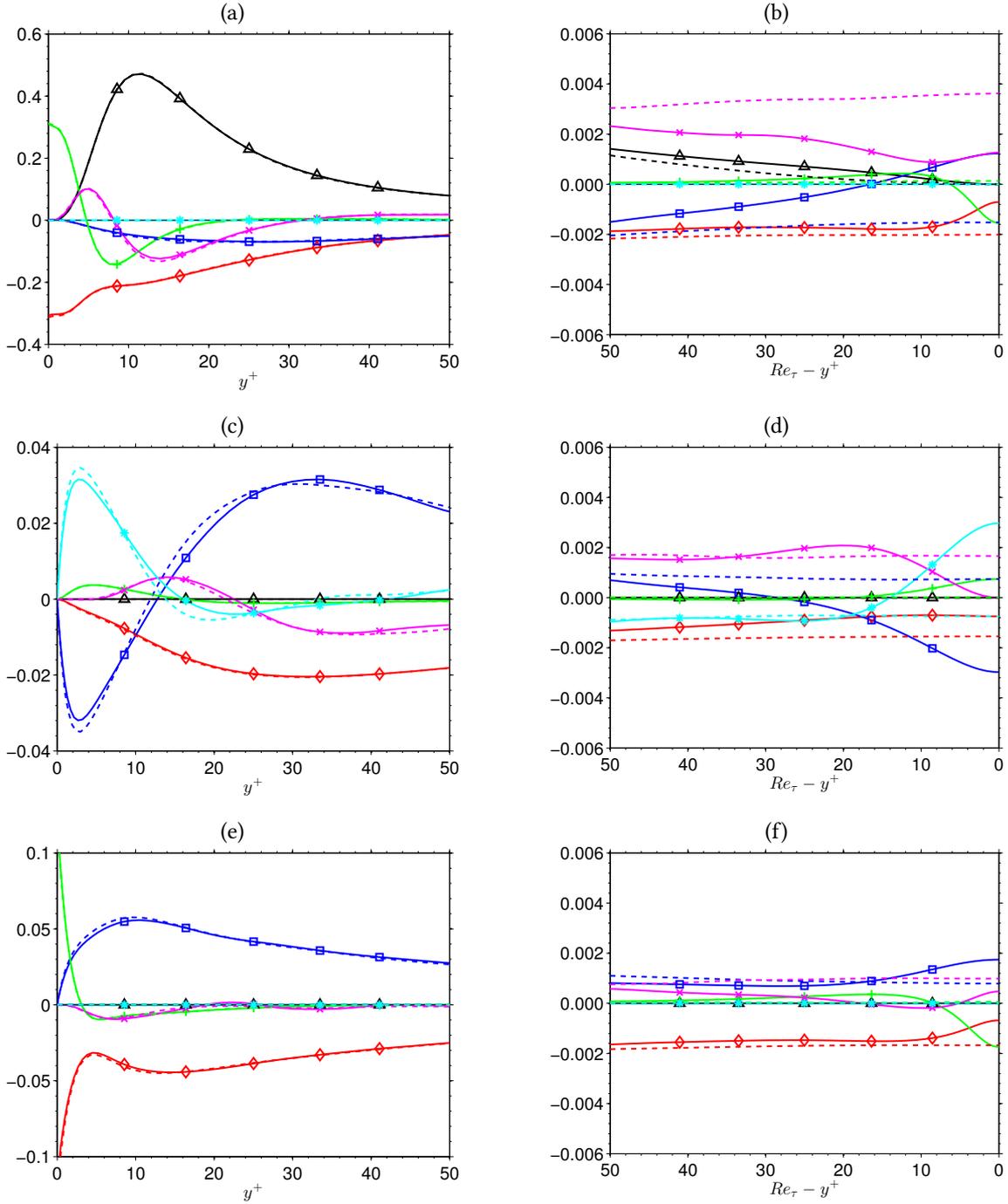

Figure A.3.: Budget terms for normal Reynolds stress transport equations for $Re_\tau = 400$:
△, production term $\mathcal{P}_{ii} = -2\langle u'_i v'\rangle \frac{d\langle u_i\rangle}{dy}$; ◇, dissipation term $-\varepsilon_{ii} = -2\nu\langle \frac{\partial u_i}{\partial x_k}\frac{\partial u_i}{\partial x_k}\rangle$;
+, viscous diffusion term $\nu\frac{d^2\langle u'_i u'_i\rangle}{dy}$; ×, turbulent convection term $\frac{d}{dy}\langle v'u'_i u'_i\rangle$;
∗, pressure diffusion term $\frac{2}{\rho}\frac{d\langle u'_i p'\rangle}{dx_i}$; □, pressure-strain term $\frac{2}{\rho}\langle p'\frac{\partial u'_i}{\partial x_i}\rangle$. Terms are normalized by viscous scales. Dashed lines indicate values from corresponding closed channel cases of Moser et al. (1999). (a,c,e): Near-wall profiles in wall units; (b,d,f): Near-surface profiles, the x-axis shows the distance from the free surface or the channel centerline respectively in wall units ($Re_\tau - y^+$).

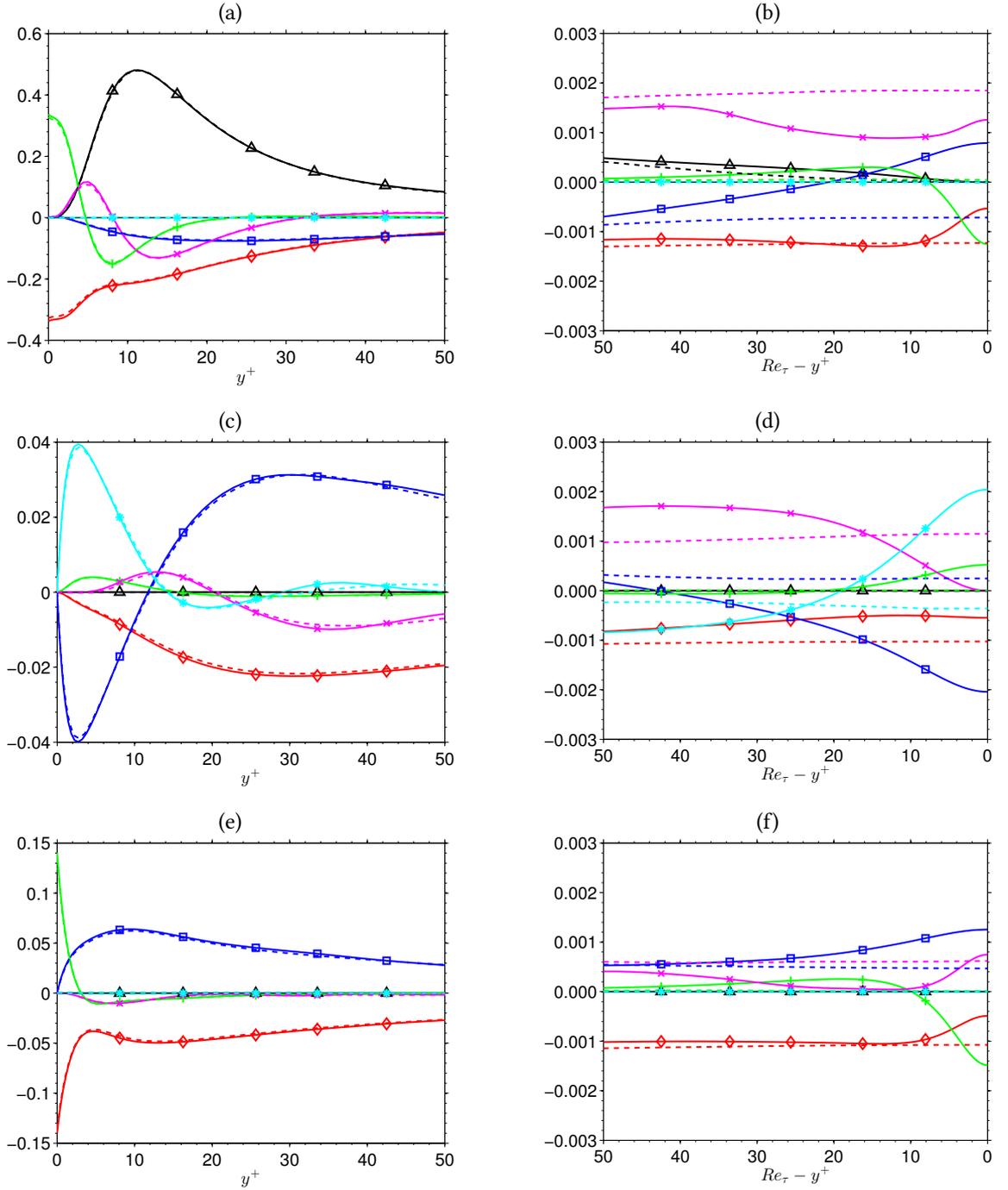

Figure A.4.: Budget terms for normal Reynolds stress transport equations for $Re_\tau = 600$: $\triangle$, production term $\mathcal{P}_{ii} = -2\langle u_i' v' \rangle \frac{d\langle u_i \rangle}{dy}$; $\diamond$, dissipation term $-\varepsilon_{ii} = -2\nu\langle \frac{\partial u_i}{\partial x_k} \frac{\partial u_i}{\partial x_k} \rangle$; $+$, viscous diffusion term $\nu\frac{d^2\langle u_i' u_i' \rangle}{dy}$; $\times$, turbulent convection term $\frac{d}{dy}\langle v' u_i' u_i' \rangle$; $\ast$, pressure diffusion term $\frac{2}{\rho}\frac{d\langle u_i' p' \rangle}{dx_i}$; $\square$, pressure-strain term $\frac{2}{\rho}\langle p' \frac{\partial u_i'}{\partial x_i} \rangle$. Terms are normalized by viscous scales. Dashed lines indicate values from corresponding closed channel cases of Moser et al. (1999). (a,c,e): Near-wall profiles in wall units; (b,d,f): Near-surface profiles, the x-axis shows the distance from the free surface or the channel centerline respectively in wall units ($Re_\tau - y^+$).

## A.7. Figures of Premultiplied Energy Spectra

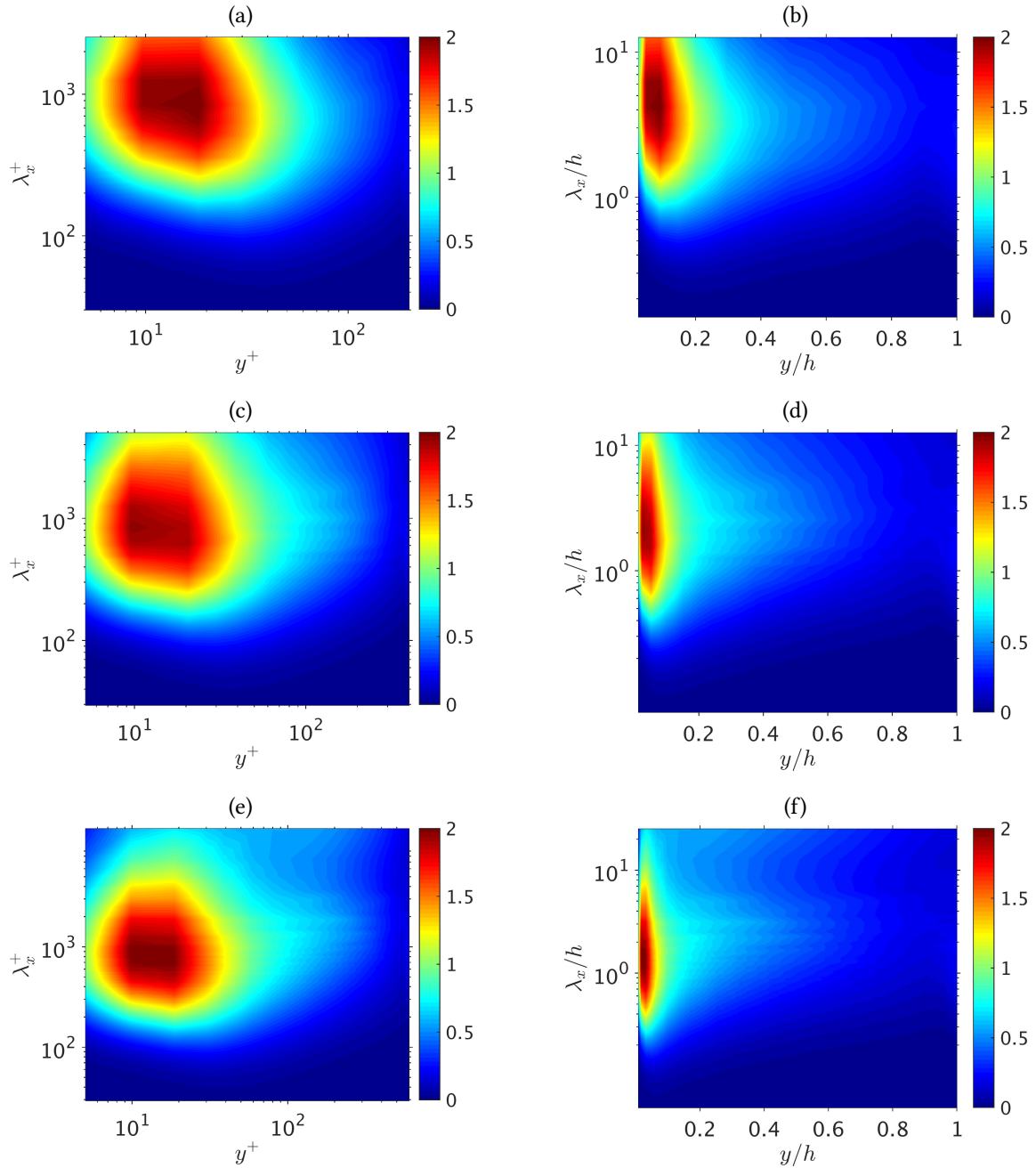

Figure A.5.: Contour maps of premultiplied one-dimensional streamwise energy spectra of streamwise velocity fluctuations $\kappa_x \phi_{uu}(\kappa_x)/u_\tau$ normalized with viscous units as a function of streamwise wavelength $\lambda_x$ and distance from the wall y. (a,b): $Re_\tau = 200$, (c,d): $Re_\tau = 400$, (e,f): $Re_\tau = 600$; (a,c,e): wall units, (b,e,f): outer flow units.

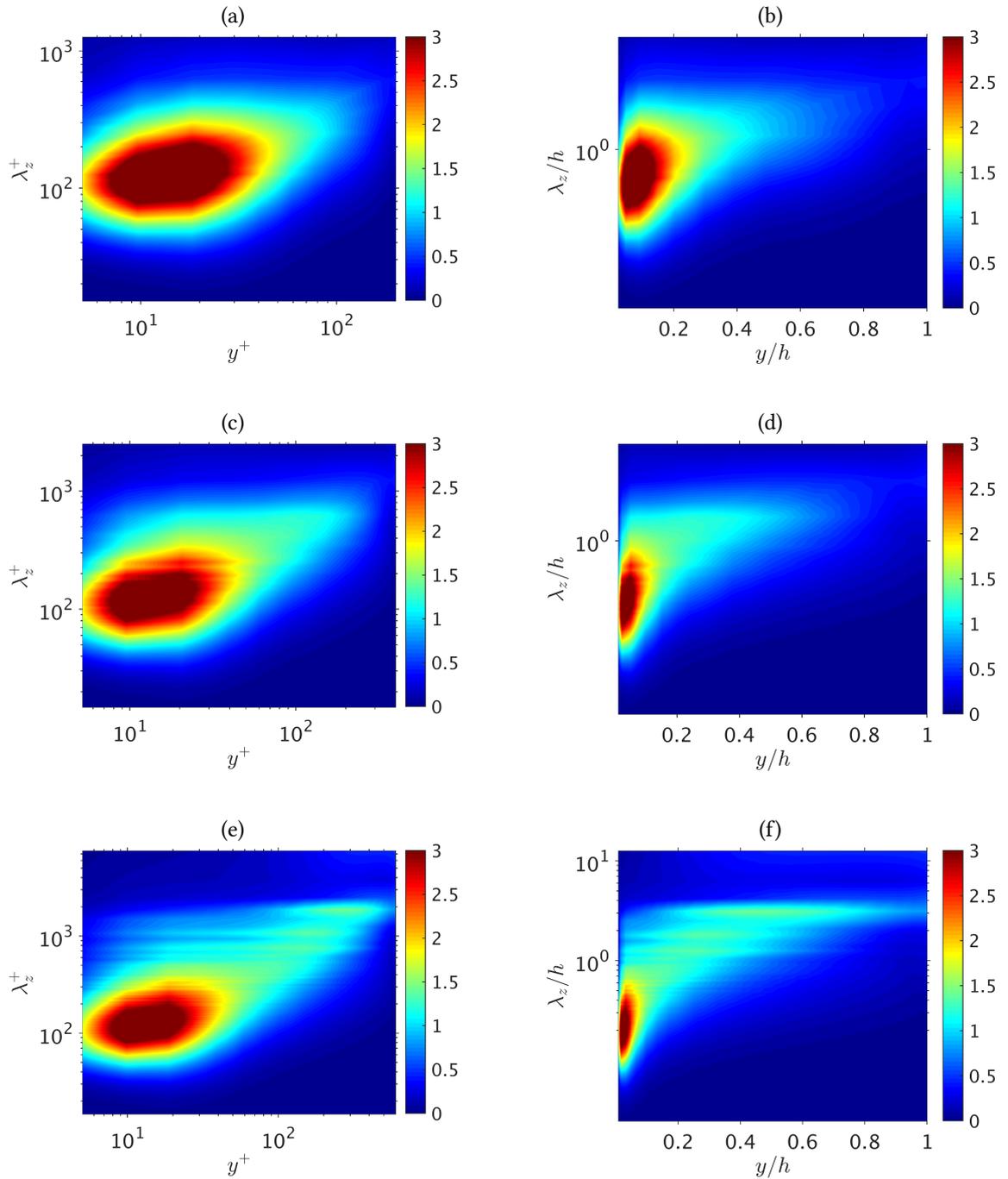

Figure A.6.: Contour maps of premultiplied one-dimensional spanwise energy spectra of streamwise velocity fluctuations $\kappa_z \phi_{uu}(\kappa_z)/u_\tau$ normalized with viscous units as a function of spanwise wavelength $\lambda_z$ and distance from the wall y. (a,b): $Re_\tau = 200$, (c,d): $Re_\tau = 400$, (e,f): $Re_\tau = 600$; (a,c,e): wall units, (b,e,f): outer flow units.

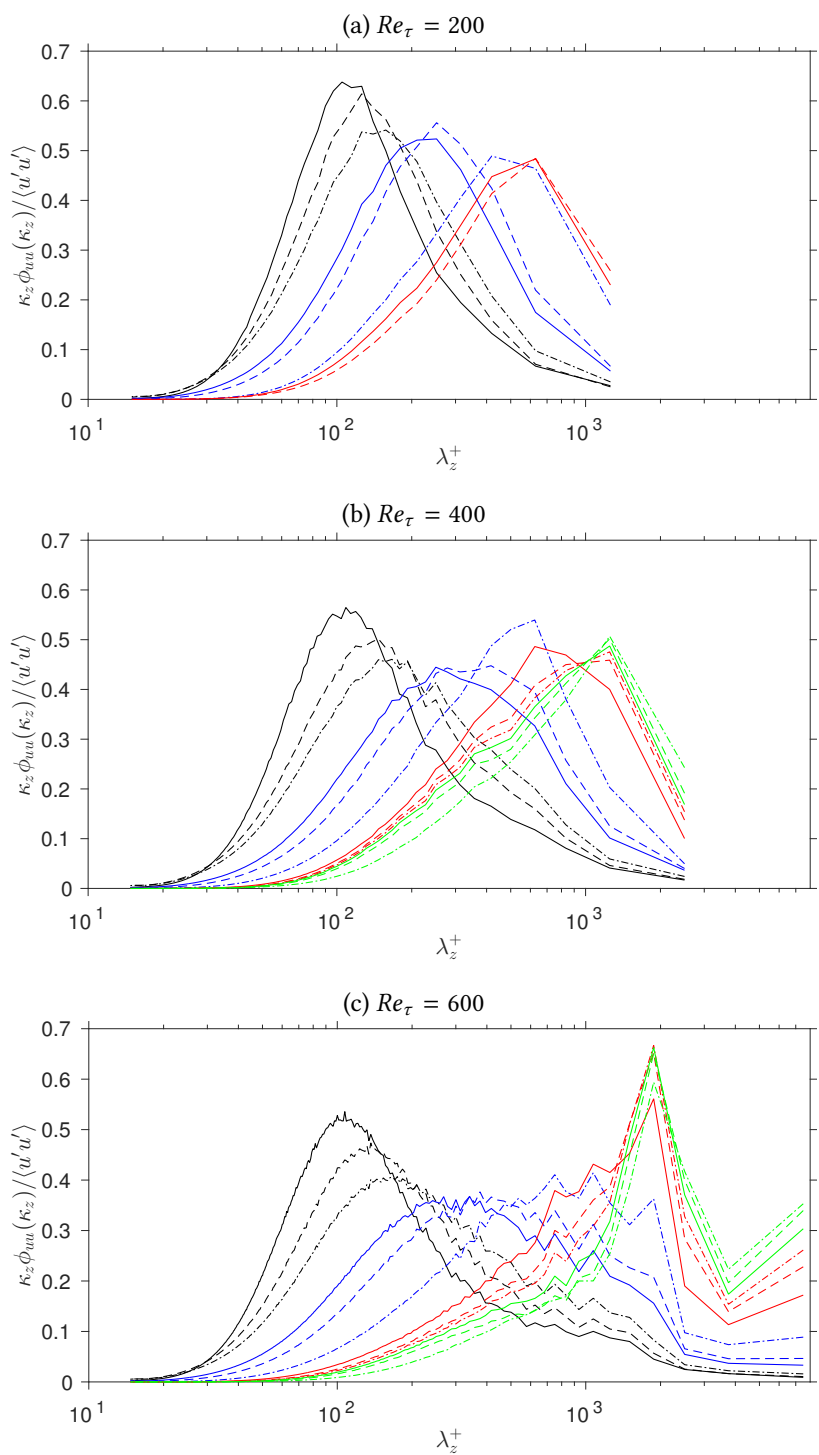

Figure A.7.: One-dimensional premultiplied spanwise power spectra $\kappa_z \phi_{uu}(\kappa_z)/\langle u'u'\rangle$ as a function of $\lambda_z^+$ at different $y^+$ positions.

(a): —, $y^+$=5; - -, $y^+$=18; ---, $y^+$=29; —, $y^+$=60; - -, $y^+$=78; -·-, $y^+$=98; —, $y^+$=178; - -, $y^+$=189; -·-, $y^+$=200.

(b): —, $y^+$=5; - -, $y^+$=20; -·-, $y^+$=30; —, $y^+$=61; - -, $y^+$=80; -·-, $y^+$=150; —, $y^+$=298; - -, $y^+$=337; -·-, $y^+$=348; —, $y^+$=356; - -, $y^+$=367; -·-, $y^+$=397.

(c): —, $y^+$=5; - -, $y^+$=19; ---, $y^+$=30; —, $y^+$=59; - -, $y^+$=80; -·-, $y^+$=151; —, $y^+$=299; - -, $y^+$=406; -·-, $y^+$=455; —, $y^+$=505; - -, $y^+$=543; -·-, $y^+$=597.

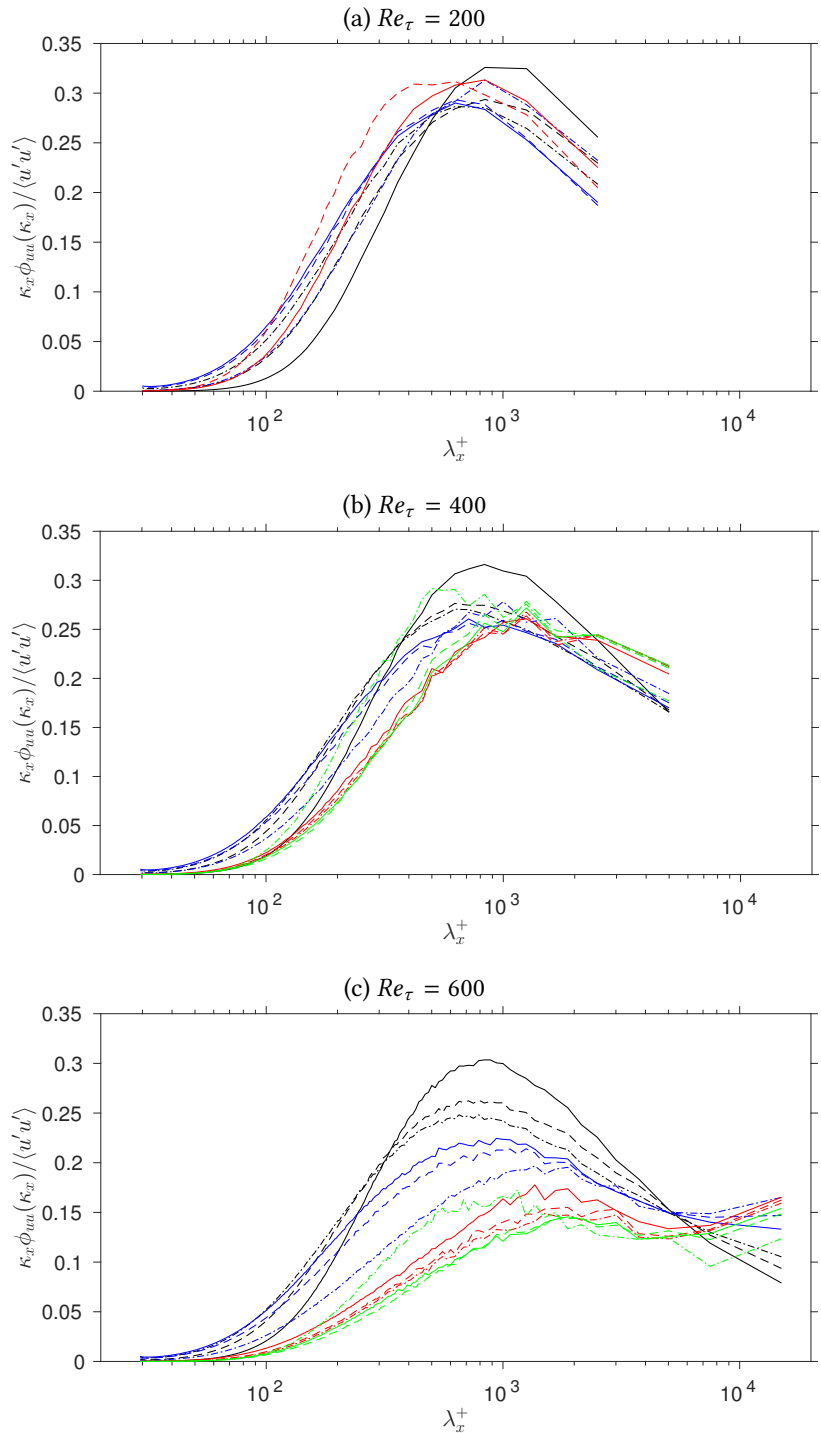

Figure A.8.: One-dimensional premultiplied streamwise power spectra $\kappa_z \phi_{uu}(\kappa_z)/\langle u'u'\rangle$ as a function of $\lambda_z^+$ at different $y^+$ positions.

(a): —, $y^+$=5; - -, $y^+$=18; ---, $y^+$=29; —, $y^+$=60; - -, $y^+$=78; ---, $y^+$=98; —, $y^+$=178; - -, $y^+$=189; ---, $y^+$=200.

(b): —, $y^+$=5; - -, $y^+$=20; ---, $y^+$=30; —, $y^+$=61; - -, $y^+$=80; ---, $y^+$=150; —, $y^+$=298; - -, $y^+$=337; ---, $y^+$=348; —, $y^+$=356; - -, $y^+$=367; ---, $y^+$=397.

(c): —, $y^+$=5; - -, $y^+$=19; ---, $y^+$=30; —, $y^+$=59; - -, $y^+$=80; ---, $y^+$=151; —, $y^+$=299; - -, $y^+$=406; ---, $y^+$=455; —, $y^+$=505; - -, $y^+$=543; ---, $y^+$=597.